\pdfoutput=1

\documentclass[11pt,twoside,a4paper,cmspaper,final,collab]{cms-tdr}
\def\svnVersion{182562}\def\cmsCernNo{2012-321}\def\cmsCernDate{\today}\def\cmsMessage{Submitted to the Journal of High Energy Physics}

\begin{document}\cmsNoteHeader{SUS-12-001}

\hyphenation{had-ron-i-za-tion}
\hyphenation{cal-or-i-me-ter}
\hyphenation{de-vices}

\RCS$Revision: 165256 $
\RCS$HeadURL: svn+ssh://svn.cern.ch/reps/tdr2/papers/SUS-12-001/trunk/SUS-12-001.tex $
\RCS$Id: SUS-12-001.tex 165256 2013-01-17 02:24:53Z paulini $
\newlength\cmsFigWidth
\ifthenelse{\boolean{cms@external}}{\setlength\cmsFigWidth{0.85\columnwidth}}{\setlength\cmsFigWidth{0.4\textwidth}}
\ifthenelse{\boolean{cms@external}}{\providecommand{\cmsLeft}{top}}{\providecommand{\cmsLeft}{left}}
\ifthenelse{\boolean{cms@external}}{\providecommand{\cmsRight}{bottom}}{\providecommand{\cmsRight}{right}}
\ifthenelse{\boolean{cms@external}}{\providecommand{\suppMaterial}{the supplemental material [URL will be inserted by publisher]}}{\providecommand{\suppMaterial}{Appendix~\ref{app:suppMat}}}
\newcommand{\todo}[1]{{\color{blue}{\textbf{To-do:} #1}}}
\newcommand{\CLs}{CL$_{\mbox{s}}$\xspace}
\cmsNoteHeader{SUS-12-001} 
\title{Search for new physics in events with photons, jets, and missing
  transverse energy in pp collisions at $\sqrt{s}=7$\TeV}

\date{\today}

\abstract{
  A search for physics beyond the standard model involving events with one
  or more photons, jets, and missing transverse energy has been
  performed by the CMS experiment.  The data sample corresponds to an
  integrated luminosity of 4.93\fbinv of proton-proton collisions at
  $\sqrt{s}=7$\TeV, produced at the Large Hadron Collider.  No excess
  of events with large missing transverse energy is observed beyond
  expectations from standard model processes, and upper limits on the
  signal production cross sections for new physics processes are set at the 95\%
  confidence level.  The results of this search are
  interpreted in the context of three models of new physics: a general
  model of gauge-mediated supersymmetry breaking, Simplified
  Models, and a theory involving universal extra
  dimensions. In the absence of evidence for new physics,
  exclusion regions are derived in the parameter spaces of the
  respective models.}

\hypersetup{%
pdfauthor={CMS Collaboration},%
pdftitle={Search for new physics in events with photons, jets, and missing
  transverse energy in pp collisions at sqrt(s) = 7 TeV},%
pdfsubject={CMS},%
pdfkeywords={CMS, physics}}

\maketitle 

\section{Introduction}
\label{sec:introduction}

The standard model (SM) of particle physics is a very successful theory
describing existing experimental data. However it is not expected to
describe physics up to the Planck scale, because of the extreme fine
tuning required to control particle masses (hierarchy
problem)~\cite{Wess:1974tw,Cheng:2003ju,Appelquist:2000nn}, nor
does it provide an explanation for dark matter. These
issues with the SM motivate a broad program of searches for physics
beyond the SM. Among the theories proposing physics beyond the SM,
supersymmetry (SUSY) is of particular interest as it resolves these
problems by introducing a symmetry between fermions and bosons resulting
in a superpartner (sparticle) for each SM particle with identical
quantum numbers except spin. Since no sparticles have been found yet,
SUSY must be a broken symmetry with the masses of the supersymmetric
particles being heavier than their SM partners.  The version of
supersymmetry based on general gauge-mediated (GGM) SUSY
breaking~\cite{GGMa,GGMd1,GGMd2,GGMd3,GGMd4,GGMd5,GGMd} is of particular
theoretical interest for new physics as it not only stabilizes the mass
of the SM Higgs boson and drives the grand unification of forces, but also
avoids the large flavor-changing neutral currents that trouble other
SUSY-breaking scenarios.  Another extension to the SM is the theory of
universal extra dimensions (UED)~\cite{Appelquist:2000nn}, which
predicts additional compactified dimensions beyond the regular four
space-time dimensions of the SM. These extra dimensions (ED), which are
accessible to standard model fields, could allow gauge coupling
unification and provide new mechanisms for the generation of fermion
mass hierarchies.

This paper describes a search for events with two signatures containing
photons, which may indicate new-physics processes in a variety of
theoretical scenarios including GGM supersymmetry and UED. Final states
with photons are experimentally interesting as photons can be identified
with relatively high purity and efficiency with the Compact Muon
Solenoid (CMS) detector. The first signature studied consists of at
least one isolated photon with high energy measured in the plane
transverse to the beam direction (\ET), at least two hadronic jets, and
large missing transverse energy~(\MET). The second signature is
characterized by at least two isolated photons with high \ET, at least
one jet, and large \MET. This search is based on a data sample recorded
with the CMS experiment corresponding to an integrated luminosity of
$4.93\pm0.11$\fbinv of \Pp\Pp~collisions at $\sqrt{s}=7\TeV$ produced
at the Large Hadron Collider (LHC).

The organization of this paper is as follows. This introductory section
is followed in Section ~\ref{sec:theory} by a discussion of the
theoretical framework used for the interpretation of this search, and
then in Section~\ref{sec:detector} by a description of the CMS
detector. The event selection criteria are detailed in
Section~\ref{sec:selection} and the description of the simulated samples
is given in Section~\ref{sec:MC}.  The methodology to estimate
backgrounds is explained in Section~\ref{sec:backgrounds}.
Sections~\ref{sec:single_photon} and \ref{sec:diphoton} discuss details
of the single-photon and diphoton analyses including the experimental
results.  Section~\ref{sec:GGM} expresses the search results in terms of
exclusion regions in the context of the GGM SUSY scenario, while in
Section~\ref{sec:SMS} and Section~\ref{sec:UED} the results are
interpreted in the context of a final state driven ``simplified'' model,
and universal extra dimensions, respectively. Conclusions are stated in
Section~\ref{sec:conclusions}.

\section{Theoretical Framework}
\label{sec:theory}

The result of this search is interpreted in the context of three models
of new physics. We discuss in this section the theoretical framework
on which these models are based.

\subsection{General Gauge-Mediated Supersymmetry Breaking}

\begin{figure*}[tbp]
\begin{center}
\includegraphics[width=0.45\textwidth]{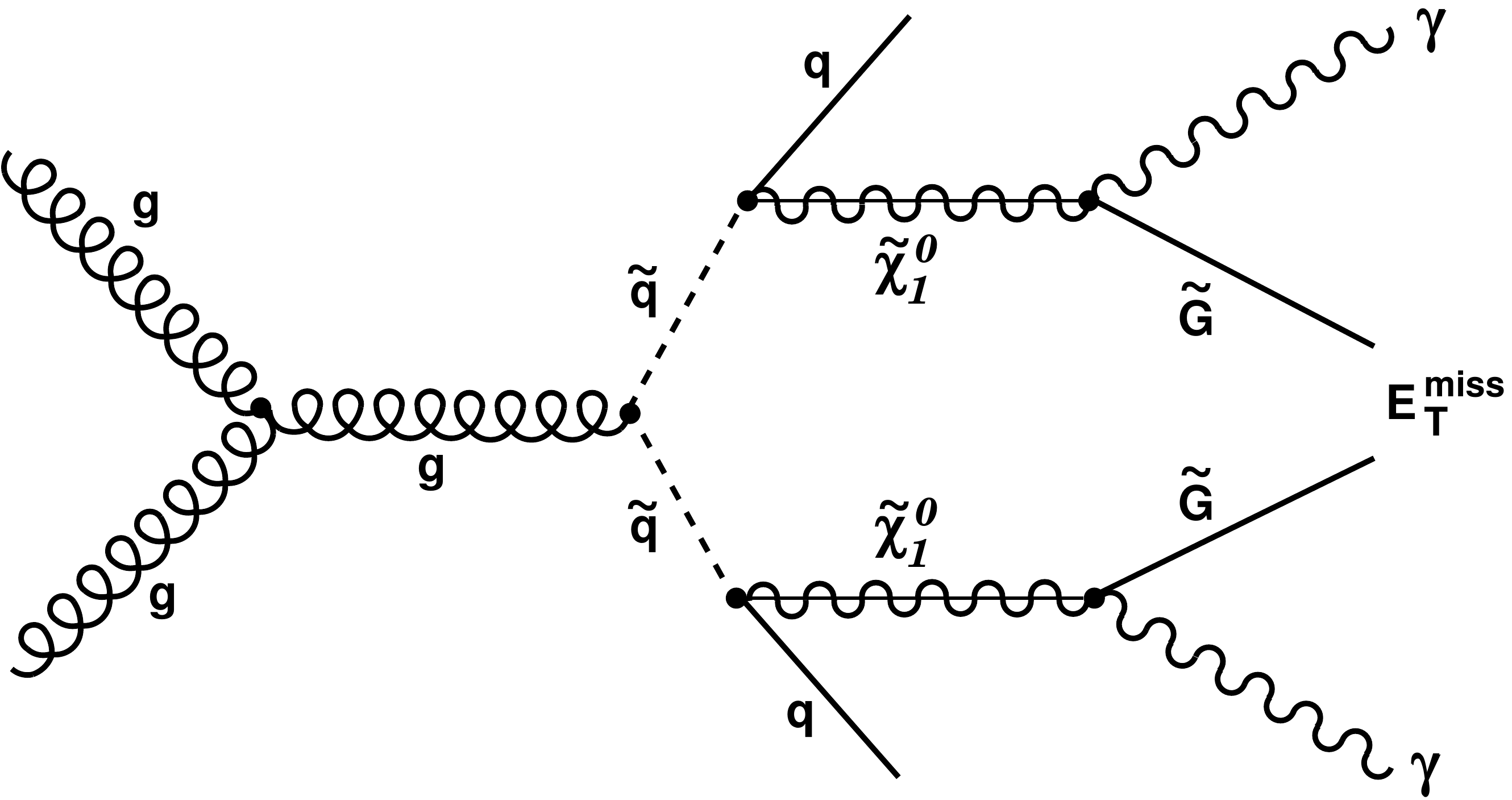}\quad
\includegraphics[width=0.45\textwidth]{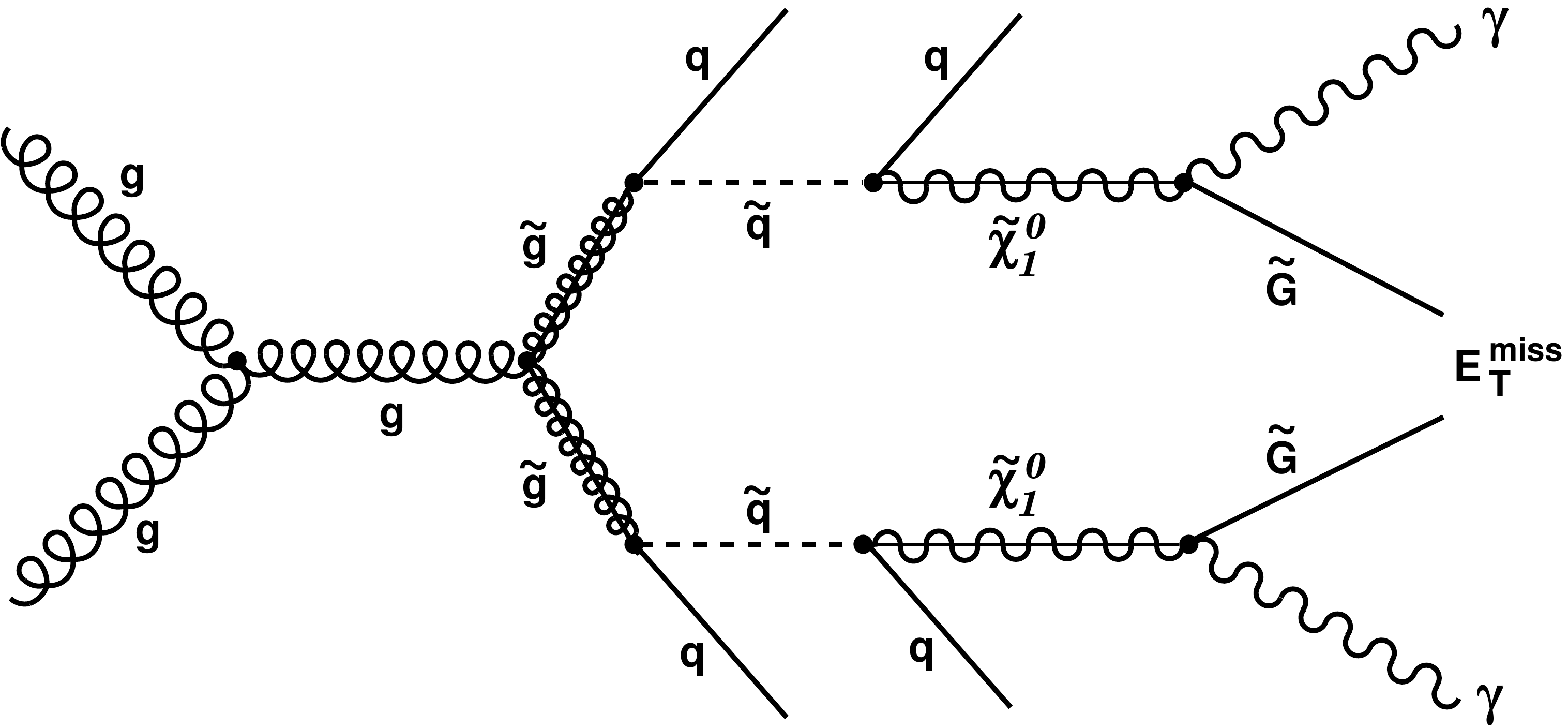}
\\ \vspace*{0.4cm}
\includegraphics[width=0.45\textwidth]{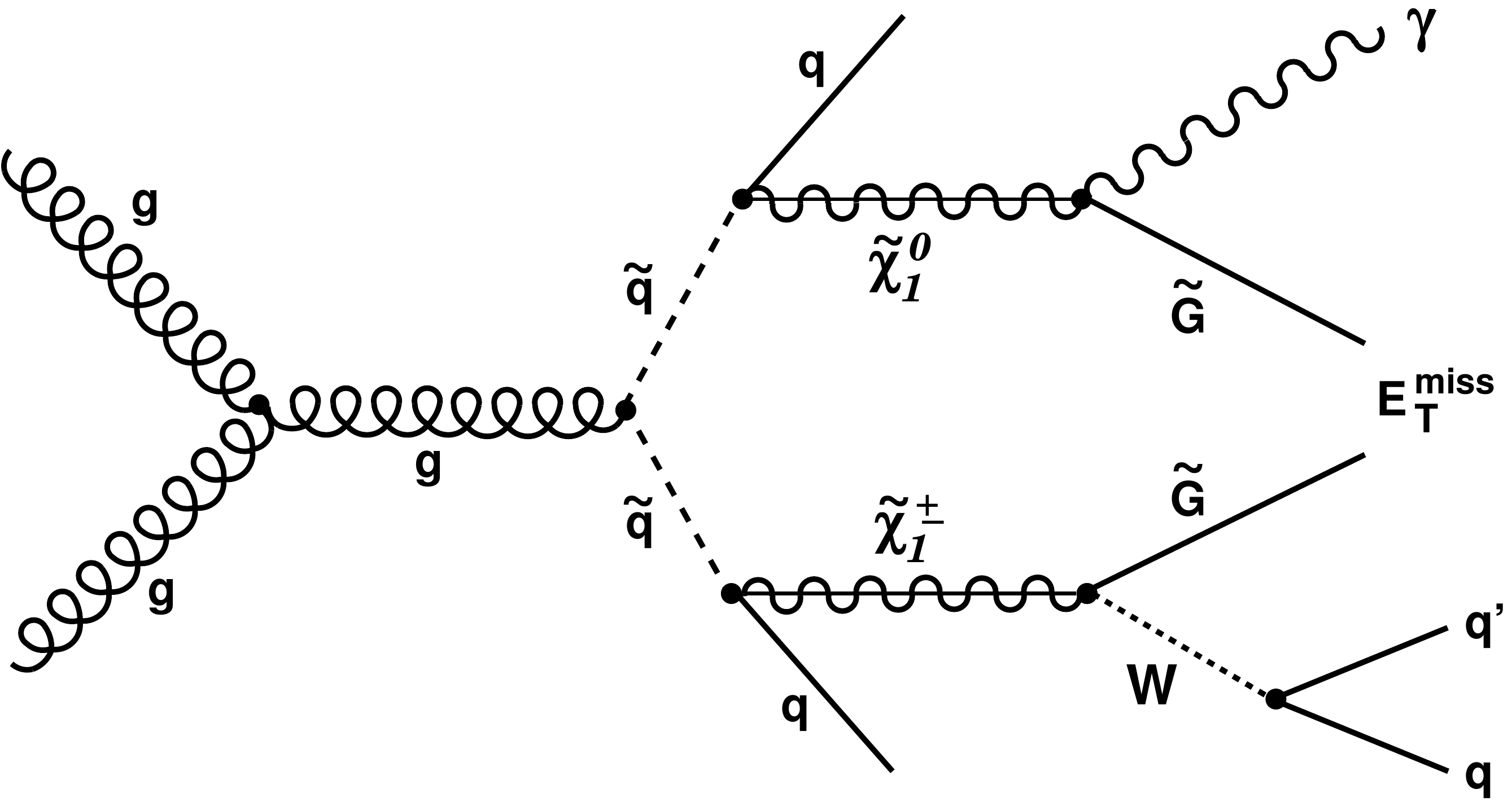}\quad
\includegraphics[width=0.45\textwidth]{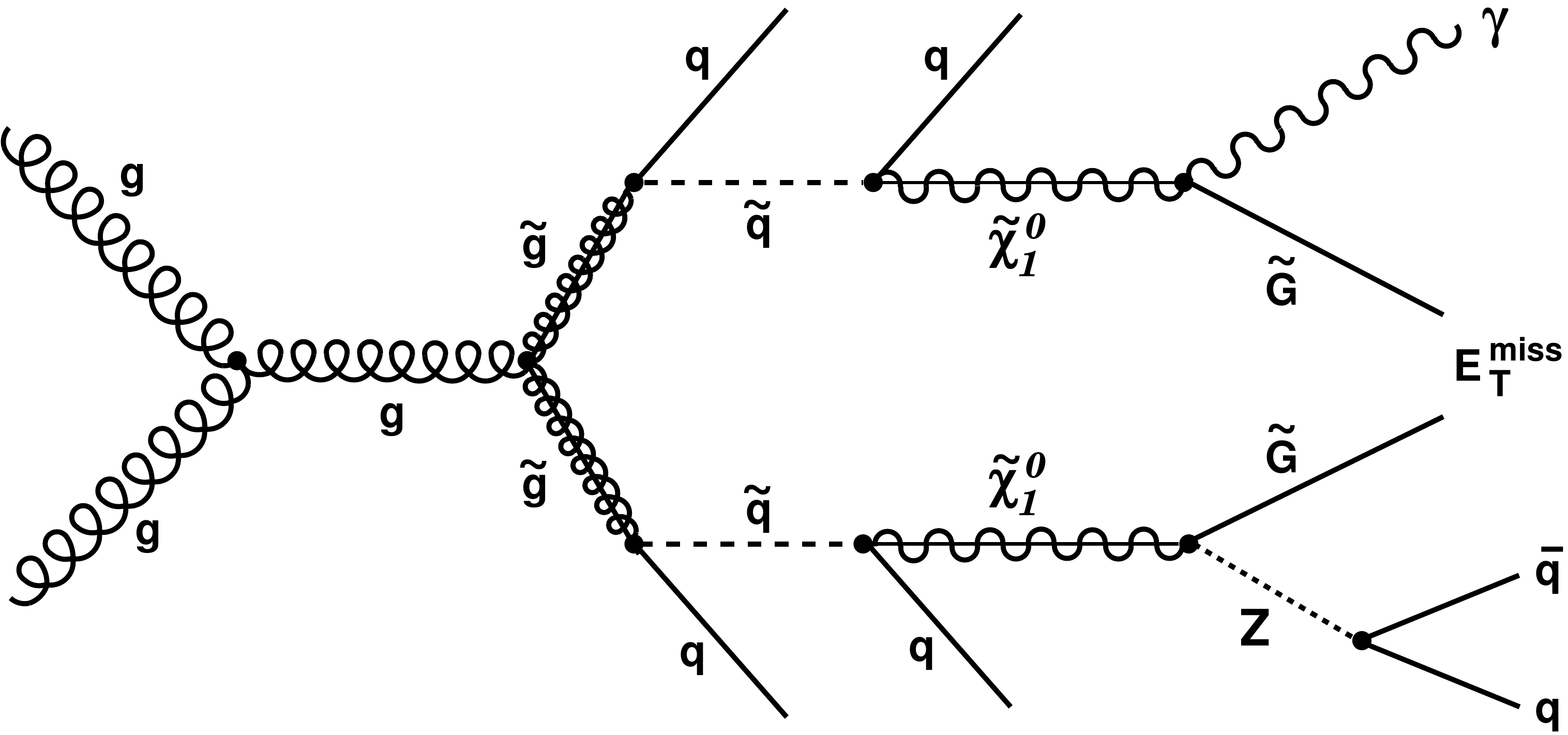}
\caption{\label{fig:Feynmans} Example diagrams for GGM SUSY
processes that result in a diphoton (top) and single-photon (bottom)
final state through squark (left) and gluino (right) production at the LHC.}
\end{center}
\end{figure*}

The first model is a gauge-mediated SUSY
scenario~\cite{Meade:2008wd,Buican:2008ws,Kats:2011qh} in which the
gravitino (\sGra) is the lightest SUSY particle (LSP) and the lightest
neutralino ($\chiz_1$) is the next-to-lightest SUSY particle (NLSP).
The gravitino escapes detection, leading to \MET in the event.  The
neutralino in the GGM models that we consider consists predominantly of
either the bino, the superpartner of the $U(1)$ gauge field, or the
wino, the superpartner of the $SU(2)$ gauge fields.
Assuming that $R$~parity~\cite{Farrar:1978xj} is conserved,
strongly-interacting SUSY particles are pair-produced at the LHC. Their
decay chain includes one or more quarks and gluons, as well as the
neutralino NLSP, which in turn decays into a gravitino and a photon or a
\Z~boson.  Figure~\ref{fig:Feynmans} shows several diagrams of possible
GGM processes that result in a single-photon or diphoton final state, in
squark and gluino pair production processes. If the NLSP is bino-like,
its branching fraction to a photon and gravitino is expected to be
large, leading to an enhancement of the diphoton final state (see
Fig.~\ref{fig:Feynmans} top). If the NLSP is wino-like, its branching
fraction to a photon and gravitino is reduced, leading to a relative
enhancement of the single-photon final state (see
Fig.~\ref{fig:Feynmans} bottom). Therefore we perform searches in both
the single-photon and diphoton final states in order to be sensitive to
models with different NLSP composition.

Table~\ref{tab_topo} provides examples of such GGM decay chains leading
to photons in the final state. The table is divided horizontally between
single-photon and diphoton final states.  The vertical direction
differentiates between bino-NLSP and wino-co-NLSP cases.  The number of
jets produced in the cascades can vary depending on whether gluinos or
squarks are produced, and the species of quarks in the final state.
This search is also sensitive to the scenario in which the NLSP is a pure
wino.  In that case, the lightest chargino ($\chipm_1$) is also a wino,
and the chargino-neutralino mass difference is too small for one to
decay into the other, resulting in the chargino to decay directly into
a gravitino and a \PW~boson (see Fig.~\ref{fig:Feynmans}).  In this
analysis we do not veto on the presence of isolated leptons since in the
wino co-NLSP case we seek to detect the neutralino decays into \Z~bosons
and chargino decays into $\PW^\pm$, both of which decay chains can
result in leptons. The NLSP lifetime is a free parameter in GGM
SUSY. Only prompt neutralino decays are considered in this analysis.

\begin{table*}[btp]
  \topcaption{\label{tab_topo}
    Examples of GGM cascades leading to the topologies of a single photon
    or diphotons in the final state.  }
\begin{center}
\begin{tabular}{c|c|c}
\hline
NLSP type & $\gamma$ + 2 jets + \MET & $\gamma\gamma$ + jet + \MET \\
\hline
Bino-like     & $ \text{jets} + \chiz_1\chiz_1 \rightarrow \text{jets} + \gamma + Z + \sGra\sGra$ &
          $ \text{jets} + \chiz_1\chiz_1 \rightarrow \text{jets} + \gamma\gamma + \sGra\sGra$ \\
\hline
\multirow{2}{*}{Wino-like}      &
            $ \text{jets} + \chiz_1\chiz_1 \rightarrow \text{jets} + \gamma + Z + \sGra\sGra$ &

\multirow{2}{*}{$ \text{jets} + \chiz_1\chiz_1 \rightarrow \text{jets} + \gamma\gamma + \sGra\sGra$} \\

         &   $ \text{jets} + \chiz_1\chipm_1 \rightarrow \text{jets} + \gamma + W^\pm + \sGra\sGra$ & \\
\hline
\end{tabular}
\end{center}
\end{table*}

Previous searches for gauge-mediated SUSY breaking were performed by the
ATLAS experiment with 36\pbinv~\cite{Aad:2011kz},
1.1\fbinv~\cite{Aad:2011zj}, and 4.8\fbinv~\cite{Aad:2012ae} of
\Pp\Pp~collision data, by CMS with 36\pbinv~\cite{Chatrchyan:2011wc},
as well as by experiments at the Tevatron~\cite{Aaltonen:2009tp,Abazov:2010us},
LEP~\cite{Heister:2002vh,Abdallah:2003np,Achard:2003tx,Abbiendi:2005gc},
and HERA~\cite{Aktas:2004cc}.

\subsection{Simplified Models}

\begin{figure*}[tbp]
\begin{center}
\includegraphics[width=0.45\textwidth]{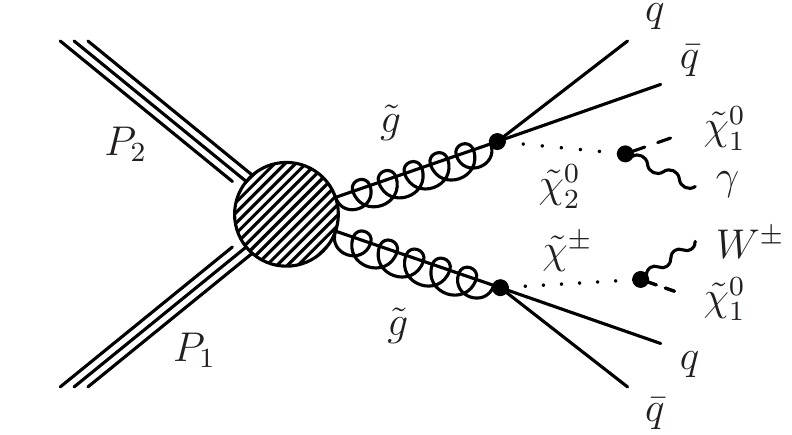}\
\includegraphics[width=0.45\textwidth]{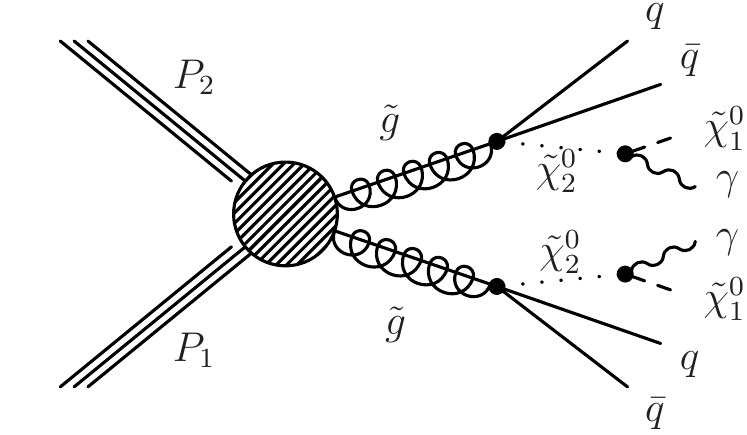}
\caption{\label{fig:SMS_Feynmans} Example diagrams of Simplified Models
  resulting in single-photon (left) and diphoton (right) final states.
}
\end{center}
\end{figure*}

The experimental results of the single-photon and diphoton analyses are
in addition interpreted in the context of Simplified Models
(SMS)~\cite{ArkaniHamed:2007fw,Alwall:2008ag,Alwall:2008va,
  Alves:2011sq,Alves:2011wf,Papucci:2011wy}. In SMS, a limited set of
hypothetical particles and decay chains are introduced to produce a
given topological signature such as the single or diphoton final state
studied in this analysis. The amplitudes describing the production and
decays of these particles are parametrized in terms of the particle
masses.
In particular, pairs of gluinos are initially produced that decay to jets
and either a neutralino, and chargino or two neutralinos as shown in
Fig.~\ref{fig:SMS_Feynmans}.  The neutralino is then forced to decay
into a photon and undetected LSP while the chargino is forced to produce
a \PW~boson resulting in either a single-photon or a diphoton final
state. Simplified Models provide a benchmark different from other
constrained models such as the GGM SUSY scenario for comparing different
search strategies on a topological level. They also facilitate limit
comparisons with other final state topologies.

\subsection{Universal Extra Dimensions}

Diphoton final states with large \MET similar to those expected from GGM
SUSY scenarios are also predicted by models based on universal extra
dimensions.  Here the
existence of additional compactified dimensions are predicted in which SM
fields can propagate. The UED scenario provides several significant
consequences including gauge-coupling unification, supersymmetry
breaking, and other phenomena beyond those predicted by the standard
model~\cite{Appelquist:2000nn,Antoniadis:1990ew}. The propagation of SM
particles through the additional dimensions leads to the existence of a
series of excitations for each SM particle, known as a Kaluza--Klein (KK)
tower, which can decay via cascades involving other KK particles until
reaching the lightest Kaluza-Klein particle (LKP), which is the first
level KK photon. SM particles such as quarks and leptons can also be
produced in the cascades.

The UED space can be embedded in a larger space that has $n$ large extra
dimensions (LED) where only the graviton propagates with a
$(4+n)$-dimensional Planck scale ($M_D$) of a few \TeVns{}. In this case the
LKP is allowed to decay gravitationally, producing a photon and a
graviton. As the dominant production mechanism at the LHC is from the
strong interaction, KK quark and gluon pairs are produced, cascading
down to two LKP decays resulting in the two photon plus jet(s) and
$\MET$ final state.
Previous UED studies have been performed by the D0 experiment at the
Tevatron~\cite{Abazov:2010us} and most recently by
ATLAS~\cite{Aad:2011kz}.

\section{The CMS Detector}
\label{sec:detector}

The central feature of the CMS detector is a superconducting solenoid,
of 6~m internal diameter, providing an axial magnetic solenoid of 3.8\unit{T}
along the beam direction. Within the field volume are a silicon pixel
and strip tracker, a crystal electromagnetic calorimeter (ECAL), and a
brass/scintillator hadron calorimeter (HCAL).  Charged particle
trajectories are measured by the silicon pixel and strip tracker system,
covering $0 \leq \phi \leq 2\pi$ in azimuth and $|\eta|<2.5$, where the
pseudorapidity is $\eta = - \ln[\tan{\theta/2}]$, and $\theta$ is the polar
angle with respect to the counterclockwise-beam direction.  Muons are
measured in gas-ionization detectors embedded in the steel return
yoke. Extensive forward calorimetry complements the coverage provided by
the barrel and endcap detectors.

The electromagnetic calorimeter, which surrounds the tracker volume,
consists of 75\,848 lead-tungstate crystals that provide coverage in
pseudorapidity $\vert \eta \vert< 1.479 $ in the barrel region (EB) and
$1.479 <\vert \eta \vert < 3.0$ in two endcap regions (EE). The EB
modules are arranged in projective towers. A preshower detector
consisting of two planes of silicon sensors interleaved with a total of
3$\,X_0$ of lead is located in front of the EE.  In the region $\vert
\eta \vert< 1.74$, the HCAL cells have widths of 0.087 in pseudorapidity
and azimuth ($\phi$). In the $(\eta,\phi)$ plane, and for
$\vert \eta \vert< 1.48$, the HCAL cells map on to $5 \times 5$ ECAL
crystal arrays to form calorimeter towers projecting radially outwards
from close to the nominal interaction point. At larger values of $\vert
\eta \vert$, the size of the towers increases and the matching ECAL
arrays contain fewer crystals. Within each tower, the energy deposits in
ECAL and HCAL cells are summed to define the calorimeter tower energies,
subsequently used to provide the energies and directions of hadronic
jets.  In the 2011 collision data, unconverted photons with energy
greater than 30\GeV are measured within the barrel ECAL with a
resolution of better than 1\% \cite{Adzic:2007mi}.  The HCAL, when combined with the ECAL,
measures jets with a resolution $\Delta E/E \approx 100\,\% /
\sqrt{E\,[\GeVns{}]} \oplus 5\,\%$.  The CMS detector is nearly hermetic,
allowing for reliable measurements of \MET.  A more detailed description
of the CMS detector can be found in Ref.~\cite{JINST}.

\section{Data Selection}
\label{sec:selection}

The data sample used in this analysis was recorded during the 2011
\Pp\Pp~run of the LHC at a center-of-mass energy of 7\TeV and
corresponds to an integrated luminosity of 4.93\fbinv.  Events were
selected using the CMS two-level trigger system requiring the presence
of at least one high-energy photon and significant hadronic activity or
at least two photons.  The first level of the CMS trigger system,
composed of custom hardware processors, uses information from the
calorimeters and muon detectors to select events in less than
3.2\unit{$\mu$s}. The High Level Trigger processor farm further decreases
the event rate from around 100\unit{kHz} to around 300\unit{Hz}, before data
storage.

Photon triggers are utilized to select both the signal candidates and
control samples used for background estimation. The efficiency for
signal events to pass the trigger requirements ranges around 40--60\%,
while the efficiency for signal events which pass the photon offline
selection is estimated to be greater than 99\%. The single-photon search
is based on the photon-$\HT$ trigger requiring the presence of one
photon with $\ET>70\GeV$ and the quantity \HT, the scalar sum of
transverse momenta of reconstructed and calibrated calorimeter jets with
$\pt>40\GeV$ and $|\eta|<3.0$ in the event. Because of the continuous
increase in the instantaneous luminosity, the trigger evolved with time
from $\HT>200$ to $\HT>400$\GeV. An inefficiency of this trigger during
a short time period of data taking restricts the single-photon analysis
to an integrated luminosity of 4.62\fbinv.  The diphoton measurement
using an integrated luminosity of 4.93\fbinv of \Pp\Pp~collisions is
based on a diphoton trigger with an \ET~threshold of 36\GeV (22\GeV)
for the leading (sub-leading) photon.

\subsection{Photon and Electron Reconstruction and Identification}
\label{photonid}

Photon candidates are reconstructed from clusters of energy in the
ECAL.
The photon identification requires the ECAL cluster shape to be
consistent with that expected from a photon, and the hadronic energy
detected in the HCAL behind the photon shower not to exceed 5\% of the
ECAL energy.  To suppress hadronic jets being misreconstructed as
photons, we require photon candidates to be isolated from other activity
in the tracker, ECAL and HCAL. A cone of $\Delta R = \sqrt{
  (\Delta\eta)^2+(\Delta\phi)^2} = 0.3$ is constructed around the
direction of the photon candidate, and the scalar sums of transverse
energies of tracks and calorimeter deposits within this $\Delta R$ cone
are determined, after excluding the contribution from the photon
candidate itself. These isolation sums for the tracker, ECAL and HCAL
are added to form $I_{\text{comb}}$. This combined isolation sum is
corrected for contributions from additional \Pp\Pp~interactions
(pileup) other than the hard scattering that
produced the photon(s) and jets of interest.

With increasing instantaneous luminosity during the 2011 LHC operation,
the number of interactions per bunch crossing has also increased,
resulting in an approximately linear rise in the occupancy of the
detector.  The energy in the photon isolation cone is sensitive to
pileup effects. In an effort to reduce the dependence on the variation
of pileup, an effective energy proportional to the amount of pileup
$E_{\text{pileup}} = \rho \times A_{\text{eff}}$ is subtracted from the
combined photon-isolation variable.  The $\rho$~variable, which is
described in detail in Ref.~\cite{Cacciari:2007fd}, quantifies the
amount of transverse momentum added to the event per unit area, \eg by
minimum bias particles. The variable $A_{\text{eff}}$ corresponds to an
effective area determined from the slope of the average isolation energy
versus $\rho$. The values of $\rho$ and the isolation compensation
factor, $\rho \times A_{\text{eff}}$, are calculated from the data on an
event by event basis.  Separate effective areas are calculated for the
ECAL and HCAL isolation.

The combined isolation sum is corrected for contributions from pileup
using $I_{\text{comb}}^{\text{corr}}=I_{\text{comb}} -
E_{\text{pileup}}$~\cite{Cacciari:2007fd}.  The corrected combined
isolation is required to be $I_{\text{comb}}^{\text{corr}}<6\GeV$, which
is based on an optimization of $S/\sqrt{B}$ as a figure of merit, where
the signal $S$ is from simulated SUSY-GGM events (see
Section~\ref{sec:MC}) and the background $B$ corresponds to a multijet
simulated sample. As a cross check, data from a multijet-enriched sample
consisting of events with low missing transverse energy $\MET <30\GeV$,
where the photon candidates passing all analysis requirements except the
isolation cut, were also used as background sample. Using the same
signal GGM sample, this test also results in an optimal value of
$I_{\text{comb}}^{\text{corr}}<6\GeV$.

The criteria above are efficient for the selection not only of photons
but also of electrons. To reliably separate them, we search for hit
patterns in the pixel detector consistent with at least a single pixel
hit matching a track from an electron. The candidates without pixel
match are considered to be photons. Otherwise they are considered to be
electrons, which are used to select control samples for background
estimation.

Photons that fail either the shower shape or combined isolation
requirement are referred to as misidentified photons. These objects are
predominantly electromagnetically-fluctuated jets and are used for the
background estimation based on data. The definition of the misidentified
photon is designed to be orthogonal to our real candidate photons, but
still similar to that of the real photon definition to provide an
accurate background estimate. An upper bound on
$I_{\text{comb}}^{\text{corr}}$ is introduced in order to avoid events
with highly non-isolated misidentified photon objects where the
resolution on \MET is expected to be different compared to events with
real photons. An upper cut of $I_{\text{comb}}^{\text{corr}}<30\GeV$
(20\GeV) was found optimal for the single-photon (diphoton) analysis.

Photons which convert in the tracker material ahead of the ECAL are
reconstructed and counted as photon objects.  These photons can have
slightly higher isolation sums that unconverted photons or, if they
convert in the pixel detector, can be counted as electrons.  Both
possibilities of contamination have been studied and found to be
negligible in this analysis.

\subsection{Jet and Missing Transverse Energy Reconstruction and Identification}
\label{ijet}

Jets and \MET are reconstructed with a particle-flow (PF)
technique~\cite{CMS-PAS-PFT-09-001,CMS-PAPERS-JME-10-011}.  The PF event
reconstruction consists of identifying every particle with an optimized
combination of all sub-detector information. The energy of photons is
obtained directly from the ECAL measurement, corrected for detector
effects. The energy of electrons is determined from a combination of the
track momentum at the primary interaction vertex, the corresponding ECAL
cluster energy, and the energy sum of all bremsstrahlung photons
attached to the track. The energy of muons is obtained from the
corresponding track momentum. The energy of charged hadrons is
determined from a combination of the track momentum and the
corresponding ECAL and HCAL energy, corrected for detector effects, and
calibrated for the non-linear response of the calorimeters. Finally, the
energy of neutral hadrons is obtained from the corresponding calibrated
ECAL and HCAL energy.

All these particles are clustered into jets using the anti-\kt
clustering algorithm~\cite{Cacciari:2008gp} with a distance parameter of
0.5.  The jet momentum is determined as the vectorial sum of all
particle momenta in this jet and is found in the simulation to be within
5\% to 10\% of the true momentum over the whole \pt spectrum and
detector acceptance. An offset correction is applied to take into
account the extra energy clustered in jets due to multiple
\Pp\Pp~interactions within the same bunch crossing, thereby reducing the
dependence of jet energies on the instantaneous luminosity. Jet energy
corrections are derived from simulation studies and are compared with in
situ measurements using the energy balance of dijet and photon+jet
events. Additional selection criteria are applied to each event. For
example,
jets identified to originate in spurious jet-like features from isolated
electronic noise patterns in HCAL and ECAL are removed from the
sample~\cite{CMS-PAPERS-JME-10-011}.

Jets selected for this analysis are required to have transverse momentum
$\pt \geq 30$\GeV, $|\eta| \leq 2.6$ and to satisfy the following
jet-selection requirements. The neutral-hadron fraction as well as the
electromagnetic fraction of energy contributing to the shower created by
the jet should each be ${<}0.99$,
and the charged hadron fraction is required to be greater than zero.
Events must contain at least one jet isolated from the photon candidates
by $\Delta R \geq 0.5$ for the events to be retained in the signal
sample.

\subsection{Single-Photon and Diphoton Event Selections}

The single-photon analysis requires the scalar sum of the transverse
energy of all jets and all photons in the event, \HT, to be larger than
$450$\GeV, where the photon-\HT trigger is fully efficient.  To closely
resemble the trigger requirement, calorimeter jets with $\pt \geq
40$\GeV and $|\eta| \leq 3.0$ are used for the \HT\ calculation, but
with the addition that these jets are pileup corrected. Both real and
misidentified photons are included in the \HT\ calculation. Since the
photon objects are also reconstructed as jets, the \pt of the jet is
used in the \HT~calculation instead of the photon object, if the
transverse momentum ratio between jet and photon object is greater than
95\% and the photon and jet are within $\Delta R\leq 0.3$. This avoids a
bias in \HT\ and \MET\ due to the different isolation requirements for
the genuine photon candidates and the misidentified photons in the
multijet control samples.  In addition, a photon with $\ET>80$\GeV
within $|\eta|<1.4$ and at least two jets with $\pt\geq30\GeV$ and
$|\eta|\leq2.6$ are required. Events with isolated leptons are not
rejected, and the lepton momenta are not included in the
\HT~determination to follow the trigger requirement.

To be within the full efficiency of the $\gamma\gamma$~trigger with an
\ET~threshold of 36\GeV (22\GeV) on the leading (sub-leading) photon,
the diphoton offline analysis
requires at least two photons with $\ET>40$\GeV (25\GeV) for the
leading (sub-leading) photon in the event and at least one jet with
$\pt\ge30\GeV$ and $|\eta| \leq 2.6$.  Table~\ref{tab:cutsummary}
contains a summary of the signal sample selection criteria for the
single-photon and diphoton analyses. It also includes information on the
background control samples described in Section~\ref{sec:backgrounds} as
well as the search region for new physics in the variable of transverse
missing energy as further discussed in Section~\ref{sec:limit}.

\begin{table}[tbp]
\begin{center}
  \topcaption{Summary of the signal and control sample selection criteria
    used for the single-photon and diphoton analyses. Electron (\Pe\Pe) and
    misidentified photon ($ff$) categories are used in background
    estimations described in Sections~\ref{sec:single_photon} and
    \ref{sec:diphoton}.  The exclusive bins in \MET are used in the
    limit setting procedure.}
\label{tab:cutsummary}
\begin{tabular}{l|c|c|c|c|c|c}
\hline
           &\multicolumn{3}{c|}{Single photon}                                                          & \multicolumn{3}{c}{Diphoton} \\
           & Signal & Multijet control & EWK control                                                         & Signal & $\Pe\Pe$ control & $ff$ control \\ \hline
$I_{\text{comb}}^{\text{corr}}$ [\GeVns{}]
& $< 6$ & $\geq 6$, $<30$ & $< 6$  & $< 6$ & $< 6$ & $\geq 6$, $<20$ \\
pixel seed & veto   & veto        & required & veto & required & veto \\ \hline

Trigger    &\multicolumn{3}{l|}{$\gamma$-\HT trigger with}                   & \multicolumn{3}{l}{$\gamma\gamma$ trigger with}\\
           &\multicolumn{3}{l|}{$\pt^\gamma\geq70$\GeV, $\HT\geq400$\GeV}   & \multicolumn{3}{l}{$\pt^{\gamma 1,2}\geq36~ (22)$\GeV}\\
           &\multicolumn{3}{l|}{(using $\pt^{\text{\tiny jets}}\geq40$\GeV, $|\eta|<3.0$)}   & \multicolumn{3}{l}{}\\[1.5ex]
Photon(s)  &\multicolumn{3}{l|}{$\pt^\gamma\geq80$\GeV, $|\eta|<1.4$ }                                  & \multicolumn{3}{l}{$\pt^{\gamma 1,2}\geq40~ (25)$\GeV, $|\eta|<1.4$}\\[1.5ex]
PF Jet(s) &\multicolumn{3}{l|}{$\pt^{\text{\tiny jets 1,2}}\geq30$\GeV, $|\eta|<2.6$ }                 & \multicolumn{3}{l}{$\pt^{\text{\tiny jet}}\geq 30$\GeV, $|\eta|<2.6$}\\[1.5ex]
\HT        &\multicolumn{3}{l|}{$\HT\geq 450$\GeV }                       & \multicolumn{3}{c}{ --- }\\
           &\multicolumn{3}{l|}{(using $\pt^{\text{\tiny jets, $\gamma$}}\geq40$\GeV, $|\eta|<3.0$)}        & \multicolumn{3}{c}{  }\\[1.5ex]
\MET       &\multicolumn{3}{l|}{$\MET\geq 100$\GeV (6 excl.~bins in \MET)}                             & \multicolumn{3}{l}{ $\MET\geq 50$\GeV (5 excl.~bins in \MET) }\\
	
\hline
\end{tabular}
\end{center}
\end{table}

\section{Simulated Samples}
\label{sec:MC}

Although this analysis uses methods based on data to estimate the
background components, simulated samples are used to evaluate less
significant backgrounds, which might be difficult to measure directly
from the data, or to model the new physics (NP) signals and to validate
the performance of the background estimation from data.

The simulated samples used in this search are produced in several
ways. Depending on the process either the
\PYTHIA~\cite{Sjostrand:2006za} or \MADGRAPH~\cite{Alwall:2011uj} Monte
Carlo (MC) event generators are used to generate event kinematics and
fragment partons into jets. For most simulated data, in particular to
study SM backgrounds, the generated events are passed through the full
\GEANTfour-based~\cite{Agostinelli:2002hh} CMS detector simulation.
Because of the large number of individual simulated samples required in
the NP parameter space scans used in the interpretation of results in
the light of NP, a fast detector simulation~\cite{FastSim1}
based on a full description of the CMS detector geometry and a
parameterization of single-particle showers and response is utilized to
reduce the computation time for those samples.  Event pileup
corresponding to the luminosity profile of the analyzed data is added to
all simulated samples and the generated events are reconstructed using
the same software program as for the collision data.

\begin{table*}[btp]
  \topcaption{\label{tab_signalscans}
    Parameters varied in GGM signal scans used in the interpretation.
    Grid values along either axis in the scan are offset by
    10--20\GeV to prevent degeneracies between the generated particles.}
\begin{center}
\begin{tabular}{l|c|c|c|c}
\hline
Scan name & Squark mass & Gluino mass & Bino mass & Wino mass \\
\hline
Squark-Gluino (Bino) & 400--2000\GeV & 400--2000\GeV & 375\GeV      & 2000\GeV  \\
Squark-Gluino (Wino) & 400--2000\GeV & 400--2000\GeV & 5000\GeV     & 375\GeV   \\
Gluino-Bino     & 5000\GeV &  300--1500\GeV    &  50--1500\GeV & 2000\GeV  \\
Gluino-Wino     & 5000\GeV &  300--1000\GeV    &  5000\GeV    & 100--1000\GeV \\
Wino-Bino     & 5000\GeV     & 5000\GeV     &  5--1000\GeV  & 115--1000\GeV \\
\hline
\end{tabular}
\end{center}
\end{table*}

In interpreting our results, multiple samples of simulated signal data
are produced by varying model parameters individually (as in the case of
the UED interpretation) or in pairs (in the case of the GGM and SMS
interpretations). General gauge-mediated SUSY breaking requires the LSP
to be a gravitino, and the NLSP needs to be a wino-like or bino-like
neutralino to produce a final state with photon(s) plus \MET.  Bino-like
neutralinos decay most of the time into a photon.  Wino-like neutralinos
decay mostly into \Z~bosons, but they also decay into a photon
${\sim}20\%$ of the time, allowing our measurement to be sensitive to this
channel. In the GGM scans, other SUSY particles are decoupled (forced to
have high mass) in order to leave only the possibility of light squarks,
gluinos and the desired neutralino NLSP or neutralino/chargino co-NLSP
as kinematically allowed production particles.
Table~\ref{tab_signalscans} shows the mass parameters varied in the five
GGM planes investigated in this analysis~\cite{Kats:2011qh}.  The masses
of these particles take values within the ranges indicated in the table
as different scan grids are produced. In particular, the SUSY mass
spectra are calculated in form of files following the SUSY Les Houches
Accord (SLHA)~\cite{Skands:2003cj} utilizing
\textsc{SuSpect}~\cite{Djouadi:2002ze} with decay tables from
\textsc{sdecay}~\cite{Muhlleitner:2003vg}. The SUSY GGM events are
generated in a three-dimensional grid of squark, gluino, and NLSP
masses.  Squarks are taken to be degenerate in mass and all other SUSY
particles are assumed to be heavy.  In the scans where the NLSP mass is
varied, the ``next-to-next to LSP" (usually a gluino) is required to have
a higher mass, resulting in scans that only span above the diagonal in
the corresponding mass plane.  This is also the case for the Simplified
Model scans
described below.  In the ``Wino-Bino" scan shown at the bottom of
Table~\ref{tab_signalscans}, we decouple the squarks and gluinos,
leaving only electroweak production of wino-like neutralino/charginos to
study our sensitivity to electroweak production of SUSY.

For the Simplified Model interpretation, more controls are exerted over
the production and decay of sparticles, which are often forced to decay
into a certain final state, \eg, 100\% of the time. Two parameter scans
referred to as the $\PW\gamma$ SMS (Fig.~\ref{fig:SMS_Feynmans} left)
and the $\gamma\gamma$ SMS (Fig.~\ref{fig:SMS_Feynmans} right) are used
in this analysis. They both span a grid in gluino and
neutralino/chargino mass space, forcing the initial pair production of
gluinos, which then decay to jets and neutralino or chargino.  In the
$\gamma\gamma$ Simplified Model, both gluinos are forced to decay to
jets and neutralinos, which in turn decay to photons.
The $\PW\gamma$ SMS forces one gluino to
decay to a chargino, which is forced to always produce a \PW~boson, and
the other gluino decays as in the $\gamma\gamma$ Simplified Model.  The
$\gamma\gamma$ scan produces final states to which both the
single-photon and diphoton analyses are sensitive, while the $\PW\gamma$
SMS scan is interpreted only through the single-photon analysis.  The
production cross sections of the GGM and SMS scans~\cite{Kramer:2012bx}
are calculated at next-to-leading order (NLO) plus next-to-leading log
in QCD using the \PROSPINO program~\cite{Beenakker:1996ch,Kulesza:2008jb,
Kulesza:2009kq,Beenakker:2009ha,Beenakker:2011fu}. Except for
the GGM "Wino-Bino" scan, the production in these scans is dominated by
gluino-gluino, gluino-squark, and squark-squark production.

Simulated signal samples for the UED interpretation are generated using
the UED model as implemented at leading order (LO) in
\PYTHIA~\cite{Sjostrand:2006za}. Parameters for the UED model
investigated in this analysis including the LO cross section are chosen
to match previous UED searches by other
experiments~\cite{Abazov:2010us,Aad:2011kz}. The UED model has two
varying parameters, the ultraviolet cutoff $\Lambda$ and the radius of
compactification $R$. In this study $R$ is chosen as a free parameter
while $\Lambda$ is set to satisfy the relation $\Lambda
R=20$~\cite{Appelquist:2000nn}.  Additional parameters that are used in
the MC generation of the signal are chosen as follows. The number of
large extra dimensions is $N=2$ or 6, the $(N + 4)$-dimensional Planck
scale $M_{D}$ is 5\TeV, while the number of KK excitation quark flavors
is five. Sample points of $1/R$ ranging from 900 to 1600\GeV are
produced in increments of 50\GeV.

\section{Background Estimation Methodology}
\label{sec:backgrounds}

The NP signature of the photon(s) plus \MET final state can be
mimicked by SM processes in several ways.  The largest backgrounds are
due to events without true \MET resulting from abundant hadronic processes,
such as direct photon plus jets processes, and
multijet production with electromagnetically rich jets misidentified as
photons, which result in events with the same topology as the NP
signal. The missing \ET in these hadronic events comes from poorly measured
hadronic activity in the event.  This background is referred to as
background with false \MET or as QCD background.  The \MET resolution
for this background is much poorer than the resolution of the total \ET
of the photon(s) and is determined by the resolution of the hadronic
energy in the event. The strategy for determining the shape of the \MET
distribution for the QCD background is to find a control sample that
reproduces the hadronic activity in the candidate sample while having no
significant true \MET that mimic a substantial missing \ET contribution.

The second kind of background comes from processes with true \MET.  It
is dominated by $\PW\gamma$ events and \PW~plus jets production where
the \PW~decays into an electron plus a neutrino, with the electron or
jet misidentified as a photon and the neutrino leading to \MET.  We
refer to this sample as background with true \MET or electroweak (EWK)
background and it is determined in the following way.  Since the photon
is expected to behave almost identically to an electron in the
electromagnetic calorimeter, electrons can be mistaken as photons except
that electrons have hits matching the particle track in the pixel
detector. We measure the electron-photon misidentification rate
$f_{\Pe\rightarrow\gamma}$ and determine the contribution of the EWK
background by applying $f_{\Pe\rightarrow\gamma}$ to our
\MET~distribution (see Section~\ref{emisid}). The rates of other
processes with true \MET that have single photon or diphotons in their
final states are quite small and are discussed for the single-photon and
diphoton analyses separately in Sections~\ref{sec:single_photon}
and~\ref{sec:diphoton}.

\subsection{Electron Misidentification Rate}
\label{emisid}

We determine the probability to misidentify an electron as a photon, by
fitting the mass of the $\Z\to \EE$~peak seen in the $\Pe\Pe$ and
$\Pe\gamma$ mass spectra, and comparing the integrals of these fits. For
this purpose we identify a sample of $\Pe\Pe$ events where pixel matches are
required on both objects that otherwise satisfy the photon selection
requirements (see details of diphoton analysis in
Section~\ref{sec:diphoton}). The $\Pe\gamma$ sample has the same requirements
imposed on it as the real $\gamma\gamma$ sample, except a pixel match is
required for one of the electromagnetic objects.

We extract the electron misidentification fraction from the $\Pe\Pe$ and
$\Pe\gamma$ spectrum using the number of observed $\Z\rightarrow \Pe\Pe$
events in the $\Pe\Pe$ mass spectrum given as
$N_{\Pe\Pe}=(1-f_{\Pe\to\gamma})^2\,N_{\Z~\text{true}}$ where
$N_{\Z~\text{true}}$ is the number of true $\Z\rightarrow \Pe\Pe$ events.
The observed $\Z\rightarrow \Pe\Pe$ peak in the $\Pe\gamma$ spectrum is
$N_{\Pe\gamma}=2\,[f_{\Pe\to\gamma}(1-f_{\Pe\to\gamma})]\,N_{\Z~\text{true}}$
leading to $f_{\Pe\to\gamma}=N_{\Pe\gamma}/(2N_{\Pe\Pe}+N_{\Pe\gamma})$.  We can
calculate the number of $\Z\rightarrow \Pe\Pe$ events expected in the
$\gamma\gamma$ spectrum using $N_{\gamma\gamma}=(f_{\Pe\to\gamma})^2
\times N_{\Pe\Pe}\,/ (1-f_{\Pe\to\gamma})^2$ and cross check the number of
observed diphoton events.

We measure $f_{\Pe\to\gamma}$ in bins of photon transverse momentum. The overall
misidentification rate integrated over the whole \pt~range is determined as
$f_{\Pe\to\gamma}=0.015\pm0.002\stat\pm0.005\syst$. This number is used
for the diphoton analysis, while for $\pt>80\GeV$ a misidentification
rate of $f_{\Pe\to\gamma}=0.0080\pm0.0025\stat$ is determined. The latter rate
is used for the single-photon search since $\pt(\gamma)>80\GeV$ is the
momentum region relevant for this analysis.
\section{Single-Photon Analysis}
\label{sec:single_photon}

The single-photon analysis targets especially SUSY scenarios in which
the lightest gaugino comprises a large non-bino-like mixture.  In this
case the branching fraction of the lightest gaugino to a photon and the
gravitino LSP is reduced and decays into other bosons like $\PW$, $\Z$,
or Higgs occur, leading to additional jets and possibly leptons in the
final state, suppressing events with more than one photon.  Events with
leptons or more than one photon are not removed in the single-photon
analysis. The potential overlap with the diphoton selection has been
studied and is found to be negligible.

\subsection{Background Estimation}
\label{sec:single_photon_background}

The dominant background in the single-photon analysis is a composition
of processes such as $\gamma$+jets and multijet QCD production with
one jet misidentified as a photon. The shape of the \MET distribution
is similar for both background contributions, as the event topologies
are very similar.  Therefore, these two contributions to the QCD
background are estimated together from the same control sample. This
background sample is selected by applying the signal selection
requirements, except that the photon candidate is required to fail the
photon identification criteria but to satisfy a loose isolation
requirement.  Such misidentified photon candidates follow a definition
orthogonal to the photon identification criteria in the signal
selection. The background control sample is weighted to correct for the
difference in $\pt$ spectra of misidentified and genuine photons. The
weights as a function of the photon transverse energy are determined in
bins of \pt from the ratio of events in the misidentified and genuine
photon samples for $\MET<100$\GeV, which is taken as a signal-depleted
region for the normalization of the QCD background to the single-photon
data.

The EWK background contribution is much smaller than the QCD
background. The dominant contributions are from \ttbar~production or
events with \PW\ or \Z~bosons with one or more neutrinos in the final
state in which the electron is misidentified as a photon.  This
background is modeled from the data using an electron control sample
selected by the same trigger as the signal dataset. The electron control
sample is weighted according to the misidentification rate,
$f_{\Pe\to\gamma}$, measured in $\Z\to \Pe\Pe$ events, as discussed in
Section~\ref{emisid}.

Additional backgrounds can contribute due to initial-state radiation
(ISR) and final-state radiation (FSR) of photons. Both ISR and FSR, in
events with electrons in the final state, are already covered by the EWK
background prediction from data. The remaining contributions from \PW,
\Z, and \ttbar~events are taken from MC simulation.

\subsection{Results}

The dominant systematic uncertainty in the background estimation arises
from the small number of events in the misidentified-photon control sample. The
statistical uncertainty associated with the $\MET<100$\GeV sample,
where the normalization of misidentified and genuine photon samples is calculated in bins of photon \pt,
is propagated as a systematic uncertainty. The uncertainty is taken to
be correlated among the \MET bins, as a given \MET bin receives contributions
from several photon \pt normalization bins.
The method assumes, that the \MET\ and the photon momentum are uncorrelated. This has been validated
in simulation up to 5\%, which is assigned as additional systematic
uncertainty.

\begin{figure*}[tbp]
\begin{center}
\includegraphics[width=0.6\textwidth]{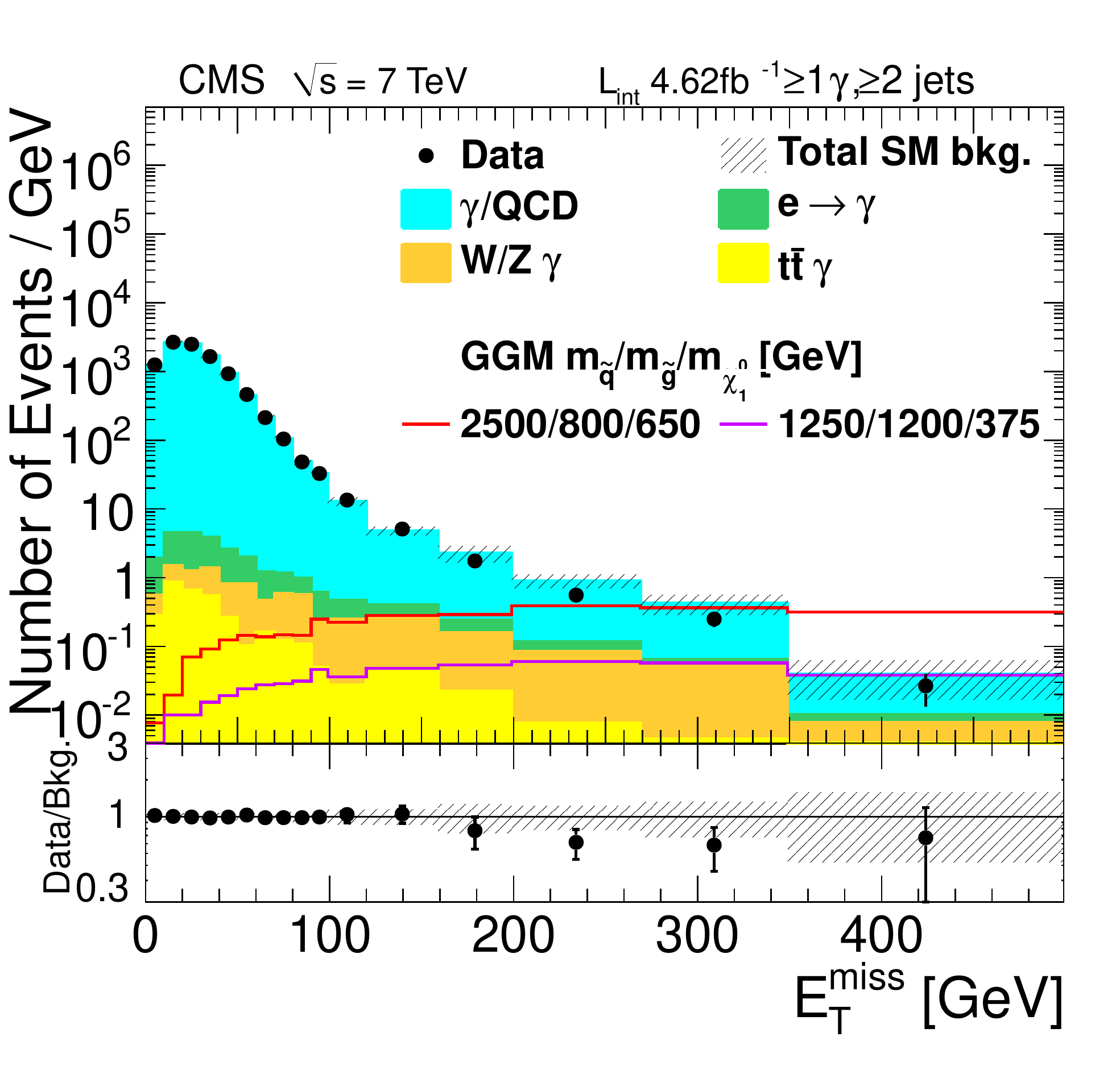}
\caption{
  Missing \ET spectrum of single-photon data (dots with error bars)
  compared to various SM background predictions (solid colored
  histograms).  The shaded area indicates the uncertainty in the total
  background prediction. The \MET~spectrum for two example GGM points
  (red upper and blue lower solid curves with masses of $m_{{\PSq}} /
  m_{\sGlu} / m_{\chiz_1}$ in\GeV) on either side of our exclusion
  boundary are also shown. At the bottom, the ratio of data over
  standard model prediction is shown as a function of \MET.  The error
  bars take into account only the statistical error of the data sample,
  while the hatched area is the uncertainty in the expected background
  from the SM processes.  }
\label{fig:preselmet}
\end{center}
\end{figure*}

In comparison, the systematic uncertainty due to the statistically
limited electron control sample used for the electroweak background
prediction is negligible. In addition, the small uncertainty in the
electron misidentification rate $f_{\Pe\to\gamma}=0.008\pm0.0025$ is
propagated resulting in small systematic uncertainties in the EWK
background prediction. Finally, a conservative uncertainty of $50\%$ on
the ISR and FSR contributions to the $\PW/\Z$ and \ttbar cross sections
is added.

\def\lumiData{4.62}
\def\UNCERTFITMC{ 25}
\def\UNCERTFIT{ 10}
\def\UNCERTEXTRAPOL{ 5}
\def\EWKFAKERATE{0.006}
\def\EWKFAKERATESYST{0.0025}
\def\jTWOLOWWJMCnoEvts{13}
\def\jTWOLOWWJMCstatErr{4.9}
\def\jTWOLOWWJMCsystErrUp{\ensuremath{\leq 0.01}}
\def\jTWOLOWWJMCsystErrDown{\ensuremath{\leq 0.01}}
\def\jTWOLOWWJMCnoEvtsBinA{2.9}
\def\jTWOLOWWJMCstatErrBinA{2.4}
\def\jTWOLOWWJMCsystErrUpBinA{\ensuremath{\leq 0.01}}
\def\jTWOLOWWJMCsystErrDownBinA{\ensuremath{\leq 0.01}}
\def\jTWOLOWWJMCcombErrBinA{2.4}
\def\jTWOLOWWJMCnoEvtsBinB{2.2}
\def\jTWOLOWWJMCstatErrBinB{1.7}
\def\jTWOLOWWJMCsystErrUpBinB{\ensuremath{\leq 0.01}}
\def\jTWOLOWWJMCsystErrDownBinB{\ensuremath{\leq 0.01}}
\def\jTWOLOWWJMCcombErrBinB{1.7}
\def\jTWOLOWWJMCnoEvtsBinC{3.0}
\def\jTWOLOWWJMCstatErrBinC{2.4}
\def\jTWOLOWWJMCsystErrUpBinC{\ensuremath{\leq 0.01}}
\def\jTWOLOWWJMCsystErrDownBinC{\ensuremath{\leq 0.01}}
\def\jTWOLOWWJMCcombErrBinC{2.4}
\def\jTWOLOWWJMCnoEvtsBinD{4.5}
\def\jTWOLOWWJMCstatErrBinD{3.2}
\def\jTWOLOWWJMCsystErrUpBinD{\ensuremath{\leq 0.01}}
\def\jTWOLOWWJMCsystErrDownBinD{\ensuremath{\leq 0.01}}
\def\jTWOLOWWJMCcombErrBinD{3.2}
\def\jTWOLOWWJMCnoEvtsBinE{\ensuremath{\leq 0.01}}
\def\jTWOLOWWJMCstatErrBinE{\ensuremath{\leq 0.01}}
\def\jTWOLOWWJMCsystErrUpBinE{\ensuremath{\leq 0.01}}
\def\jTWOLOWWJMCsystErrDownBinE{\ensuremath{\leq 0.01}}
\def\jTWOLOWWJMCcombErrBinE{\ensuremath{\leq 0.01}}
\def\jTWOLOWWJMCnoEvtsBinF{\ensuremath{\leq 0.01}}
\def\jTWOLOWWJMCstatErrBinF{\ensuremath{\leq 0.01}}
\def\jTWOLOWWJMCsystErrUpBinF{\ensuremath{\leq 0.01}}
\def\jTWOLOWWJMCsystErrDownBinF{\ensuremath{\leq 0.01}}
\def\jTWOLOWWJMCcombErrBinF{\ensuremath{\leq 0.01}}
\def\jTWOLOWTTMCnoEvts{10.0}
\def\jTWOLOWTTMCstatErr{1.4}
\def\jTWOLOWTTMCsystErrUp{\ensuremath{\leq 0.01}}
\def\jTWOLOWTTMCsystErrDown{\ensuremath{\leq 0.01}}
\def\jTWOLOWTTMCnoEvtsBinA{2.6}
\def\jTWOLOWTTMCstatErrBinA{0.7}
\def\jTWOLOWTTMCsystErrUpBinA{\ensuremath{\leq 0.01}}
\def\jTWOLOWTTMCsystErrDownBinA{\ensuremath{\leq 0.01}}
\def\jTWOLOWTTMCcombErrBinA{0.7}
\def\jTWOLOWTTMCnoEvtsBinB{4.1}
\def\jTWOLOWTTMCstatErrBinB{0.9}
\def\jTWOLOWTTMCsystErrUpBinB{\ensuremath{\leq 0.01}}
\def\jTWOLOWTTMCsystErrDownBinB{\ensuremath{\leq 0.01}}
\def\jTWOLOWTTMCcombErrBinB{0.9}
\def\jTWOLOWTTMCnoEvtsBinC{1.8}
\def\jTWOLOWTTMCstatErrBinC{0.6}
\def\jTWOLOWTTMCsystErrUpBinC{\ensuremath{\leq 0.01}}
\def\jTWOLOWTTMCsystErrDownBinC{\ensuremath{\leq 0.01}}
\def\jTWOLOWTTMCcombErrBinC{0.6}
\def\jTWOLOWTTMCnoEvtsBinD{1.0}
\def\jTWOLOWTTMCstatErrBinD{0.4}
\def\jTWOLOWTTMCsystErrUpBinD{\ensuremath{\leq 0.01}}
\def\jTWOLOWTTMCsystErrDownBinD{\ensuremath{\leq 0.01}}
\def\jTWOLOWTTMCcombErrBinD{0.4}
\def\jTWOLOWTTMCnoEvtsBinE{0.6}
\def\jTWOLOWTTMCstatErrBinE{0.4}
\def\jTWOLOWTTMCsystErrUpBinE{\ensuremath{\leq 0.01}}
\def\jTWOLOWTTMCsystErrDownBinE{\ensuremath{\leq 0.01}}
\def\jTWOLOWTTMCcombErrBinE{0.4}
\def\jTWOLOWTTMCnoEvtsBinF{\ensuremath{\leq 0.01}}
\def\jTWOLOWTTMCstatErrBinF{\ensuremath{\leq 0.01}}
\def\jTWOLOWTTMCsystErrUpBinF{\ensuremath{\leq 0.01}}
\def\jTWOLOWTTMCsystErrDownBinF{\ensuremath{\leq 0.01}}
\def\jTWOLOWTTMCcombErrBinF{\ensuremath{\leq 0.01}}
\def\jTWOLOWPhotonVMCAllnoEvts{40}
\def\jTWOLOWPhotonVMCAllstatErr{4.0}
\def\jTWOLOWPhotonVMCAllsystErrUp{\ensuremath{\leq 0.01}}
\def\jTWOLOWPhotonVMCAllsystErrDown{\ensuremath{\leq 0.01}}
\def\jTWOLOWPhotonVMCAllnoEvtsBinA{9.0}
\def\jTWOLOWPhotonVMCAllstatErrBinA{1.9}
\def\jTWOLOWPhotonVMCAllsystErrUpBinA{\ensuremath{\leq 0.01}}
\def\jTWOLOWPhotonVMCAllsystErrDownBinA{\ensuremath{\leq 0.01}}
\def\jTWOLOWPhotonVMCAllcombErrBinA{1.9}
\def\jTWOLOWPhotonVMCAllnoEvtsBinB{10}
\def\jTWOLOWPhotonVMCAllstatErrBinB{2.0}
\def\jTWOLOWPhotonVMCAllsystErrUpBinB{\ensuremath{\leq 0.01}}
\def\jTWOLOWPhotonVMCAllsystErrDownBinB{\ensuremath{\leq 0.01}}
\def\jTWOLOWPhotonVMCAllcombErrBinB{2.0}
\def\jTWOLOWPhotonVMCAllnoEvtsBinC{6.4}
\def\jTWOLOWPhotonVMCAllstatErrBinC{1.6}
\def\jTWOLOWPhotonVMCAllsystErrUpBinC{\ensuremath{\leq 0.01}}
\def\jTWOLOWPhotonVMCAllsystErrDownBinC{\ensuremath{\leq 0.01}}
\def\jTWOLOWPhotonVMCAllcombErrBinC{1.6}
\def\jTWOLOWPhotonVMCAllnoEvtsBinD{7.4}
\def\jTWOLOWPhotonVMCAllstatErrBinD{1.6}
\def\jTWOLOWPhotonVMCAllsystErrUpBinD{\ensuremath{\leq 0.01}}
\def\jTWOLOWPhotonVMCAllsystErrDownBinD{\ensuremath{\leq 0.01}}
\def\jTWOLOWPhotonVMCAllcombErrBinD{1.6}
\def\jTWOLOWPhotonVMCAllnoEvtsBinE{5.1}
\def\jTWOLOWPhotonVMCAllstatErrBinE{1.5}
\def\jTWOLOWPhotonVMCAllsystErrUpBinE{\ensuremath{\leq 0.01}}
\def\jTWOLOWPhotonVMCAllsystErrDownBinE{\ensuremath{\leq 0.01}}
\def\jTWOLOWPhotonVMCAllcombErrBinE{1.5}
\def\jTWOLOWPhotonVMCAllnoEvtsBinF{2.1}
\def\jTWOLOWPhotonVMCAllstatErrBinF{1.0}
\def\jTWOLOWPhotonVMCAllsystErrUpBinF{\ensuremath{\leq 0.01}}
\def\jTWOLOWPhotonVMCAllsystErrDownBinF{\ensuremath{\leq 0.01}}
\def\jTWOLOWPhotonVMCAllcombErrBinF{1.0}
\def\jTWOLOWQCDMCnoEvts{129}
\def\jTWOLOWQCDMCstatErr{33}
\def\jTWOLOWQCDMCsystErrUp{\ensuremath{\leq 0.01}}
\def\jTWOLOWQCDMCsystErrDown{\ensuremath{\leq 0.01}}
\def\jTWOLOWQCDMCnoEvtsBinA{73}
\def\jTWOLOWQCDMCstatErrBinA{31}
\def\jTWOLOWQCDMCsystErrUpBinA{\ensuremath{\leq 0.01}}
\def\jTWOLOWQCDMCsystErrDownBinA{\ensuremath{\leq 0.01}}
\def\jTWOLOWQCDMCcombErrBinA{31}
\def\jTWOLOWQCDMCnoEvtsBinB{36}
\def\jTWOLOWQCDMCstatErrBinB{7.8}
\def\jTWOLOWQCDMCsystErrUpBinB{\ensuremath{\leq 0.01}}
\def\jTWOLOWQCDMCsystErrDownBinB{\ensuremath{\leq 0.01}}
\def\jTWOLOWQCDMCcombErrBinB{7.8}
\def\jTWOLOWQCDMCnoEvtsBinC{11}
\def\jTWOLOWQCDMCstatErrBinC{3.4}
\def\jTWOLOWQCDMCsystErrUpBinC{\ensuremath{\leq 0.01}}
\def\jTWOLOWQCDMCsystErrDownBinC{\ensuremath{\leq 0.01}}
\def\jTWOLOWQCDMCcombErrBinC{3.4}
\def\jTWOLOWQCDMCnoEvtsBinD{7.2}
\def\jTWOLOWQCDMCstatErrBinD{2.6}
\def\jTWOLOWQCDMCsystErrUpBinD{\ensuremath{\leq 0.01}}
\def\jTWOLOWQCDMCsystErrDownBinD{\ensuremath{\leq 0.01}}
\def\jTWOLOWQCDMCcombErrBinD{2.6}
\def\jTWOLOWQCDMCnoEvtsBinE{1.0}
\def\jTWOLOWQCDMCstatErrBinE{0.3}
\def\jTWOLOWQCDMCsystErrUpBinE{\ensuremath{\leq 0.01}}
\def\jTWOLOWQCDMCsystErrDownBinE{\ensuremath{\leq 0.01}}
\def\jTWOLOWQCDMCcombErrBinE{0.3}
\def\jTWOLOWQCDMCnoEvtsBinF{0.6}
\def\jTWOLOWQCDMCstatErrBinF{0.2}
\def\jTWOLOWQCDMCsystErrUpBinF{\ensuremath{\leq 0.01}}
\def\jTWOLOWQCDMCsystErrDownBinF{\ensuremath{\leq 0.01}}
\def\jTWOLOWQCDMCcombErrBinF{0.2}
\def\jTWOLOWPJMCnoEvts{68}
\def\jTWOLOWPJMCstatErr{1.3}
\def\jTWOLOWPJMCsystErrUp{\ensuremath{\leq 0.01}}
\def\jTWOLOWPJMCsystErrDown{\ensuremath{\leq 0.01}}
\def\jTWOLOWPJMCnoEvtsBinA{34}
\def\jTWOLOWPJMCstatErrBinA{1.1}
\def\jTWOLOWPJMCsystErrUpBinA{\ensuremath{\leq 0.01}}
\def\jTWOLOWPJMCsystErrDownBinA{\ensuremath{\leq 0.01}}
\def\jTWOLOWPJMCcombErrBinA{1.1}
\def\jTWOLOWPJMCnoEvtsBinB{22}
\def\jTWOLOWPJMCstatErrBinB{0.8}
\def\jTWOLOWPJMCsystErrUpBinB{\ensuremath{\leq 0.01}}
\def\jTWOLOWPJMCsystErrDownBinB{\ensuremath{\leq 0.01}}
\def\jTWOLOWPJMCcombErrBinB{0.8}
\def\jTWOLOWPJMCnoEvtsBinC{6.6}
\def\jTWOLOWPJMCstatErrBinC{0.3}
\def\jTWOLOWPJMCsystErrUpBinC{\ensuremath{\leq 0.01}}
\def\jTWOLOWPJMCsystErrDownBinC{\ensuremath{\leq 0.01}}
\def\jTWOLOWPJMCcombErrBinC{0.3}
\def\jTWOLOWPJMCnoEvtsBinD{3.4}
\def\jTWOLOWPJMCstatErrBinD{0.2}
\def\jTWOLOWPJMCsystErrUpBinD{\ensuremath{\leq 0.01}}
\def\jTWOLOWPJMCsystErrDownBinD{\ensuremath{\leq 0.01}}
\def\jTWOLOWPJMCcombErrBinD{0.2}
\def\jTWOLOWPJMCnoEvtsBinE{0.7}
\def\jTWOLOWPJMCstatErrBinE{0.1}
\def\jTWOLOWPJMCsystErrUpBinE{\ensuremath{\leq 0.01}}
\def\jTWOLOWPJMCsystErrDownBinE{\ensuremath{\leq 0.01}}
\def\jTWOLOWPJMCcombErrBinE{0.1}
\def\jTWOLOWPJMCnoEvtsBinF{0.2}
\def\jTWOLOWPJMCstatErrBinF{\ensuremath{\leq 0.01}}
\def\jTWOLOWPJMCsystErrUpBinF{\ensuremath{\leq 0.01}}
\def\jTWOLOWPJMCsystErrDownBinF{\ensuremath{\leq 0.01}}
\def\jTWOLOWPJMCcombErrBinF{\ensuremath{\leq 0.01}}
\def\jTWOLOWEWKDatanoEvts{17}
\def\jTWOLOWEWKDatastatErr{0.3}
\def\jTWOLOWEWKDatasystErrUp{7.2}
\def\jTWOLOWEWKDatasystErrDown{7.2}
\def\jTWOLOWEWKDatanoEvtsBinA{4.5}
\def\jTWOLOWEWKDatastatErrBinA{0.2}
\def\jTWOLOWEWKDatasystErrUpBinA{1.9}
\def\jTWOLOWEWKDatasystErrDownBinA{1.9}
\def\jTWOLOWEWKDatacombErrBinA{1.9}
\def\jTWOLOWEWKDatanoEvtsBinB{6.0}
\def\jTWOLOWEWKDatastatErrBinB{0.2}
\def\jTWOLOWEWKDatasystErrUpBinB{2.5}
\def\jTWOLOWEWKDatasystErrDownBinB{2.5}
\def\jTWOLOWEWKDatacombErrBinB{2.5}
\def\jTWOLOWEWKDatanoEvtsBinC{3.2}
\def\jTWOLOWEWKDatastatErrBinC{0.1}
\def\jTWOLOWEWKDatasystErrUpBinC{1.3}
\def\jTWOLOWEWKDatasystErrDownBinC{1.3}
\def\jTWOLOWEWKDatacombErrBinC{1.3}
\def\jTWOLOWEWKDatanoEvtsBinD{2.3}
\def\jTWOLOWEWKDatastatErrBinD{0.1}
\def\jTWOLOWEWKDatasystErrUpBinD{1.0}
\def\jTWOLOWEWKDatasystErrDownBinD{1.0}
\def\jTWOLOWEWKDatacombErrBinD{1.0}
\def\jTWOLOWEWKDatanoEvtsBinE{0.8}
\def\jTWOLOWEWKDatastatErrBinE{0.1}
\def\jTWOLOWEWKDatasystErrUpBinE{0.4}
\def\jTWOLOWEWKDatasystErrDownBinE{0.4}
\def\jTWOLOWEWKDatacombErrBinE{0.4}
\def\jTWOLOWEWKDatanoEvtsBinF{0.4}
\def\jTWOLOWEWKDatastatErrBinF{0.1}
\def\jTWOLOWEWKDatasystErrUpBinF{0.2}
\def\jTWOLOWEWKDatasystErrDownBinF{0.2}
\def\jTWOLOWEWKDatacombErrBinF{0.2}
\def\jTWOLOWEWKDatasystErrUpRel{1.1}
\def\jTWOLOWEWKDatasystErrDownRel{1.1}
\def\jTWOLOWQCDPJDatanoEvts{608}
\def\jTWOLOWQCDPJDatastatErr{47}
\def\jTWOLOWQCDPJDatasystErrUp{54}
\def\jTWOLOWQCDPJDatasystErrDown{54}
\def\jTWOLOWQCDPJDatanoEvtsBinA{262}
\def\jTWOLOWQCDPJDatastatErrBinA{30}
\def\jTWOLOWQCDPJDatasystErrUpBinA{22}
\def\jTWOLOWQCDPJDatasystErrDownBinA{22}
\def\jTWOLOWQCDPJDatacombErrBinA{37}
\def\jTWOLOWQCDPJDatanoEvtsBinB{173}
\def\jTWOLOWQCDPJDatastatErrBinB{22}
\def\jTWOLOWQCDPJDatasystErrUpBinB{15}
\def\jTWOLOWQCDPJDatasystErrDownBinB{15}
\def\jTWOLOWQCDPJDatacombErrBinB{27}
\def\jTWOLOWQCDPJDatanoEvtsBinC{82}
\def\jTWOLOWQCDPJDatastatErrBinC{23}
\def\jTWOLOWQCDPJDatasystErrUpBinC{6.9}
\def\jTWOLOWQCDPJDatasystErrDownBinC{6.9}
\def\jTWOLOWQCDPJDatacombErrBinC{24}
\def\jTWOLOWQCDPJDatanoEvtsBinD{55}
\def\jTWOLOWQCDPJDatastatErrBinD{12}
\def\jTWOLOWQCDPJDatasystErrUpBinD{6.7}
\def\jTWOLOWQCDPJDatasystErrDownBinD{6.7}
\def\jTWOLOWQCDPJDatacombErrBinD{14}
\def\jTWOLOWQCDPJDatanoEvtsBinE{29}
\def\jTWOLOWQCDPJDatastatErrBinE{10}
\def\jTWOLOWQCDPJDatasystErrUpBinE{2.4}
\def\jTWOLOWQCDPJDatasystErrDownBinE{2.4}
\def\jTWOLOWQCDPJDatacombErrBinE{11}
\def\jTWOLOWQCDPJDatanoEvtsBinF{6.8}
\def\jTWOLOWQCDPJDatastatErrBinF{4.1}
\def\jTWOLOWQCDPJDatasystErrUpBinF{0.8}
\def\jTWOLOWQCDPJDatasystErrDownBinF{0.8}
\def\jTWOLOWQCDPJDatacombErrBinF{4.2}
\def\jTWOLOWQCDPJDatasystErrUpRel{8.2}
\def\jTWOLOWQCDPJDatasystErrDownRel{8.2}
\def\jTWOLOWPhotonVMCnoEvts{30}
\def\jTWOLOWPhotonVMCstatErr{3.4}
\def\jTWOLOWPhotonVMCsystErrUp{15}
\def\jTWOLOWPhotonVMCsystErrDown{15}
\def\jTWOLOWPhotonVMCnoEvtsBinA{4.7}
\def\jTWOLOWPhotonVMCstatErrBinA{1.3}
\def\jTWOLOWPhotonVMCsystErrUpBinA{2.4}
\def\jTWOLOWPhotonVMCsystErrDownBinA{2.4}
\def\jTWOLOWPhotonVMCcombErrBinA{1.3}
\def\jTWOLOWPhotonVMCnoEvtsBinB{8.2}
\def\jTWOLOWPhotonVMCstatErrBinB{1.8}
\def\jTWOLOWPhotonVMCsystErrUpBinB{4.1}
\def\jTWOLOWPhotonVMCsystErrDownBinB{4.1}
\def\jTWOLOWPhotonVMCcombErrBinB{1.8}
\def\jTWOLOWPhotonVMCnoEvtsBinC{5.5}
\def\jTWOLOWPhotonVMCstatErrBinC{1.5}
\def\jTWOLOWPhotonVMCsystErrUpBinC{2.7}
\def\jTWOLOWPhotonVMCsystErrDownBinC{2.7}
\def\jTWOLOWPhotonVMCcombErrBinC{1.5}
\def\jTWOLOWPhotonVMCnoEvtsBinD{5.4}
\def\jTWOLOWPhotonVMCstatErrBinD{1.3}
\def\jTWOLOWPhotonVMCsystErrUpBinD{2.7}
\def\jTWOLOWPhotonVMCsystErrDownBinD{2.7}
\def\jTWOLOWPhotonVMCcombErrBinD{1.3}
\def\jTWOLOWPhotonVMCnoEvtsBinE{4.0}
\def\jTWOLOWPhotonVMCstatErrBinE{1.3}
\def\jTWOLOWPhotonVMCsystErrUpBinE{2.0}
\def\jTWOLOWPhotonVMCsystErrDownBinE{2.0}
\def\jTWOLOWPhotonVMCcombErrBinE{1.3}
\def\jTWOLOWPhotonVMCnoEvtsBinF{1.7}
\def\jTWOLOWPhotonVMCstatErrBinF{0.9}
\def\jTWOLOWPhotonVMCsystErrUpBinF{0.8}
\def\jTWOLOWPhotonVMCsystErrDownBinF{0.8}
\def\jTWOLOWPhotonVMCcombErrBinF{0.9}
\def\jTWOLOWPhotonVMCsystErrUpRel{2.2}
\def\jTWOLOWPhotonVMCsystErrDownRel{2.2}
\def\jTWOLOWTTFSRMCnoEvts{4.1}
\def\jTWOLOWTTFSRMCstatErr{0.9}
\def\jTWOLOWTTFSRMCsystErrUp{2.1}
\def\jTWOLOWTTFSRMCsystErrDown{2.1}
\def\jTWOLOWTTFSRMCnoEvtsBinA{0.6}
\def\jTWOLOWTTFSRMCstatErrBinA{0.3}
\def\jTWOLOWTTFSRMCsystErrUpBinA{0.3}
\def\jTWOLOWTTFSRMCsystErrDownBinA{0.3}
\def\jTWOLOWTTFSRMCcombErrBinA{0.3}
\def\jTWOLOWTTFSRMCnoEvtsBinB{1.7}
\def\jTWOLOWTTFSRMCstatErrBinB{0.6}
\def\jTWOLOWTTFSRMCsystErrUpBinB{0.9}
\def\jTWOLOWTTFSRMCsystErrDownBinB{0.9}
\def\jTWOLOWTTFSRMCcombErrBinB{0.6}
\def\jTWOLOWTTFSRMCnoEvtsBinC{0.9}
\def\jTWOLOWTTFSRMCstatErrBinC{0.4}
\def\jTWOLOWTTFSRMCsystErrUpBinC{0.5}
\def\jTWOLOWTTFSRMCsystErrDownBinC{0.5}
\def\jTWOLOWTTFSRMCcombErrBinC{0.4}
\def\jTWOLOWTTFSRMCnoEvtsBinD{0.5}
\def\jTWOLOWTTFSRMCstatErrBinD{0.4}
\def\jTWOLOWTTFSRMCsystErrUpBinD{0.3}
\def\jTWOLOWTTFSRMCsystErrDownBinD{0.3}
\def\jTWOLOWTTFSRMCcombErrBinD{0.4}
\def\jTWOLOWTTFSRMCnoEvtsBinE{0.4}
\def\jTWOLOWTTFSRMCstatErrBinE{0.3}
\def\jTWOLOWTTFSRMCsystErrUpBinE{0.2}
\def\jTWOLOWTTFSRMCsystErrDownBinE{0.2}
\def\jTWOLOWTTFSRMCcombErrBinE{0.3}
\def\jTWOLOWTTFSRMCnoEvtsBinF{\ensuremath{\leq 0.01}}
\def\jTWOLOWTTFSRMCstatErrBinF{\ensuremath{\leq 0.01}}
\def\jTWOLOWTTFSRMCsystErrUpBinF{\ensuremath{\leq 0.01}}
\def\jTWOLOWTTFSRMCsystErrDownBinF{\ensuremath{\leq 0.01}}
\def\jTWOLOWTTFSRMCcombErrBinF{\ensuremath{\leq 0.01}}
\def\jTWOLOWTTFSRMCsystErrUpRel{0.3}
\def\jTWOLOWTTFSRMCsystErrDownRel{0.3}
\def\jTWOLOWEWKDatanoEvts{17}
\def\jTWOLOWEWKDatastatErr{0.3}
\def\jTWOLOWEWKDatasystErrUp{7.2}
\def\jTWOLOWEWKDatasystErrDown{7.2}
\def\jTWOLOWEWKDatanoEvtsBinA{4.5}
\def\jTWOLOWEWKDatastatErrBinA{0.2}
\def\jTWOLOWEWKDatasystErrUpBinA{1.9}
\def\jTWOLOWEWKDatasystErrDownBinA{1.9}
\def\jTWOLOWEWKDatacombErrBinA{1.9}
\def\jTWOLOWEWKDatanoEvtsBinB{6.0}
\def\jTWOLOWEWKDatastatErrBinB{0.2}
\def\jTWOLOWEWKDatasystErrUpBinB{2.5}
\def\jTWOLOWEWKDatasystErrDownBinB{2.5}
\def\jTWOLOWEWKDatacombErrBinB{2.5}
\def\jTWOLOWEWKDatanoEvtsBinC{3.2}
\def\jTWOLOWEWKDatastatErrBinC{0.1}
\def\jTWOLOWEWKDatasystErrUpBinC{1.3}
\def\jTWOLOWEWKDatasystErrDownBinC{1.3}
\def\jTWOLOWEWKDatacombErrBinC{1.3}
\def\jTWOLOWEWKDatanoEvtsBinD{2.3}
\def\jTWOLOWEWKDatastatErrBinD{0.1}
\def\jTWOLOWEWKDatasystErrUpBinD{1.0}
\def\jTWOLOWEWKDatasystErrDownBinD{1.0}
\def\jTWOLOWEWKDatacombErrBinD{1.0}
\def\jTWOLOWEWKDatanoEvtsBinE{0.8}
\def\jTWOLOWEWKDatastatErrBinE{0.1}
\def\jTWOLOWEWKDatasystErrUpBinE{0.4}
\def\jTWOLOWEWKDatasystErrDownBinE{0.4}
\def\jTWOLOWEWKDatacombErrBinE{0.4}
\def\jTWOLOWEWKDatanoEvtsBinF{0.4}
\def\jTWOLOWEWKDatastatErrBinF{0.1}
\def\jTWOLOWEWKDatasystErrUpBinF{0.2}
\def\jTWOLOWEWKDatasystErrDownBinF{0.2}
\def\jTWOLOWEWKDatacombErrBinF{0.2}
\def\jTWOLOWEWKDatasystErrUpRel{1.1}
\def\jTWOLOWEWKDatasystErrDownRel{1.1}
\def\jTWOLOWQCDPJDatanoEvts{611}
\def\jTWOLOWQCDPJDatastatErr{48}
\def\jTWOLOWQCDPJDatasystErrUp{57}
\def\jTWOLOWQCDPJDatasystErrDown{57}
\def\jTWOLOWQCDPJDatanoEvtsBinA{265}
\def\jTWOLOWQCDPJDatastatErrBinA{31}
\def\jTWOLOWQCDPJDatasystErrUpBinA{24}
\def\jTWOLOWQCDPJDatasystErrDownBinA{24}
\def\jTWOLOWQCDPJDatacombErrBinA{39}
\def\jTWOLOWQCDPJDatanoEvtsBinB{172}
\def\jTWOLOWQCDPJDatastatErrBinB{22}
\def\jTWOLOWQCDPJDatasystErrUpBinB{16}
\def\jTWOLOWQCDPJDatasystErrDownBinB{16}
\def\jTWOLOWQCDPJDatacombErrBinB{27}
\def\jTWOLOWQCDPJDatanoEvtsBinC{82}
\def\jTWOLOWQCDPJDatastatErrBinC{24}
\def\jTWOLOWQCDPJDatasystErrUpBinC{7.1}
\def\jTWOLOWQCDPJDatasystErrDownBinC{7.1}
\def\jTWOLOWQCDPJDatacombErrBinC{25}
\def\jTWOLOWQCDPJDatanoEvtsBinD{56}
\def\jTWOLOWQCDPJDatastatErrBinD{12}
\def\jTWOLOWQCDPJDatasystErrUpBinD{7.2}
\def\jTWOLOWQCDPJDatasystErrDownBinD{7.2}
\def\jTWOLOWQCDPJDatacombErrBinD{14}
\def\jTWOLOWQCDPJDatanoEvtsBinE{29}
\def\jTWOLOWQCDPJDatastatErrBinE{10}
\def\jTWOLOWQCDPJDatasystErrUpBinE{2.5}
\def\jTWOLOWQCDPJDatasystErrDownBinE{2.5}
\def\jTWOLOWQCDPJDatacombErrBinE{11}
\def\jTWOLOWQCDPJDatanoEvtsBinF{7.0}
\def\jTWOLOWQCDPJDatastatErrBinF{4.3}
\def\jTWOLOWQCDPJDatasystErrUpBinF{0.9}
\def\jTWOLOWQCDPJDatasystErrDownBinF{0.9}
\def\jTWOLOWQCDPJDatacombErrBinF{4.4}
\def\jTWOLOWQCDPJDatasystErrUpRel{8.6}
\def\jTWOLOWQCDPJDatasystErrDownRel{8.6}
\def\jTWOLOWPhotonVMCnoEvts{30}
\def\jTWOLOWPhotonVMCstatErr{3.4}
\def\jTWOLOWPhotonVMCsystErrUp{15}
\def\jTWOLOWPhotonVMCsystErrDown{15}
\def\jTWOLOWPhotonVMCnoEvtsBinA{4.7}
\def\jTWOLOWPhotonVMCstatErrBinA{1.3}
\def\jTWOLOWPhotonVMCsystErrUpBinA{2.4}
\def\jTWOLOWPhotonVMCsystErrDownBinA{2.4}
\def\jTWOLOWPhotonVMCcombErrBinA{1.3}
\def\jTWOLOWPhotonVMCnoEvtsBinB{8.2}
\def\jTWOLOWPhotonVMCstatErrBinB{1.8}
\def\jTWOLOWPhotonVMCsystErrUpBinB{4.1}
\def\jTWOLOWPhotonVMCsystErrDownBinB{4.1}
\def\jTWOLOWPhotonVMCcombErrBinB{1.8}
\def\jTWOLOWPhotonVMCnoEvtsBinC{5.5}
\def\jTWOLOWPhotonVMCstatErrBinC{1.5}
\def\jTWOLOWPhotonVMCsystErrUpBinC{2.7}
\def\jTWOLOWPhotonVMCsystErrDownBinC{2.7}
\def\jTWOLOWPhotonVMCcombErrBinC{1.5}
\def\jTWOLOWPhotonVMCnoEvtsBinD{5.4}
\def\jTWOLOWPhotonVMCstatErrBinD{1.3}
\def\jTWOLOWPhotonVMCsystErrUpBinD{2.7}
\def\jTWOLOWPhotonVMCsystErrDownBinD{2.7}
\def\jTWOLOWPhotonVMCcombErrBinD{1.3}
\def\jTWOLOWPhotonVMCnoEvtsBinE{4.0}
\def\jTWOLOWPhotonVMCstatErrBinE{1.3}
\def\jTWOLOWPhotonVMCsystErrUpBinE{2.0}
\def\jTWOLOWPhotonVMCsystErrDownBinE{2.0}
\def\jTWOLOWPhotonVMCcombErrBinE{1.3}
\def\jTWOLOWPhotonVMCnoEvtsBinF{1.7}
\def\jTWOLOWPhotonVMCstatErrBinF{0.9}
\def\jTWOLOWPhotonVMCsystErrUpBinF{0.8}
\def\jTWOLOWPhotonVMCsystErrDownBinF{0.8}
\def\jTWOLOWPhotonVMCcombErrBinF{0.9}
\def\jTWOLOWPhotonVMCsystErrUpRel{2.2}
\def\jTWOLOWPhotonVMCsystErrDownRel{2.2}
\def\jTWOLOWTTFSRMCnoEvts{4.1}
\def\jTWOLOWTTFSRMCstatErr{0.9}
\def\jTWOLOWTTFSRMCsystErrUp{2.1}
\def\jTWOLOWTTFSRMCsystErrDown{2.1}
\def\jTWOLOWTTFSRMCnoEvtsBinA{0.6}
\def\jTWOLOWTTFSRMCstatErrBinA{0.3}
\def\jTWOLOWTTFSRMCsystErrUpBinA{0.3}
\def\jTWOLOWTTFSRMCsystErrDownBinA{0.3}
\def\jTWOLOWTTFSRMCcombErrBinA{0.3}
\def\jTWOLOWTTFSRMCnoEvtsBinB{1.7}
\def\jTWOLOWTTFSRMCstatErrBinB{0.6}
\def\jTWOLOWTTFSRMCsystErrUpBinB{0.9}
\def\jTWOLOWTTFSRMCsystErrDownBinB{0.9}
\def\jTWOLOWTTFSRMCcombErrBinB{0.6}
\def\jTWOLOWTTFSRMCnoEvtsBinC{0.9}
\def\jTWOLOWTTFSRMCstatErrBinC{0.4}
\def\jTWOLOWTTFSRMCsystErrUpBinC{0.5}
\def\jTWOLOWTTFSRMCsystErrDownBinC{0.5}
\def\jTWOLOWTTFSRMCcombErrBinC{0.4}
\def\jTWOLOWTTFSRMCnoEvtsBinD{0.5}
\def\jTWOLOWTTFSRMCstatErrBinD{0.4}
\def\jTWOLOWTTFSRMCsystErrUpBinD{0.3}
\def\jTWOLOWTTFSRMCsystErrDownBinD{0.3}
\def\jTWOLOWTTFSRMCcombErrBinD{0.4}
\def\jTWOLOWTTFSRMCnoEvtsBinE{0.4}
\def\jTWOLOWTTFSRMCstatErrBinE{0.3}
\def\jTWOLOWTTFSRMCsystErrUpBinE{0.2}
\def\jTWOLOWTTFSRMCsystErrDownBinE{0.2}
\def\jTWOLOWTTFSRMCcombErrBinE{0.3}
\def\jTWOLOWTTFSRMCnoEvtsBinF{\ensuremath{\leq 0.01}}
\def\jTWOLOWTTFSRMCstatErrBinF{\ensuremath{\leq 0.01}}
\def\jTWOLOWTTFSRMCsystErrUpBinF{\ensuremath{\leq 0.01}}
\def\jTWOLOWTTFSRMCsystErrDownBinF{\ensuremath{\leq 0.01}}
\def\jTWOLOWTTFSRMCcombErrBinF{\ensuremath{\leq 0.01}}
\def\jTWOLOWTTFSRMCsystErrUpRel{0.3}
\def\jTWOLOWTTFSRMCsystErrDownRel{0.3}
\def\jTWOLOWDatanoEvts{ 615}
\def\jTWOLOWDatastatErr{25}
\def\jTWOLOWDatasystErrUp{\ensuremath{\leq 0.01}}
\def\jTWOLOWDatasystErrDown{\ensuremath{\leq 0.01}}
\def\jTWOLOWDatanoEvtsBinA{ 283}
\def\jTWOLOWDatastatErrBinA{17}
\def\jTWOLOWDatasystErrUpBinA{\ensuremath{\leq 0.01}}
\def\jTWOLOWDatasystErrDownBinA{\ensuremath{\leq 0.01}}
\def\jTWOLOWDatacombErrBinA{17}
\def\jTWOLOWDatanoEvtsBinB{ 199}
\def\jTWOLOWDatastatErrBinB{14}
\def\jTWOLOWDatasystErrUpBinB{\ensuremath{\leq 0.01}}
\def\jTWOLOWDatasystErrDownBinB{\ensuremath{\leq 0.01}}
\def\jTWOLOWDatacombErrBinB{14}
\def\jTWOLOWDatanoEvtsBinC{ 70}
\def\jTWOLOWDatastatErrBinC{8.4}
\def\jTWOLOWDatasystErrUpBinC{\ensuremath{\leq 0.01}}
\def\jTWOLOWDatasystErrDownBinC{\ensuremath{\leq 0.01}}
\def\jTWOLOWDatacombErrBinC{8.4}
\def\jTWOLOWDatanoEvtsBinD{ 39}
\def\jTWOLOWDatastatErrBinD{6.2}
\def\jTWOLOWDatasystErrUpBinD{\ensuremath{\leq 0.01}}
\def\jTWOLOWDatasystErrDownBinD{\ensuremath{\leq 0.01}}
\def\jTWOLOWDatacombErrBinD{6.2}
\def\jTWOLOWDatanoEvtsBinE{ 20}
\def\jTWOLOWDatastatErrBinE{4.5}
\def\jTWOLOWDatasystErrUpBinE{\ensuremath{\leq 0.01}}
\def\jTWOLOWDatasystErrDownBinE{\ensuremath{\leq 0.01}}
\def\jTWOLOWDatacombErrBinE{4.5}
\def\jTWOLOWDatanoEvtsBinF{ 4}
\def\jTWOLOWDatastatErrBinF{2.0}
\def\jTWOLOWDatasystErrUpBinF{\ensuremath{\leq 0.01}}
\def\jTWOLOWDatasystErrDownBinF{\ensuremath{\leq 0.01}}
\def\jTWOLOWDatacombErrBinF{2.0}
\def\jTWOLOWSMBkgTotnoEvts{659}
\def\jTWOLOWSMBkgTotstatErr{47}
\def\jTWOLOWSMBkgTotsystErrUp{78}
\def\jTWOLOWSMBkgTotsystErrDown{78}
\def\jTWOLOWSMBkgTotnoEvtsBinA{272}
\def\jTWOLOWSMBkgTotstatErrBinA{30}
\def\jTWOLOWSMBkgTotsystErrUpBinA{27}
\def\jTWOLOWSMBkgTotsystErrDownBinA{27}
\def\jTWOLOWSMBkgTotcombErrBinA{40}
\def\jTWOLOWSMBkgTotnoEvtsBinB{189}
\def\jTWOLOWSMBkgTotstatErrBinB{22}
\def\jTWOLOWSMBkgTotsystErrUpBinB{22}
\def\jTWOLOWSMBkgTotsystErrDownBinB{22}
\def\jTWOLOWSMBkgTotcombErrBinB{32}
\def\jTWOLOWSMBkgTotnoEvtsBinC{91}
\def\jTWOLOWSMBkgTotstatErrBinC{23}
\def\jTWOLOWSMBkgTotsystErrUpBinC{11}
\def\jTWOLOWSMBkgTotsystErrDownBinC{11}
\def\jTWOLOWSMBkgTotcombErrBinC{26}
\def\jTWOLOWSMBkgTotnoEvtsBinD{63}
\def\jTWOLOWSMBkgTotstatErrBinD{12}
\def\jTWOLOWSMBkgTotsystErrUpBinD{11}
\def\jTWOLOWSMBkgTotsystErrDownBinD{11}
\def\jTWOLOWSMBkgTotcombErrBinD{16}
\def\jTWOLOWSMBkgTotnoEvtsBinE{34}
\def\jTWOLOWSMBkgTotstatErrBinE{11}
\def\jTWOLOWSMBkgTotsystErrUpBinE{4.9}
\def\jTWOLOWSMBkgTotsystErrDownBinE{4.9}
\def\jTWOLOWSMBkgTotcombErrBinE{12}
\def\jTWOLOWSMBkgTotnoEvtsBinF{8.8}
\def\jTWOLOWSMBkgTotstatErrBinF{4.2}
\def\jTWOLOWSMBkgTotsystErrUpBinF{1.8}
\def\jTWOLOWSMBkgTotsystErrDownBinF{1.8}
\def\jTWOLOWSMBkgTotcombErrBinF{4.6}
\def\jTWOQCDRELUNCERTFIT{5.5}
\def\jTWOQCDRELUNCERTEXTRAPOL{2.7}
\def\jTWOLOWQCDPJDatanoEvts{608}
\def\jTWOLOWQCDPJDatastatErr{47}
\def\jTWOLOWQCDPJDatasystErrUp{54}
\def\jTWOLOWQCDPJDatasystErrDown{54}
\def\jTWOLOWQCDPJDatanoEvtsBinA{262}
\def\jTWOLOWQCDPJDatastatErrBinA{30}
\def\jTWOLOWQCDPJDatasystErrUpBinA{22}
\def\jTWOLOWQCDPJDatasystErrDownBinA{22}
\def\jTWOLOWQCDPJDatacombErrBinA{37}
\def\jTWOLOWQCDPJDatanoEvtsBinB{173}
\def\jTWOLOWQCDPJDatastatErrBinB{22}
\def\jTWOLOWQCDPJDatasystErrUpBinB{15}
\def\jTWOLOWQCDPJDatasystErrDownBinB{15}
\def\jTWOLOWQCDPJDatacombErrBinB{27}
\def\jTWOLOWQCDPJDatanoEvtsBinC{82}
\def\jTWOLOWQCDPJDatastatErrBinC{23}
\def\jTWOLOWQCDPJDatasystErrUpBinC{6.9}
\def\jTWOLOWQCDPJDatasystErrDownBinC{6.9}
\def\jTWOLOWQCDPJDatacombErrBinC{24}
\def\jTWOLOWQCDPJDatanoEvtsBinD{55}
\def\jTWOLOWQCDPJDatastatErrBinD{12}
\def\jTWOLOWQCDPJDatasystErrUpBinD{6.7}
\def\jTWOLOWQCDPJDatasystErrDownBinD{6.7}
\def\jTWOLOWQCDPJDatacombErrBinD{14}
\def\jTWOLOWQCDPJDatanoEvtsBinE{29}
\def\jTWOLOWQCDPJDatastatErrBinE{10}
\def\jTWOLOWQCDPJDatasystErrUpBinE{2.4}
\def\jTWOLOWQCDPJDatasystErrDownBinE{2.4}
\def\jTWOLOWQCDPJDatacombErrBinE{11}
\def\jTWOLOWQCDPJDatanoEvtsBinF{6.8}
\def\jTWOLOWQCDPJDatastatErrBinF{4.1}
\def\jTWOLOWQCDPJDatasystErrUpBinF{0.8}
\def\jTWOLOWQCDPJDatasystErrDownBinF{0.8}
\def\jTWOLOWQCDPJDatacombErrBinF{4.2}
\def\jTWOLOWQCDPJMCDnoEvts{203}
\def\jTWOLOWQCDPJMCDstatErr{42}
\def\jTWOLOWQCDPJMCDsystErrUp{45}
\def\jTWOLOWQCDPJMCDsystErrDown{45}
\def\jTWOLOWQCDPJMCDnoEvtsBinA{91}
\def\jTWOLOWQCDPJMCDstatErrBinA{22}
\def\jTWOLOWQCDPJMCDsystErrUpBinA{17}
\def\jTWOLOWQCDPJMCDsystErrDownBinA{17}
\def\jTWOLOWQCDPJMCDcombErrBinA{28}
\def\jTWOLOWQCDPJMCDnoEvtsBinB{76}
\def\jTWOLOWQCDPJMCDstatErrBinB{32}
\def\jTWOLOWQCDPJMCDsystErrUpBinB{16}
\def\jTWOLOWQCDPJMCDsystErrDownBinB{16}
\def\jTWOLOWQCDPJMCDcombErrBinB{36}
\def\jTWOLOWQCDPJMCDnoEvtsBinC{27}
\def\jTWOLOWQCDPJMCDstatErrBinC{17}
\def\jTWOLOWQCDPJMCDsystErrUpBinC{9.0}
\def\jTWOLOWQCDPJMCDsystErrDownBinC{9.0}
\def\jTWOLOWQCDPJMCDcombErrBinC{19}
\def\jTWOLOWQCDPJMCDnoEvtsBinD{7.0}
\def\jTWOLOWQCDPJMCDstatErrBinD{1.8}
\def\jTWOLOWQCDPJMCDsystErrUpBinD{1.6}
\def\jTWOLOWQCDPJMCDsystErrDownBinD{1.6}
\def\jTWOLOWQCDPJMCDcombErrBinD{2.4}
\def\jTWOLOWQCDPJMCDnoEvtsBinE{1.8}
\def\jTWOLOWQCDPJMCDstatErrBinE{0.9}
\def\jTWOLOWQCDPJMCDsystErrUpBinE{0.8}
\def\jTWOLOWQCDPJMCDsystErrDownBinE{0.8}
\def\jTWOLOWQCDPJMCDcombErrBinE{1.2}
\def\jTWOLOWQCDPJMCDnoEvtsBinF{0.4}
\def\jTWOLOWQCDPJMCDstatErrBinF{0.1}
\def\jTWOLOWQCDPJMCDsystErrUpBinF{0.1}
\def\jTWOLOWQCDPJMCDsystErrDownBinF{0.1}
\def\jTWOLOWQCDPJMCDcombErrBinF{0.2}
\def\jTWOLOWQCDPJMCnoEvts{197}
\def\jTWOLOWQCDPJMCstatErr{33}
\def\jTWOLOWQCDPJMCsystErrUp{\ensuremath{\leq 0.01}}
\def\jTWOLOWQCDPJMCsystErrDown{\ensuremath{\leq 0.01}}
\def\jTWOLOWQCDPJMCnoEvtsBinA{108}
\def\jTWOLOWQCDPJMCstatErrBinA{31}
\def\jTWOLOWQCDPJMCsystErrUpBinA{\ensuremath{\leq 0.01}}
\def\jTWOLOWQCDPJMCsystErrDownBinA{\ensuremath{\leq 0.01}}
\def\jTWOLOWQCDPJMCcombErrBinA{31}
\def\jTWOLOWQCDPJMCnoEvtsBinB{58}
\def\jTWOLOWQCDPJMCstatErrBinB{7.9}
\def\jTWOLOWQCDPJMCsystErrUpBinB{\ensuremath{\leq 0.01}}
\def\jTWOLOWQCDPJMCsystErrDownBinB{\ensuremath{\leq 0.01}}
\def\jTWOLOWQCDPJMCcombErrBinB{7.9}
\def\jTWOLOWQCDPJMCnoEvtsBinC{18}
\def\jTWOLOWQCDPJMCstatErrBinC{3.4}
\def\jTWOLOWQCDPJMCsystErrUpBinC{\ensuremath{\leq 0.01}}
\def\jTWOLOWQCDPJMCsystErrDownBinC{\ensuremath{\leq 0.01}}
\def\jTWOLOWQCDPJMCcombErrBinC{3.4}
\def\jTWOLOWQCDPJMCnoEvtsBinD{11}
\def\jTWOLOWQCDPJMCstatErrBinD{2.6}
\def\jTWOLOWQCDPJMCsystErrUpBinD{\ensuremath{\leq 0.01}}
\def\jTWOLOWQCDPJMCsystErrDownBinD{\ensuremath{\leq 0.01}}
\def\jTWOLOWQCDPJMCcombErrBinD{2.6}
\def\jTWOLOWQCDPJMCnoEvtsBinE{1.8}
\def\jTWOLOWQCDPJMCstatErrBinE{0.3}
\def\jTWOLOWQCDPJMCsystErrUpBinE{\ensuremath{\leq 0.01}}
\def\jTWOLOWQCDPJMCsystErrDownBinE{\ensuremath{\leq 0.01}}
\def\jTWOLOWQCDPJMCcombErrBinE{0.3}
\def\jTWOLOWQCDPJMCnoEvtsBinF{0.8}
\def\jTWOLOWQCDPJMCstatErrBinF{0.2}
\def\jTWOLOWQCDPJMCsystErrUpBinF{\ensuremath{\leq 0.01}}
\def\jTWOLOWQCDPJMCsystErrDownBinF{\ensuremath{\leq 0.01}}
\def\jTWOLOWQCDPJMCcombErrBinF{0.2}
\def\jTWOLOWEWKDatanoEvts{17}
\def\jTWOLOWEWKDatastatErr{0.3}
\def\jTWOLOWEWKDatasystErrUp{7.2}
\def\jTWOLOWEWKDatasystErrDown{7.2}
\def\jTWOLOWEWKDatanoEvtsBinA{4.5}
\def\jTWOLOWEWKDatastatErrBinA{0.2}
\def\jTWOLOWEWKDatasystErrUpBinA{1.9}
\def\jTWOLOWEWKDatasystErrDownBinA{1.9}
\def\jTWOLOWEWKDatacombErrBinA{1.9}
\def\jTWOLOWEWKDatanoEvtsBinB{6.0}
\def\jTWOLOWEWKDatastatErrBinB{0.2}
\def\jTWOLOWEWKDatasystErrUpBinB{2.5}
\def\jTWOLOWEWKDatasystErrDownBinB{2.5}
\def\jTWOLOWEWKDatacombErrBinB{2.5}
\def\jTWOLOWEWKDatanoEvtsBinC{3.2}
\def\jTWOLOWEWKDatastatErrBinC{0.1}
\def\jTWOLOWEWKDatasystErrUpBinC{1.3}
\def\jTWOLOWEWKDatasystErrDownBinC{1.3}
\def\jTWOLOWEWKDatacombErrBinC{1.3}
\def\jTWOLOWEWKDatanoEvtsBinD{2.3}
\def\jTWOLOWEWKDatastatErrBinD{0.1}
\def\jTWOLOWEWKDatasystErrUpBinD{1.0}
\def\jTWOLOWEWKDatasystErrDownBinD{1.0}
\def\jTWOLOWEWKDatacombErrBinD{1.0}
\def\jTWOLOWEWKDatanoEvtsBinE{0.8}
\def\jTWOLOWEWKDatastatErrBinE{0.1}
\def\jTWOLOWEWKDatasystErrUpBinE{0.4}
\def\jTWOLOWEWKDatasystErrDownBinE{0.4}
\def\jTWOLOWEWKDatacombErrBinE{0.4}
\def\jTWOLOWEWKDatanoEvtsBinF{0.4}
\def\jTWOLOWEWKDatastatErrBinF{0.1}
\def\jTWOLOWEWKDatasystErrUpBinF{0.2}
\def\jTWOLOWEWKDatasystErrDownBinF{0.2}
\def\jTWOLOWEWKDatacombErrBinF{0.2}
\def\jTWOLOWEWKMCDnoEvts{12}
\def\jTWOLOWEWKMCDstatErr{0.3}
\def\jTWOLOWEWKMCDsystErrUp{4.9}
\def\jTWOLOWEWKMCDsystErrDown{4.9}
\def\jTWOLOWEWKMCDnoEvtsBinA{3.2}
\def\jTWOLOWEWKMCDstatErrBinA{0.2}
\def\jTWOLOWEWKMCDsystErrUpBinA{1.3}
\def\jTWOLOWEWKMCDsystErrDownBinA{1.3}
\def\jTWOLOWEWKMCDcombErrBinA{1.4}
\def\jTWOLOWEWKMCDnoEvtsBinB{4.0}
\def\jTWOLOWEWKMCDstatErrBinB{0.2}
\def\jTWOLOWEWKMCDsystErrUpBinB{1.7}
\def\jTWOLOWEWKMCDsystErrDownBinB{1.7}
\def\jTWOLOWEWKMCDcombErrBinB{1.7}
\def\jTWOLOWEWKMCDnoEvtsBinC{2.1}
\def\jTWOLOWEWKMCDstatErrBinC{0.1}
\def\jTWOLOWEWKMCDsystErrUpBinC{0.9}
\def\jTWOLOWEWKMCDsystErrDownBinC{0.9}
\def\jTWOLOWEWKMCDcombErrBinC{0.9}
\def\jTWOLOWEWKMCDnoEvtsBinD{1.5}
\def\jTWOLOWEWKMCDstatErrBinD{0.1}
\def\jTWOLOWEWKMCDsystErrUpBinD{0.6}
\def\jTWOLOWEWKMCDsystErrDownBinD{0.6}
\def\jTWOLOWEWKMCDcombErrBinD{0.7}
\def\jTWOLOWEWKMCDnoEvtsBinE{0.6}
\def\jTWOLOWEWKMCDstatErrBinE{0.1}
\def\jTWOLOWEWKMCDsystErrUpBinE{0.3}
\def\jTWOLOWEWKMCDsystErrDownBinE{0.3}
\def\jTWOLOWEWKMCDcombErrBinE{0.3}
\def\jTWOLOWEWKMCDnoEvtsBinF{0.2}
\def\jTWOLOWEWKMCDstatErrBinF{0.1}
\def\jTWOLOWEWKMCDsystErrUpBinF{0.1}
\def\jTWOLOWEWKMCDsystErrDownBinF{0.1}
\def\jTWOLOWEWKMCDcombErrBinF{0.1}
\def\jTWOLOWEWKMCnoEvts{9.5}
\def\jTWOLOWEWKMCstatErr{2.7}
\def\jTWOLOWEWKMCsystErrUp{\ensuremath{\leq 0.01}}
\def\jTWOLOWEWKMCsystErrDown{\ensuremath{\leq 0.01}}
\def\jTWOLOWEWKMCnoEvtsBinA{2.0}
\def\jTWOLOWEWKMCstatErrBinA{0.6}
\def\jTWOLOWEWKMCsystErrUpBinA{\ensuremath{\leq 0.01}}
\def\jTWOLOWEWKMCsystErrDownBinA{\ensuremath{\leq 0.01}}
\def\jTWOLOWEWKMCcombErrBinA{0.6}
\def\jTWOLOWEWKMCnoEvtsBinB{3.0}
\def\jTWOLOWEWKMCstatErrBinB{0.9}
\def\jTWOLOWEWKMCsystErrUpBinB{\ensuremath{\leq 0.01}}
\def\jTWOLOWEWKMCsystErrDownBinB{\ensuremath{\leq 0.01}}
\def\jTWOLOWEWKMCcombErrBinB{0.9}
\def\jTWOLOWEWKMCnoEvtsBinC{1.5}
\def\jTWOLOWEWKMCstatErrBinC{0.8}
\def\jTWOLOWEWKMCsystErrUpBinC{\ensuremath{\leq 0.01}}
\def\jTWOLOWEWKMCsystErrDownBinC{\ensuremath{\leq 0.01}}
\def\jTWOLOWEWKMCcombErrBinC{0.8}
\def\jTWOLOWEWKMCnoEvtsBinD{2.8}
\def\jTWOLOWEWKMCstatErrBinD{2.4}
\def\jTWOLOWEWKMCsystErrUpBinD{\ensuremath{\leq 0.01}}
\def\jTWOLOWEWKMCsystErrDownBinD{\ensuremath{\leq 0.01}}
\def\jTWOLOWEWKMCcombErrBinD{2.4}
\def\jTWOLOWEWKMCnoEvtsBinE{0.3}
\def\jTWOLOWEWKMCstatErrBinE{0.3}
\def\jTWOLOWEWKMCsystErrUpBinE{\ensuremath{\leq 0.01}}
\def\jTWOLOWEWKMCsystErrDownBinE{\ensuremath{\leq 0.01}}
\def\jTWOLOWEWKMCcombErrBinE{0.3}
\def\jTWOLOWEWKMCnoEvtsBinF{\ensuremath{\leq 0.01}}
\def\jTWOLOWEWKMCstatErrBinF{\ensuremath{\leq 0.01}}
\def\jTWOLOWEWKMCsystErrUpBinF{\ensuremath{\leq 0.01}}
\def\jTWOLOWEWKMCsystErrDownBinF{\ensuremath{\leq 0.01}}
\def\jTWOLOWEWKMCcombErrBinF{\ensuremath{\leq 0.01}}

\begin{table} 
\begin{center}
\caption{Resulting event yields for the $\geq$1 photon and $\geq$2 jet
  selection in \lumiData \fbinv of data for six distinct signal search bins.}
\label{tab:complresult2j}
{
\begin{tabular}{l|cccccc}
\hline
$\MET$ bins $[\GeV]$  &  100--120 & 120--160 & 160--200 & 200--270 & 270--350 &$>350$
\\ \hline
QCD {\footnotesize(from data)}
& $\jTWOLOWQCDPJDatanoEvtsBinA$  $\pm\jTWOLOWQCDPJDatacombErrBinA$ 
&$\jTWOLOWQCDPJDatanoEvtsBinB$  $\pm\jTWOLOWQCDPJDatacombErrBinB$ 
&$\jTWOLOWQCDPJDatanoEvtsBinC$  $\pm\jTWOLOWQCDPJDatacombErrBinC$ 
&$\jTWOLOWQCDPJDatanoEvtsBinD$ $\pm\jTWOLOWQCDPJDatacombErrBinD$ 
&$\jTWOLOWQCDPJDatanoEvtsBinE$  $\pm\jTWOLOWQCDPJDatacombErrBinE$ 
&$\jTWOLOWQCDPJDatanoEvtsBinF$  $\pm\jTWOLOWQCDPJDatacombErrBinF$ 
 \\ 
e$\rightarrow\gamma$ {\footnotesize(from data)}
& $\jTWOLOWEWKDatanoEvtsBinA$  $\pm\jTWOLOWEWKDatacombErrBinA$ 
&$\jTWOLOWEWKDatanoEvtsBinB$  $\pm\jTWOLOWEWKDatacombErrBinB$ 
&$\jTWOLOWEWKDatanoEvtsBinC$  $\pm\jTWOLOWEWKDatacombErrBinC$ 
&$\jTWOLOWEWKDatanoEvtsBinD$  $\pm\jTWOLOWEWKDatacombErrBinD$ 
&$\jTWOLOWEWKDatanoEvtsBinE$  $\pm\jTWOLOWEWKDatacombErrBinE$ 
&$\jTWOLOWEWKDatanoEvtsBinF$  $\pm\jTWOLOWEWKDatacombErrBinF$ 

\\
FSR/ISR {\footnotesize($W$,$Z$)}
& $\jTWOLOWPhotonVMCnoEvtsBinA$  $\pm\jTWOLOWPhotonVMCcombErrBinA$ 
&$\jTWOLOWPhotonVMCnoEvtsBinB$  $\pm\jTWOLOWPhotonVMCcombErrBinB$ 
&$\jTWOLOWPhotonVMCnoEvtsBinC$  $\pm\jTWOLOWPhotonVMCcombErrBinC$ 
&$\jTWOLOWPhotonVMCnoEvtsBinD$  $\pm\jTWOLOWPhotonVMCcombErrBinD$ 
&$\jTWOLOWPhotonVMCnoEvtsBinE$  $\pm\jTWOLOWPhotonVMCcombErrBinE$ 
&$\jTWOLOWPhotonVMCnoEvtsBinF$  $\pm\jTWOLOWPhotonVMCcombErrBinF$ 
\\
FSR/ISR {\footnotesize($\ttbar$)}
& $\jTWOLOWTTFSRMCnoEvtsBinA$  $\pm\jTWOLOWTTFSRMCcombErrBinA$ 
&$\jTWOLOWTTFSRMCnoEvtsBinB$  $\pm\jTWOLOWTTFSRMCcombErrBinB$ 
&$\jTWOLOWTTFSRMCnoEvtsBinC$  $\pm\jTWOLOWTTFSRMCcombErrBinC$ 
&$\jTWOLOWTTFSRMCnoEvtsBinD$  $\pm\jTWOLOWTTFSRMCcombErrBinD$ 
&$\jTWOLOWTTFSRMCnoEvtsBinE$  $\pm\jTWOLOWTTFSRMCcombErrBinE$ 
&$\jTWOLOWTTFSRMCnoEvtsBinF$  
\\
\hline
Total SM estimation
& $\jTWOLOWSMBkgTotnoEvtsBinA$ $\pm 37$ 
&$\jTWOLOWSMBkgTotnoEvtsBinB$  $\pm 27$ 
&$\jTWOLOWSMBkgTotnoEvtsBinC$  $\pm 24$ 
&$\jTWOLOWSMBkgTotnoEvtsBinD$  $\pm 14$ 
&$\jTWOLOWSMBkgTotnoEvtsBinE$  $\pm 11$ 
&$\jTWOLOWSMBkgTotnoEvtsBinF$  $\pm 4.3$ 

\\
\hline
Data
& $\jTWOLOWDatanoEvtsBinA$ 
& $\jTWOLOWDatanoEvtsBinB$ 
& $\jTWOLOWDatanoEvtsBinC$ 
& $\jTWOLOWDatanoEvtsBinD$ 
& $\jTWOLOWDatanoEvtsBinE$ 
& $\jTWOLOWDatanoEvtsBinF$  

\\ 
\hline
\end{tabular}
}
\end{center}
\end{table}

All background components are shown in Fig.~\ref{fig:preselmet} together
with the data (points with errors bars) and two GGM benchmark signal
samples, one excluded (red line) and one not excluded (violet line) by
this analysis. The same information is summarized in
Table~\ref{tab:complresult2j}. No excess beyond standard model
predictions is observed.

\section{Diphoton Analysis}
\label{sec:diphoton}

The diphoton analysis is most sensitive to SUSY scenarios in which the
lightest neutralino is bino-like decaying into a photon and the
gravitino as LSP, as well as models predicting universal extra
dimensions. To keep the analysis as inclusive as possible, no veto is
applied on additional leptons in the event.

\subsection{Background Estimation}

To estimate the QCD background from data in the diphoton analysis, two
different datasets are utilized. The first sample contains two
misidentified photons, and in what follows referred to as the $ff$
(``fake-fake'') sample, comprising multijet events. This is the main
dataset to estimate the QCD background. The second data sample contains
events with two electrons ($\Pe\Pe$) with an invariant mass between 81
and 101\GeV, and is dominated by $\Z\rightarrow \Pe\Pe$ decays. The
$\Pe\Pe$~sample is used to study systematic effects on our background
estimate.  We do not utilize a sample consisting of a real and a
misidentified photon (``photon-fake'' sample) for our background
estimate. Since only one of the photons is misidentified, such a sample
would still contain real diphoton events, giving rise to a potentially
large contamination from signal events. In addition, a ``photon-fake''
sample includes events from photon-jet QCD production. Such events
have kinematic properties (``back-to-back'') that are quite
different from our expected signal events and thus ``photon-fake''
events do not constitute a good choice for a background sample.

Comparing the \MET~resolution between diphoton signal and background
events, the $\ET$ resolution for electrons and misidentified photons is
similar to the resolution for true photons. It is negligible compared
with the resolution for the hadronic energy, which dominates the \MET
resolution.  The events in both control samples are reweighted to
reproduce the diphoton transverse energy distribution in the signal data
sample, and, therefore, the transverse energy of hadronic recoil against
the photons. The \MET distributions in the reweighted control samples
show good agreement with the diphoton signal samples within
uncertainties as shown for the $ff$~sample in
Fig.~\ref{fig:METcomp}. The shape of the \MET~distribution for the
$ff$~sample is used to determine the magnitude of the QCD background
after normalizing the $ff$~background shape to the diphoton data in the
region of low missing transverse energy $\MET<20\GeV$, which is
dominated by QCD background.  We choose to use the prediction from the
$ff$~sample as the estimator of the QCD background while the difference
from the sideband-subtracted $\Pe\Pe$~sample to the $ff$~estimate is
taken as an estimate of the systematic uncertainty in the determination
of the QCD background. The $\Pe\Pe$~sample has been corrected for a
small contribution from diboson production ($\PW\Z$ and $\Z\Z$) using
\PYTHIA with NLO cross section resulting in a correction of $0.2$--$18\%$
depending on \MET bins, and $\Pe\Pe$~events with true \MET.  As an
illustration of the reliability of the QCD background estimate, in the
\MET control region from 30 to 50\GeV, 3443~candidate diphoton events
are observed in the sample requiring ${\ge}1$~jet in the event. In the
same \MET~region the prediction from the $ff$ and $\Pe\Pe$~sample yields
$3636\pm79\stat$ and $3045\pm26\stat$ events, respectively.

The estimated EWK background is determined with the $\Pe\Pe$ and $\Pe\gamma$
samples as described in Section~\ref{sec:backgrounds} and is calculated to
be much smaller than the QCD background. Other backgrounds such as
$\Z\gamma\gamma \rightarrow \nu \nu \gamma \gamma$, $\PW\gamma\gamma
\rightarrow \ell \nu \gamma \gamma$, $\ttbar\gamma\gamma$, or
$\Z\gamma\gamma$ events where the $\Z\rightarrow\tau\tau$ is followed by a
$\tau$ decay such as $\tau\rightarrow\pi\nu$ or $\tau\rightarrow
\Pe(\mu)\nu\nu$ have been found to be ${<}0.1\%$ using simulations.

Drell--Yan events can also contribute as background if both electrons are
misidentified as photons. While the Drell--Yan process does not have true
$\MET$, it can have mismeasured \MET due to resolution effects in the
accompanying hadronic activity. Given the high expected electron pixel
match efficiency, and the relatively low cross section for Drell--Yan
production, the contribution from this background is also negligible.

\subsection{Results}

\begin{figure*}[tbp]
\begin{center}
\includegraphics[width=0.6\textwidth]{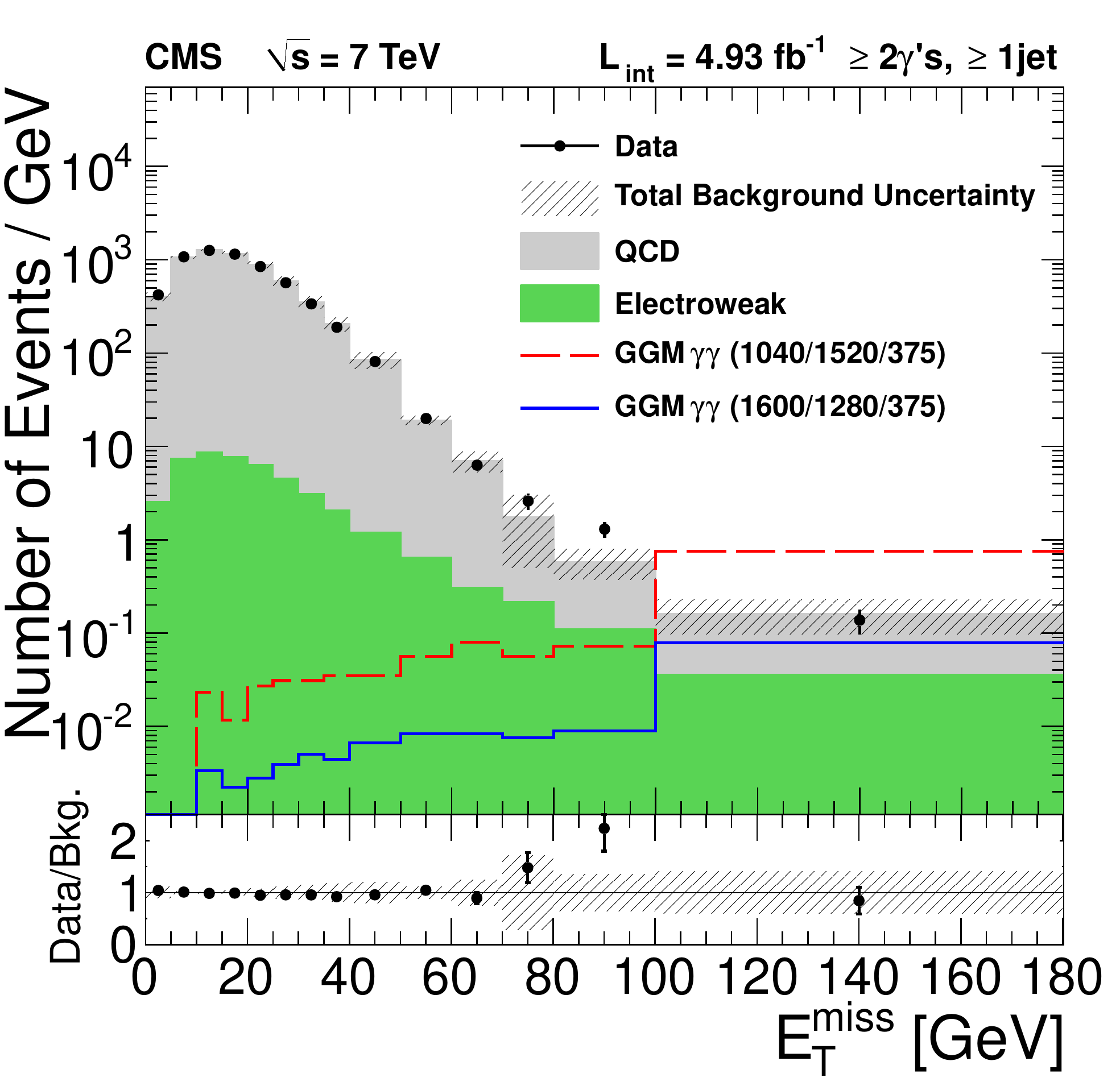}
\caption{\label{fig:METcomp}
  The \MET spectrum of $\gamma\gamma$ data compared to QCD prediction
  together with small EWK background for events with at least one jet.
  The hatched areas indicate the total background uncertainties. Two
  example GGM points (dashed red upper and solid blue lower curves with
  masses of $m_{{\PSq}} / m_{\sGlu} / m_{\chiz_1}$ in\GeV) on either
  side of our exclusion boundary are also shown. At the bottom, the
  ratio of data over standard model prediction is shown as a function of
  \MET.  The error bars take into account only the statistical error of
  the data sample, while the hatched area is the error on the expected
  background from the SM processes.  }
\end{center}
\end{figure*}

The \MET distribution in the $\gamma\gamma$ sample requiring $\ge1$~jet
in the event is presented in Fig.~\ref{fig:METcomp} as points with
errors bars. The green shaded area shows the estimated amount of the EWK
background while the QCD background prediction from the $ff$~sample is
shown in grey after normalization to the $\gamma\gamma$ sample minus the
estimated EWK contribution in the region $\MET\leq20\GeV$.  The
hatched areas indicate the total background uncertainties.

Table~\ref{closurevt} summarizes the observed number of $\gamma\gamma$
events in bins of \MET as well as the expected QCD and EWK background
with statistical and systematic uncertainty. The systematic error is
determined from the difference between the $ff$ sample used to predict
the QCD background and the $\Pe\Pe$~sample utilized as an alternative
background estimate after the $\Pe\Pe$~data are sideband subtracted and
corrected for a small diboson contributions. For the region of large
missing transverse energy, no excess of data over the SM expectation is
found. We observe 11 diphoton events with $\MET\ge100$\GeV while the total
background expectation is calculated to be $13.0\pm4.2\stat\pm1.7\syst$
events.

\begin{table*}[tbp]
  \topcaption{\label{closurevt}
    Number of diphoton candidates from data as well as estimates of QCD
    and EWK background in bins of \MET.  The first error is statistical and the second is systematic for each entry.}
\begin{center}
\begin{tabular}{l|ccccc}
\hline
\MET bins [\GeVns{}] & 50--60 & 60--70 & 70--80 & 80--100 & $>100$ \\
\hline
QCD background & \scriptsize $183.8 \pm 17.7 \pm 12.5$
& \scriptsize$67.3 \pm 10.7 \pm 13.6$
& \scriptsize $15.4 \pm 5.1 \pm 11.5$
& \scriptsize$9.4 \pm 4.0 \pm 0.7$
& \scriptsize$10.1 \pm 4.2 \pm 1.4$ \\
EWK background & \scriptsize $6.5 \pm 0.3 \pm 2.2$
& \scriptsize$3.1 \pm 0.2 \pm 1.0$
& \scriptsize $2.2 \pm 0.2 \pm 0.7$
& \scriptsize$2.2 \pm 0.2 \pm 0.8$
& \scriptsize$2.9 \pm 0.2 \pm 1.0$ \\
\hline
Total background & \scriptsize$190.3\pm17.7\pm12.7$
& \scriptsize$70.4\pm10.7\pm13.7$
& \scriptsize$17.6\pm5.1\pm11.5$
& \scriptsize$11.6\pm4.0\pm1.0$
& \scriptsize$13.0\pm4.2\pm1.7$ \\
\hline
Data & 199 & 63 & 26 & 26 & 11 \\
\hline
\end{tabular}
\end{center}
\end{table*}

\section{Interpretation in Models of New Physics}
\label{sec:limit}

We determine the efficiency for NP signal events to pass our analysis
selections by applying correction factors derived from data to the MC
simulation of the signal. Since there is no large clean sample of
genuine photons in the data, we rely on the similarities between the
detector response to electrons and photons to extract the photon
identification efficiency. A scale factor is obtained and applied to the
photon efficiency in MC simulation by forming a ratio between the
electron efficiency from $\Z\rightarrow \Pe\Pe$ events that pass all
photon selections (except for the pixel match) and the corresponding
electron efficiencies from simulation. The obtained data-to-MC scale
factor $0.994\pm0.002\stat\pm0.035\syst$ is applied to the photon
efficiencies obtained from MC simulation. Other sources of the larger
systematic uncertainties in the signal yield include the error on
integrated luminosity (2.2\%)~\cite{CMS-PAS-SMP-12-008}, pileup effects
on photon identification (2.5\%), and small parton distribution
functions (PDF) uncertainties in the acceptance. Systematic
uncertainties in the theoretical cross section prediction consist of the
PDF uncertainty (4--66\%) and renormalization scale (4--28\%)
uncertainty depending on the parameters of the NP~signal.

The goal of this analysis is to find evidence for the production of NP
by observing an excess of events above the SM background in the
high \MET region of the single-photon and diphoton signal samples. Since
no such excess is observed, upper limits are derived on potential
signals of various NP models. The statistical approach used to derive
limits constructs a test statistic as the product of likelihood ratios
in bins of \MET. These likelihoods are functions of the
predicted signal and background yields in each bin. Systematic
uncertainties are introduced as nuisance parameters in the signal and
background models. Log-normal distributions are taken as a suitable
choice for the probability density distributions of the nuisance
parameters in order to incorporate uncertainties in the background
rates, integrated luminosity, and the signal acceptance times efficiency.

In order to compare the compatibility of the observed data with a NP signal
hypothesis, we use a LHC-style profiled likelihood test
statistics~\cite{ATLAS:1379837}.  In particular, for the comparison of
the data to a signal-plus-background hypothesis, where the
signal and background expectations are functions of nuisance
parameters~$\theta$ and the signal is scaled by a signal strength
parameter~$\mu$, we construct a one-sided test statistic $-2\ln
\widetilde{q}_\mu$ based on the profile likelihood ratio
$\widetilde{q}_\mu = \mathcal{L}(\text{data} | \mu, \widehat{\theta}_\mu) /
\mathcal{L}(\text{data} | \widehat{\mu},\widehat{\theta})$ with constraint
$0\le\widehat{\mu}\le\mu$~\cite{ATLAS:1379837}.
Here, $\widehat{\theta}_\mu$ refers to the conditional maximum likelihood
estimators of $\theta$ given the signal strength parameter $\mu$ and the
actual data. The pair of parameter estimators $\widehat{\theta}$ and
$\widehat{\mu}$ correspond to the global maximum of the likelihood.  The
modified frequentist \CLs criterion \cite{Junk:1999kv,Read:2002hq} is
used to determine upper limits on the cross section of a possible NP
signal at the 95\% confidence level (CL).

To achieve optimal sensitivity, the limits are calculated in distinct
bins and multiple exclusive search channels in \MET\ are combined into
one test statistic considering the bin-to-bin correlations of the
systematic uncertainties.  For the single-photon analysis, six distinct
bins for $\MET \geq 100\GeV$ are used, [100,120), [120,160), [160,200),
[200,270), [270,350), and [350,$\infty$) given in\GeV, while the
diphoton analysis uses the following \MET\ ranges given in \GeV:
[50,60), [60,70), [70,80), [80,100), and [100,$\infty$). These bins in
\MET correspond to the event yields given in
Tables~\ref{tab:complresult2j} and~\ref{closurevt}. In general, the
sensitivity is dominated by the highest \MET\ bin.  Since in both
searches the estimated background exceeds slightly the observed data in
the highest \MET\ bin, the observed limits are generally slightly
stronger than the expected limits. Some regions of the possible signal
phase space, e.g. where the LSP receives only a small amount of
transverse momentum, resulting in small \MET, also benefit from other
search bins and therefore from the combination.

A possible contamination by signal in the control samples used for the
background estimation has been studied and was found to be negligible
for the diphoton final state. For the single-photon analysis the
expected contamination for a given signal is considered in the limit
calculation in the signal-plus-background hypothesis. The background overestimation due to the contamination is
typically a few percent, if the signal cross section is of the same order
than the cross section limits.

\subsection{General Gauge Mediated SUSY Breaking}
\label{sec:GGM}

Since the physical neutralinos \chiz\ and charginos \chipm\ are an
admixture of gaugino eigenstates, different scenarios of gaugino mixing
have been studied. In the first case, referred to as bino-like, the
lightest neutralino is assumed to be pure bino-like, while the lightest
chargino is assumed to be heavy and decoupled. In this case, the
production of the neutralino occurs mostly in the cascade decays of the
squarks and gluinos, since the neutralino pair production cross section
is very small. In the second case, referred to as wino-like, the
neutralino and chargino have comparable mass and are assumed to be pure
wino-like. In this case, both the neutralino and chargino are produced
in squark and gluino decays, but direct chargino-neutralino production
may also contribute. Furthermore, in the wino-like case, the expected
event yields for the single-photon and diphoton analyses are reduced
since the chargino (neutralino) may decay to a \PW\ (\Z) and the
gravitino~(see Fig.~\ref{fig:Feynmans}).

\begin{figure*}[tbp]
\begin{center}
\includegraphics[width=0.49\textwidth]{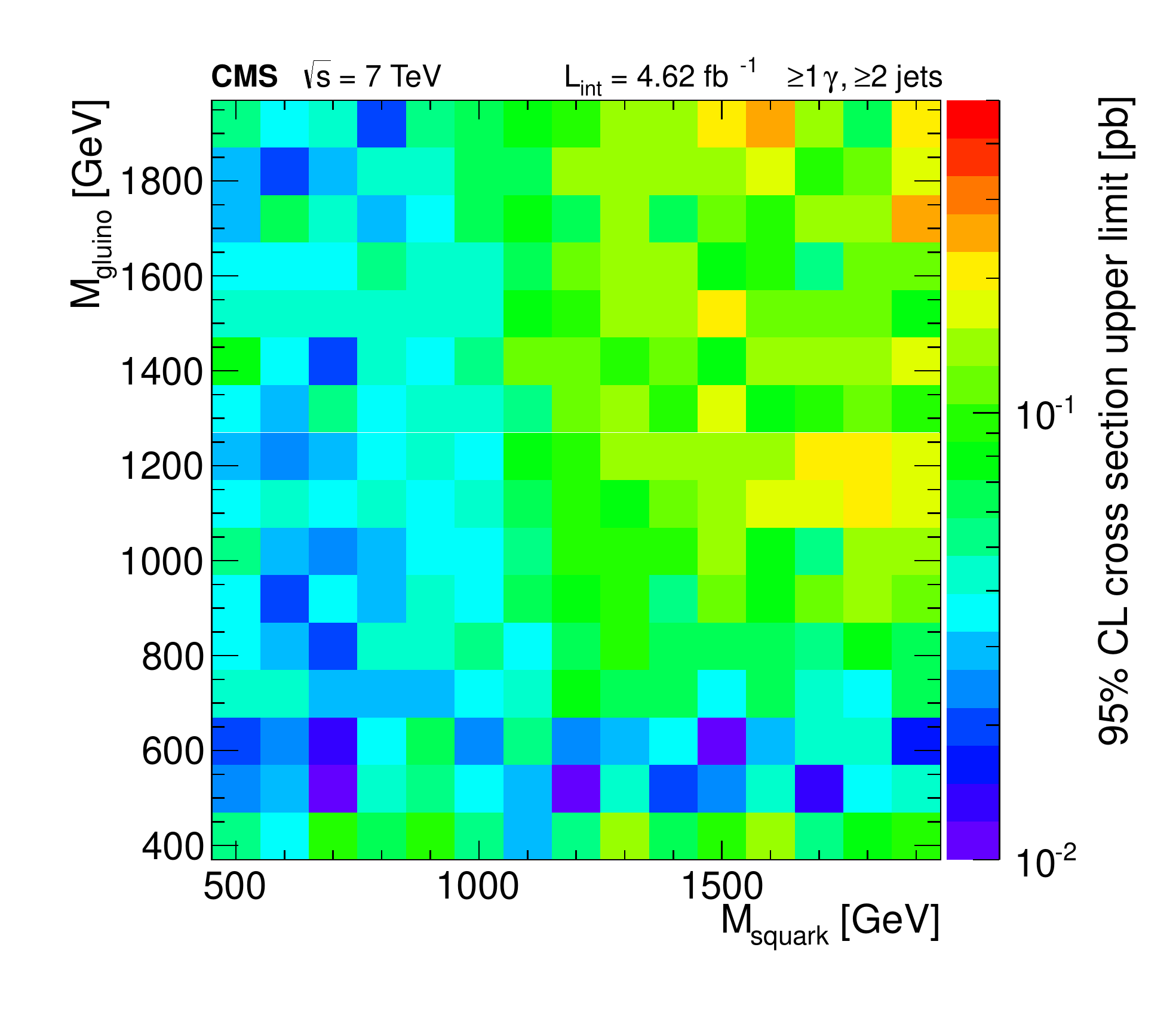}\
\includegraphics[width=0.49\textwidth]{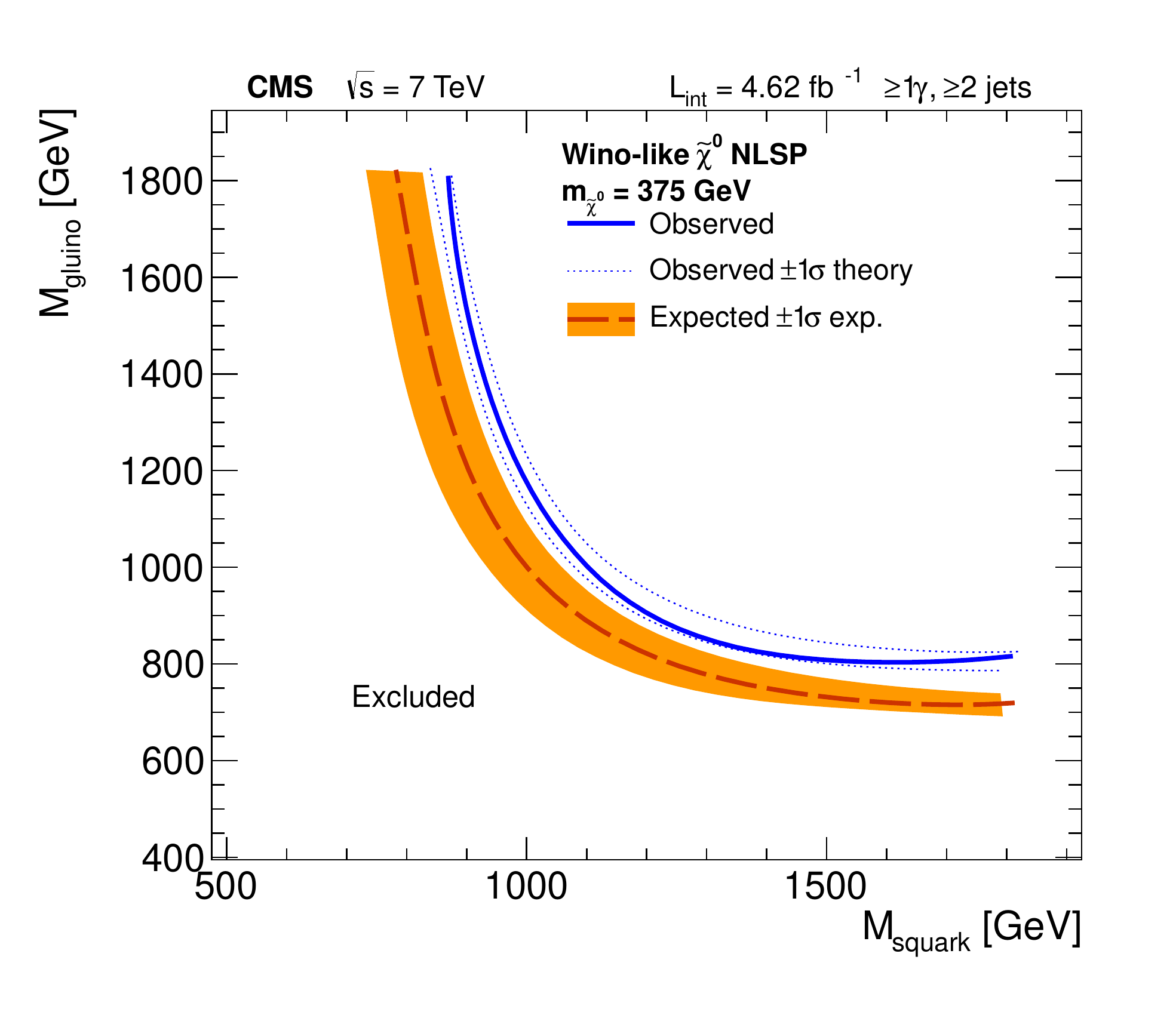}
\includegraphics[width=0.49\textwidth]{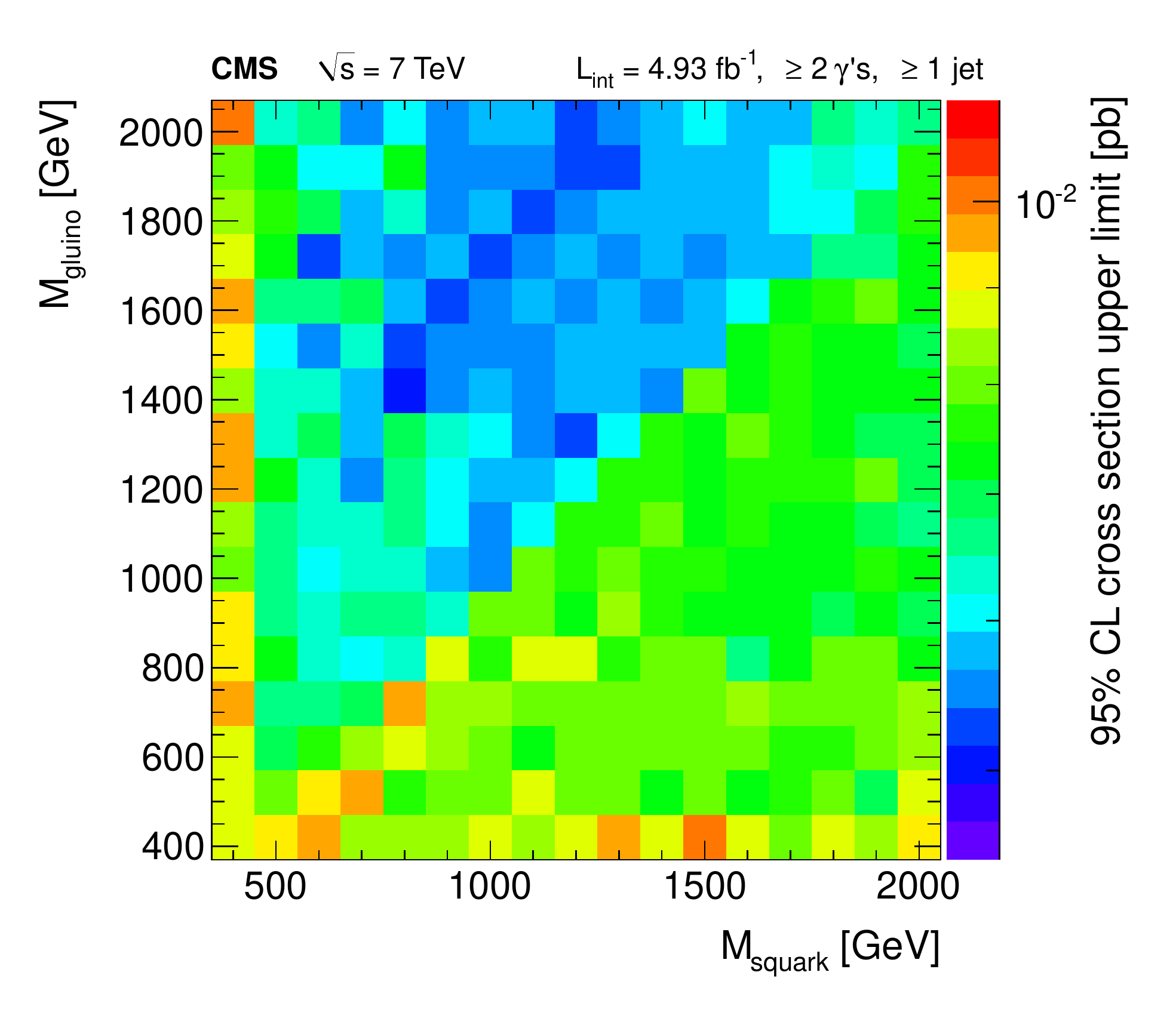}\
\includegraphics[width=0.49\textwidth]{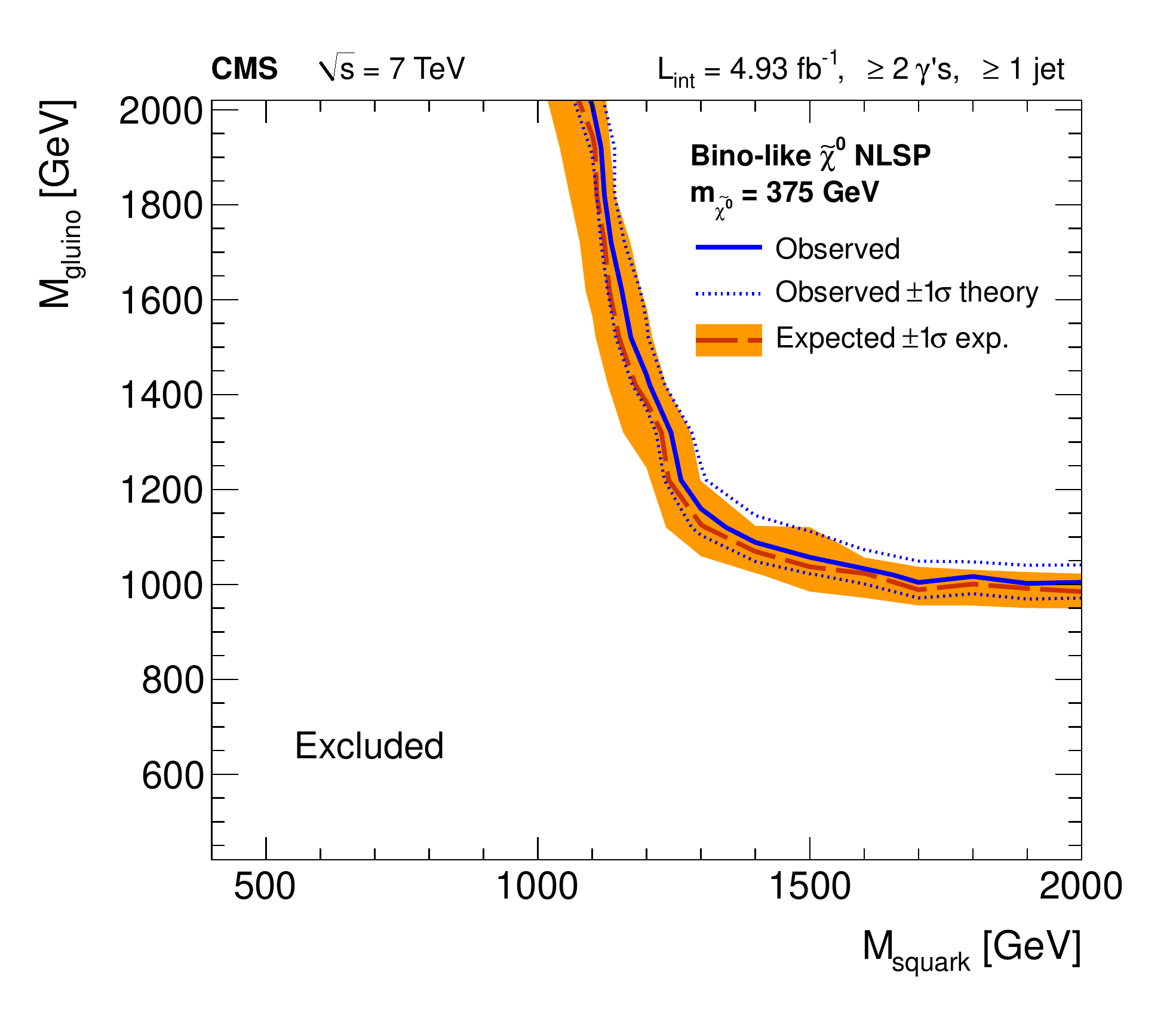}
\caption{\label{fig:GGM_sq_gl}
  Observed upper limits at 95\% CL on the signal cross section (left) and
  corresponding exclusion contours (right) in gluino-squark mass space
  for the single-photon search in the wino-like scenario (top) and the
  diphoton analysis for a bino-like neutralino (bottom).  The shaded
  uncertainty bands around the expected exclusion contours correspond to
  experimental uncertainties, while the NLO renormalization and PDF
  uncertainties of the signal cross section are indicated by dotted lines around the
  observed limit contour.
}
\end{center}
\end{figure*}

The resulting upper limits on the GGM production cross section, at 95\%
CL, as well as exclusion contours are shown in
Fig.~\ref{fig:GGM_sq_gl} for the gluino versus squark mass plane from
400 to 2000\GeV in squark and gluino mass, with the neutralino mass set
at $375\GeV$.  This mass value is chosen to represent a reasonably
light NLSP, but high enough to be outside current exclusion limits.  For
the wino-like scenario, the single-photon cross section upper limit is of order
$0.003$--$0.1$\unit{pb} at 95\% CL with a typical acceptance of ${\sim}7\%$.
For the bino-like scenario, the diphoton cross section limit is of order
$0.003$--$0.01$\unit{pb} at 95\% CL  with a typical acceptance of ${\sim}30\%$ for
$\MET>100\GeV$.  Squark and gluino masses up to about 800\GeV are
excluded in the wino-like scenario by the single-photon search, while
the diphoton analysis excludes squarks and gluinos up to masses of
${\sim}1\TeV$ for a bino-like neutralino, both limits at 95\% CL
The corresponding 95\% CL limits on the signal cross section and
exclusion contours for the single-photon (diphoton) analysis in the
bino-like (wino-like) scenario are
available in \suppMaterial.

\begin{figure*}[tbp]
\begin{center}
\includegraphics[width=0.49\textwidth]{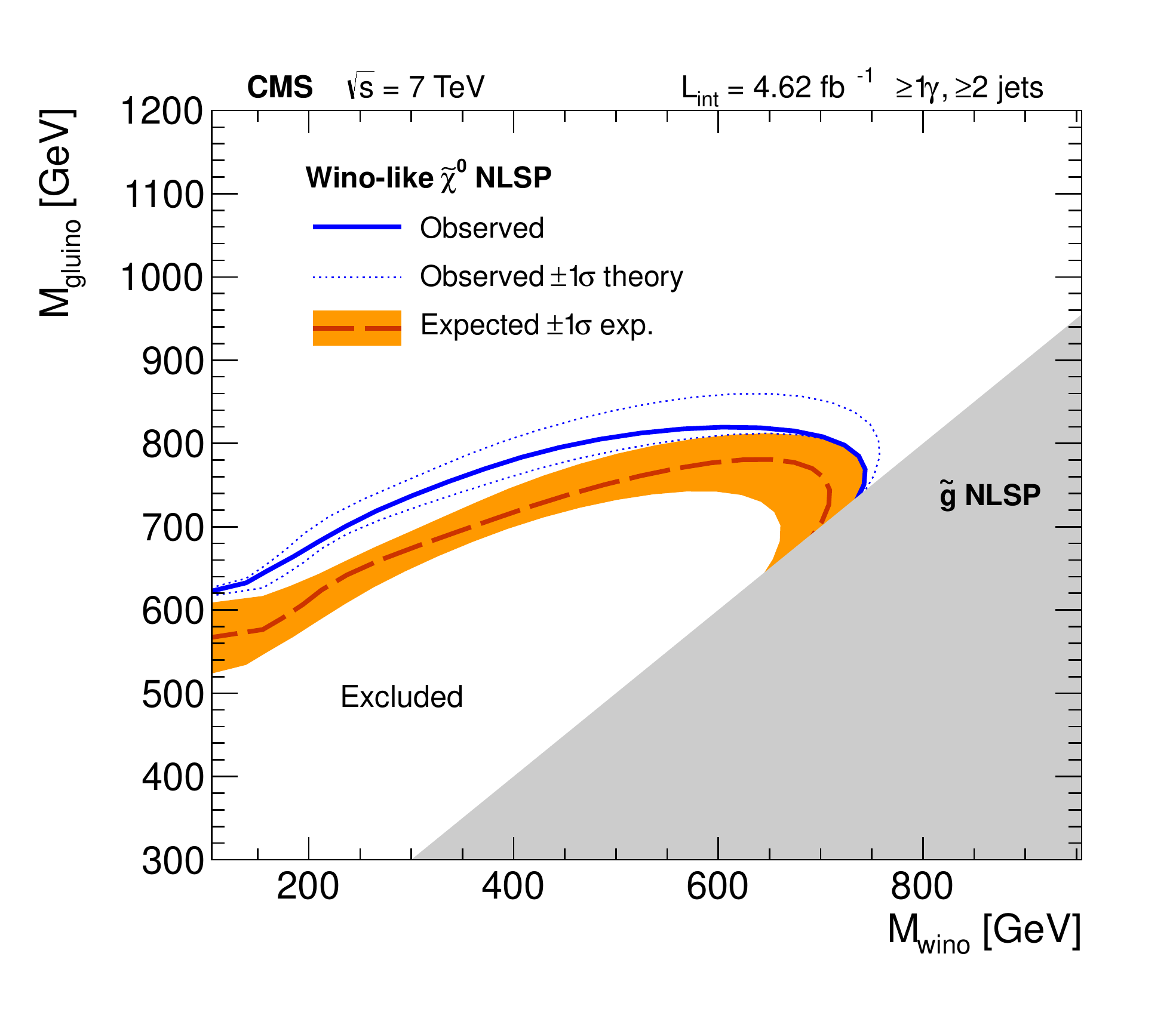}
\includegraphics[width=0.49\textwidth]{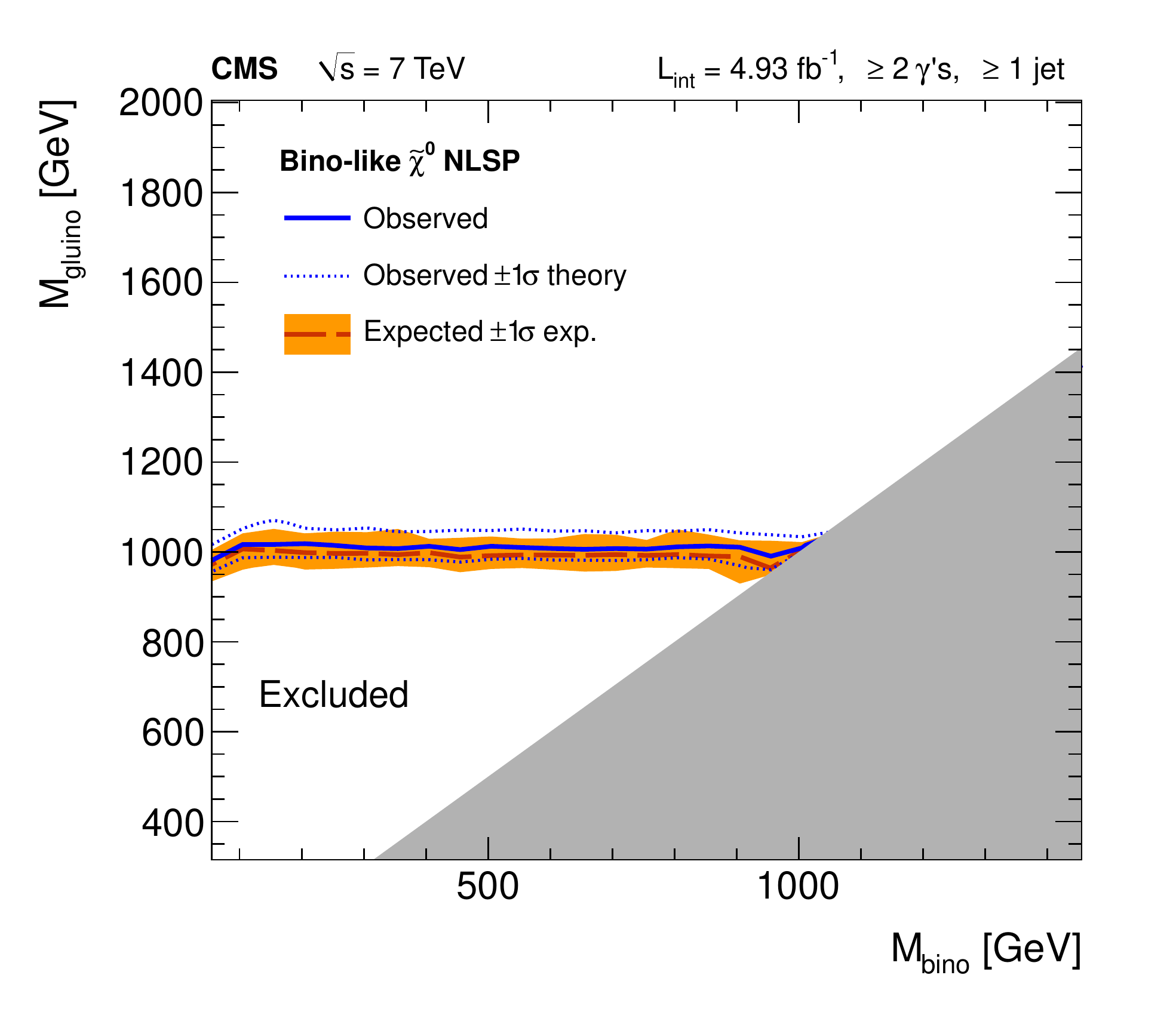}
\caption{\label{fig:GGM_chi1_gl}
  Exclusion contours at 95\% CL in the plane of gluino versus neutralino mass for
  the single-photon search in the wino-like scenario (left) and the
  diphoton analysis for a bino-like neutralino (right).
}
\end{center}
\end{figure*}

As further interpretation of the single-photon and diphoton results,
Fig.~\ref{fig:GGM_chi1_gl} shows the exclusion contours in the plane of
gluino versus neutralino mass for the single-photon wino-like and the
diphoton bino-like scenarios.  The diphoton search excludes gluino
production for a bino-like neutralino for gluino masses up to about 1\TeV rather
independent of the neutralino mass.
The 95\% CL upper limits on the signal cross section for the
single-photon (diphoton) wino-like (bino-like) scenario in the
gluino-neutralino mass plane as well as the corresponding single-photon
bino-like and diphoton wino-like 95\% CL limit plots and contours can
be found in \suppMaterial.

\begin{figure*}[tbp]
\begin{center}
\includegraphics[width=0.49\textwidth]{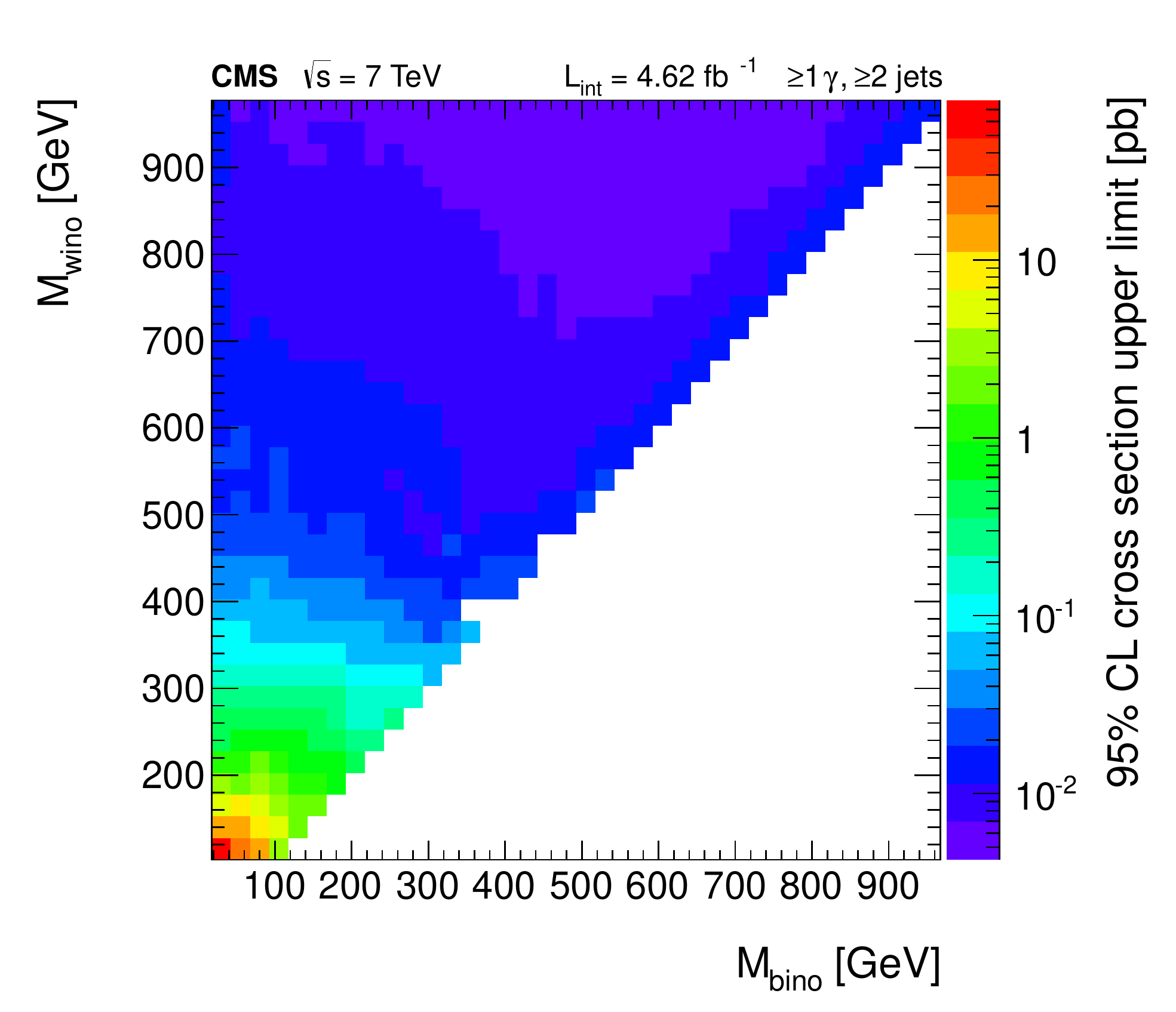}
\includegraphics[width=0.49\textwidth]{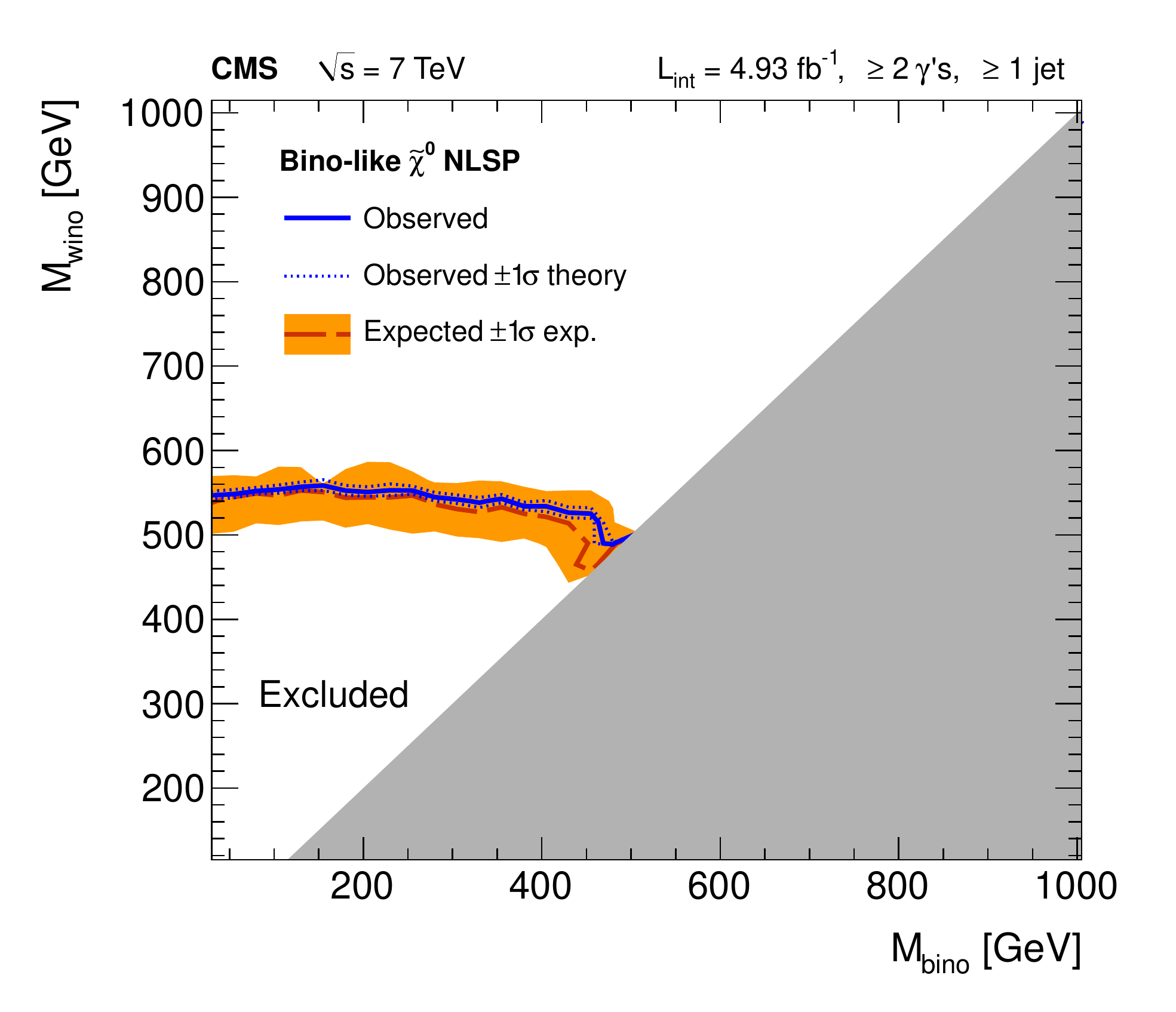}
\caption{\label{fig:GGM_W_B}
  95\% CL exclusion contour and corresponding observed and expected
  contours in the bino-like versus wino-like gaugino mass for the
  diphoton (right) and the cross section limit for the single-photon
  analysis (left).  }
\end{center}
\end{figure*}

Finally, we study for the first time in the final state with photons the electroweak production of winos, i.e. the pair
and associated production of wino-like neutralinos and charginos,
that decay to a bino-like NLSP by decoupling the squarks and gluinos leaving
only electroweak production in the simulated samples.
Figure~\ref{fig:GGM_W_B} shows limits on the signal cross section and exclusion contours in the plane of wino-like
versus bino-like gaugino mass for the single-photon and diphoton
analyses, where the diphoton search excludes wino masses up to about
500\GeV almost independent of the bino mass.  Since no continuous
exclusion contour line can be drawn for the single-photon analysis, we
can only present the 95\% CL upper limits on the signal cross section.
The corresponding 95\% CL~upper limits on the signal cross section in
the plane of wino-like versus bino-like gaugino mass for the
diphoton analysis are available
in \suppMaterial.

\subsection{Simplified Models}
\label{sec:SMS}

In this section we interpret the results of our single-photon and
diphoton search in terms of Simplified Models, which allow a
presentation of our exclusion potential in the context of a larger
variety of fundamental models, not necessarily in the GGM framework. For
the SMS interpretation, we force the initial pair production of gluinos,
which decay to jets and a neutralino or chargino. Two cases are
studied. Firstly, in the $\gamma\gamma$ Simplified Model both gluinos
decay to jets and neutralino, which are forced to decay to photons plus
gravitino (see Fig.~\ref{fig:SMS_Feynmans}) producing a final state with
two photons.  This model is sensitive to both the diphoton and
single-photon analyses. Secondly, in the $\PW\gamma$ SMS, one gluino is
forced to decay to a chargino, which always produces a \PW~boson, and
the other gluino decays as in the $\gamma\gamma$ SMS scan resulting in a
photon, allowing only the single-photon analysis to be interpreted
within this Simplified Model.

\begin{figure*}[tbp]
\begin{center}
\includegraphics[width=0.49\textwidth]{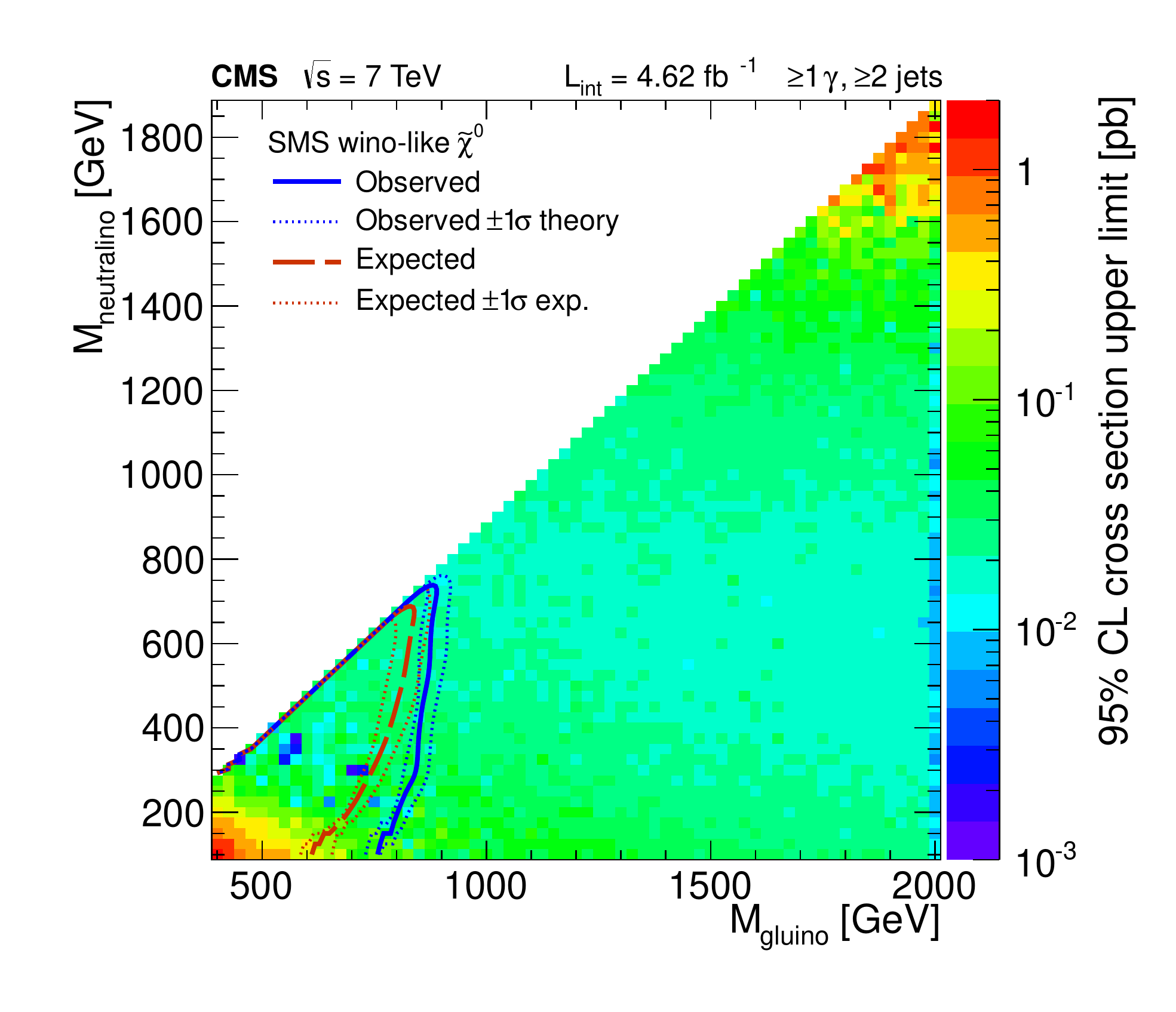}\
\includegraphics[width=0.49\textwidth]{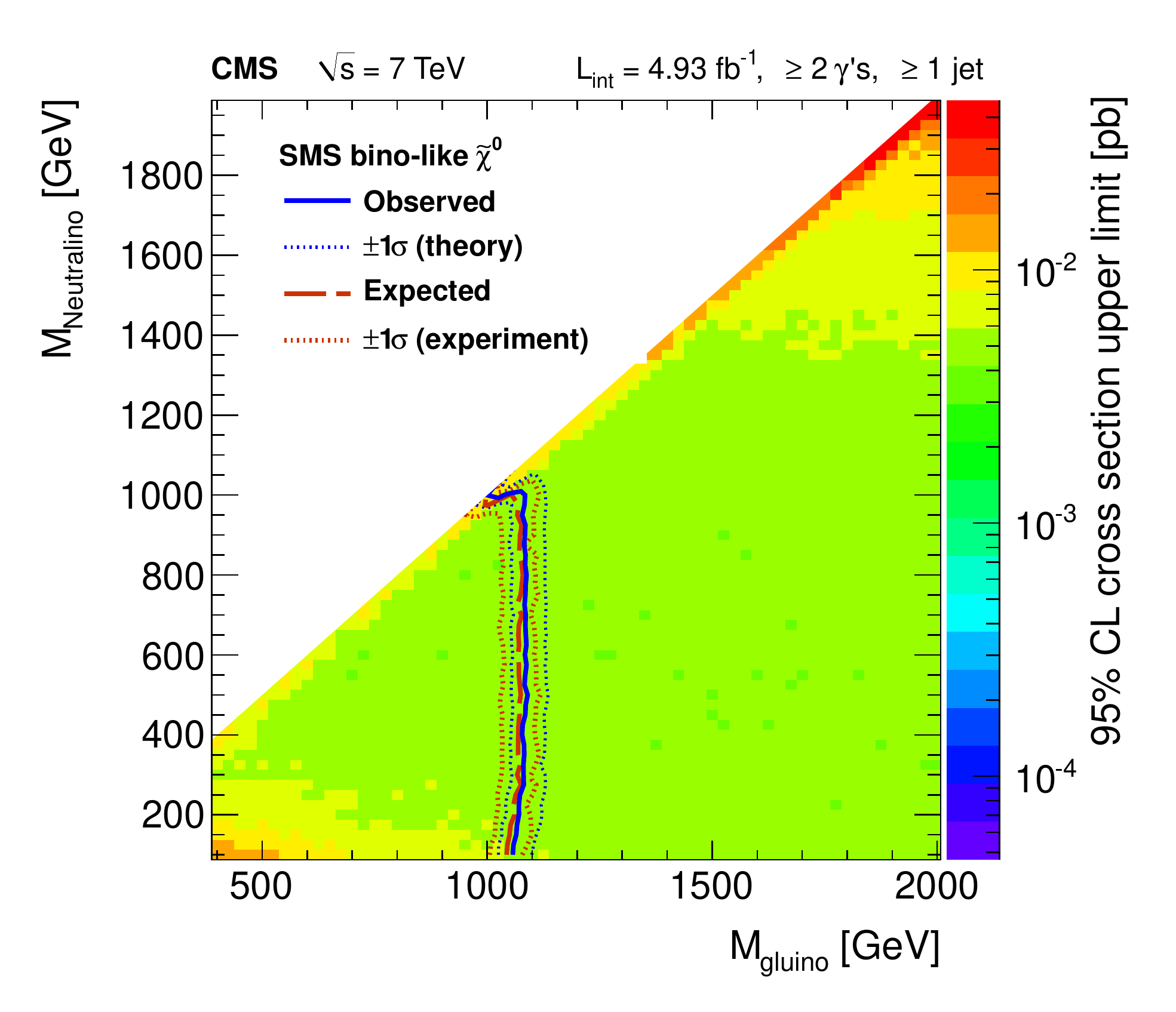}
\caption{\label{fig:SMSInterpretation}
  Results for Simplified Models in form of 95\%~CL upper limits on the
  cross section plus overlaid exclusion contours for the single-photon
  analysis in the $\PW\gamma$ Simplified Model (left) and for the
  diphoton analysis in the $\gamma\gamma$ SMS interpretation (right).  }
\end{center}
\end{figure*}

The results in the form of upper limits on the cross section and
overlaid exclusion contours, at 95\% CL, in the neutralino versus
gluino mass plane are shown in Fig.~\ref{fig:SMSInterpretation} for the
single-photon analysis in the case of the $\PW\gamma$ Simplified Model, and for
the diphoton analysis in the $\gamma\gamma$ SMS interpretation. The
Simplified Model
results in the gluino-neutralino mass plane are similar to the GGM
interpretation resulting in slightly more stringent but similar limits
as compared to the single-photon and diphoton contours shown in
Fig.~\ref{fig:GGM_chi1_gl}.  This is not unexpected since both processes
probe very similar production and decay chains and by construction, the
SMS captures the main features of the full GGM model well.
Additional figures such as the corresponding acceptances in the
gluino-neutralino mass plane for the single-photon (diphoton) analysis
in the $\PW\gamma$ ($\gamma\gamma$) SMS interpretation and
corresponding
results from the single-photon analysis in the $\gamma\gamma$ Simplified Model
are available in \suppMaterial.

\subsection{Universal Extra Dimensions}
\label{sec:UED}

Diphoton final states with large \MET are also predicted by UED
models~\cite{Appelquist:2000nn} postulating the existence of additional
spatial dimensions of compactification radius $R$.  For the investigated
model the UED space is embedded in an
additional space of large extra dimensions where only the graviton propagates and the LKP
decays gravitationally, producing a photon and a graviton.
The diphoton analysis results can thus be interpreted in the context of
the UED model. The
model parameters are chosen to match a study by the D0 collaboration,
which excludes $1/R < 477$\GeV~\cite{Abazov:2010us} and a more recent
result by the ATLAS experiment excluding $1/R <
728$\GeV~\cite{Aad:2011kz}.  To determine the effect of the number of
large ED, $n$ on the potential limit for UED,
$n$ was varied.  By changing the number of large ED, the
branching ratios of the different decay channels are changed but the
overall UED production cross section remains the same. For
$n\ge 3$ decays involving a heavy graviton with mass of
order $(1/R)$ dominate while for $n=2$ decays involving
light gravitons are more prevalent~\cite{Macesanu:2002db}. For
$n$ equal to 4 and 6, the $\MET$ distributions are very
similar allowing the comparison only for $n=6$ to
$n=2$ where the \MET~distribution is flatter resulting
in a slightly lower efficiency.

\begin{figure*}[tbp]
\begin{center}
\includegraphics[width=0.45\textwidth]{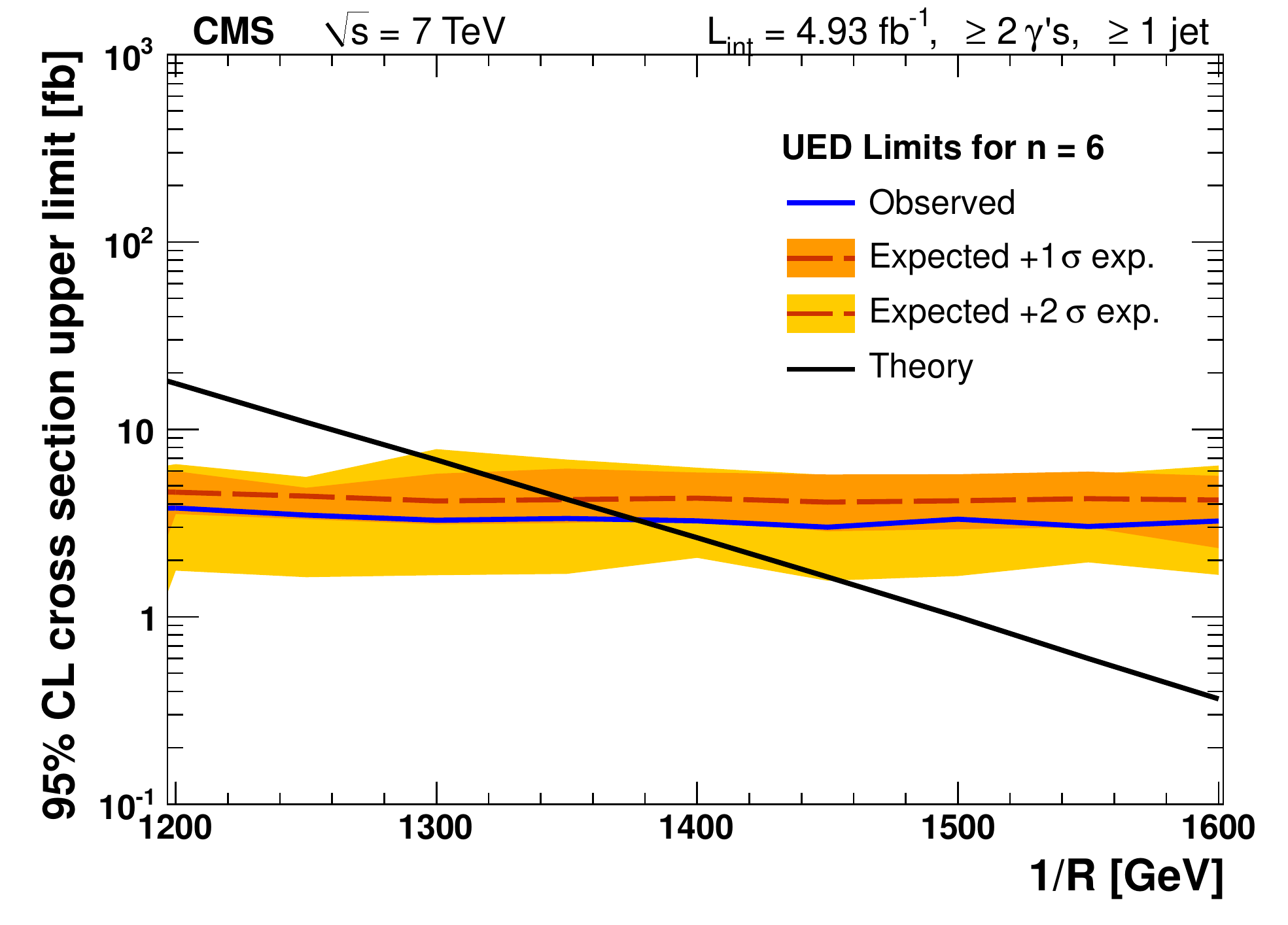}
\caption{\label{fig:UED}
  Upper limit on the UED model cross section for n=6 at 95\% CL compared
  with expected UED production cross sections (black diagonal line). The
  shaded region shows the uncertainty due to PDFs and renormalization
  scale on the expected limit.  }
\end{center}
\end{figure*}

To determine the acceptance times efficiency, UED signal simulated
samples generated with $1/R$ ranging from 900 to 1600\GeV as described
in Section~\ref{sec:MC} are analyzed adopting the same selection
criteria as used for the GGM diphoton analysis. The cross section upper
limit for the production of KK particles, which would indicate the
presence of UED, can be calculated in the same way as for the GGM limit
calculation. The maximum UED production cross section is computed using
the acceptance times efficiency from signal Monte Carlo simulations and
the same luminosity, background estimate, and number of observed
$\gamma\gamma$ signal events as for the GGM limit calculation.  The
signal acceptance times efficiency is rather flat in the region of
interest ranging from about 0.42 at $1/R\sim900$\GeV to 0.46 at
$1/R\sim1600$\GeV.  The UED cross sections and the 95\%~CL upper limit
on the signal cross section are interpolated and their intersection is
determined and shown in Fig.~\ref{fig:UED}.  Uncertainties due to PDFs
and renormalization scale are shown as the shaded region, while the
intersection of the central value implies that the range of $1/R <
1380$\GeV for $n=6$ is excluded with an expected limit of
1350\GeV. This is the best UED limit to date.  For $n= 2$ the exclusion
limit is reduced to 1350\GeV for an expected limit of 1340\GeV.
The corresponding UED acceptance times efficiency distributions for
$n= 2$ and 6 as well as the 95\%~CL cross section
upper limit for $n= 2$ are
available in \suppMaterial.

\section{Conclusions}
\label{sec:conclusions}

In summary, a search for physics beyond the standard model has been
performed in single-photon and diphoton events using the \MET spectrum
comparing data and SM background expectations.  This search is based on
2011 CMS data comprising 4.93~\fbinv of \Pp\Pp~collisions at
$\sqrt{s}=7~\TeV$.  No evidence of NP is found and upper limits are
derived for three theoretical interpretations.  First, in the SUSY GGM
model the single-photon (diphoton) analysis derives exclusion regions
for the production cross section in the parameter space of squark and
gluino masses of order 0.03 -- 0.1~pb (0.003 -- 0.01~pb) at the 95\% CL
for a wino-like (bino-like) scenario, corresponding to the exclusion of
squark and gluino masses up to masses of order 800~\GeV
(1~\TeV). Exclusion contours at the 95\%~CL are presented in the plane
of gluino versus neutralino mass for a wino-like (bino-like) neutralino
with the single-photon (diphoton) analysis. In addition, for the first
time, electroweak production is studied in the plane of wino-like versus
bino-like gaugino mass where the diphoton search excludes wino masses up
to $\sim\!500~\GeV$.

The single-photon and diphoton analyses are in addition interpreted in
the context of Simplified Models resulting in similar exclusion limits
and contours. Finally, the diphoton analysis is reinterpreted as a
search for universal extra dimensions, leading to 95\% exclusion values
of the inverse compactification radius $1/R < 1380$~\GeV for $n=6$ large
extra dimensions constituting the currently best limit on the considered
UED model.

\section*{Acknowledgements}

We would like to thank David Shih for stimulating discussions and help
with the production of our SUSY simulated samples for which he kindly
provided SLHA files.

\hyphenation{Bundes-ministerium Forschungs-gemeinschaft
  Forschungs-zentren} We congratulate our colleagues in the CERN
accelerator departments for the excellent performance of the LHC and
thank the technical and administrative staffs at CERN and at other CMS
institutes for their contributions to the success of the CMS effort. In
addition, we gratefully acknowledge the computing centres and personnel
of the Worldwide LHC Computing Grid for delivering so effectively the
computing infrastructure essential to our analyses. Finally, we
acknowledge the enduring support for the construction and operation of
the LHC and the CMS detector provided by the following funding agencies:
the Austrian Federal Ministry of Science and Research; the Belgian Fonds
de la Recherche Scientifique, and Fonds voor Wetenschappelijk Onderzoek;
the Brazilian Funding Agencies (CNPq, CAPES, FAPERJ, and FAPESP); the
Bulgarian Ministry of Education, Youth and Science; CERN; the Chinese
Academy of Sciences, Ministry of Science and Technology, and National
Natural Science Foundation of China; the Colombian Funding Agency
(COLCIENCIAS); the Croatian Ministry of Science, Education and Sport;
the Research Promotion Foundation, Cyprus; the Ministry of Education and
Research, Recurrent financing contract SF0690030s09 and European
Regional Development Fund, Estonia; the Academy of Finland, Finnish
Ministry of Education and Culture, and Helsinki Institute of Physics;
the Institut National de Physique Nucl\'eaire et de Physique des
Particules~/~CNRS, and Commissariat \`a l'\'Energie Atomique et aux
\'Energies Alternatives~/~CEA, France; the Bundesministerium f\"ur
Bildung und Forschung, Deutsche Forschungsgemeinschaft, and
Helmholtz-Gemeinschaft Deutscher Forschungszentren, Germany; the General
Secretariat for Research and Technology, Greece; the National Scientific
Research Foundation, and National Office for Research and Technology,
Hungary; the Department of Atomic Energy and the Department of Science
and Technology, India; the Institute for Studies in Theoretical Physics
and Mathematics, Iran; the Science Foundation, Ireland; the Istituto
Nazionale di Fisica Nucleare, Italy; the Korean Ministry of Education,
Science and Technology and the World Class University program of NRF,
Korea; the Lithuanian Academy of Sciences; the Mexican Funding Agencies
(CINVESTAV, CONACYT, SEP, and UASLP-FAI); the Ministry of Science and
Innovation, New Zealand; the Pakistan Atomic Energy Commission; the
Ministry of Science and Higher Education and the National Science
Centre, Poland; the Funda\c{c}\~ao para a Ci\^encia e a Tecnologia,
Portugal; JINR (Armenia, Belarus, Georgia, Ukraine, Uzbekistan); the
Ministry of Education and Science of the Russian Federation, the Federal
Agency of Atomic Energy of the Russian Federation, Russian Academy of
Sciences, and the Russian Foundation for Basic Research; the Ministry of
Science and Technological Development of Serbia; the Secretar\'{\i}a de
Estado de Investigaci\'on, Desarrollo e Innovaci\'on and Programa
Consolider-Ingenio 2010, Spain; the Swiss Funding Agencies (ETH Board,
ETH Zurich, PSI, SNF, UniZH, Canton Zurich, and SER); the National
Science Council, Taipei; the Thailand Center of Excellence in Physics,
the Institute for the Promotion of Teaching Science and Technology and
National Electronics and Computer Technology Center; the Scientific and
Technical Research Council of Turkey, and Turkish Atomic Energy
Authority; the Science and Technology Facilities Council, UK; the US
Department of Energy, and the US National Science Foundation.

Individuals have received support from the Marie-Curie programme and the
European Research Council (European Union); the Leventis Foundation; the
A. P. Sloan Foundation; the Alexander von Humboldt Foundation; the
Belgian Federal Science Policy Office; the Fonds pour la Formation \`a
la Recherche dans l'Industrie et dans l'Agriculture (FRIA-Belgium); the
Agentschap voor Innovatie door Wetenschap en Technologie (IWT-Belgium);
the Ministry of Education, Youth and Sports (MEYS) of Czech Republic;
the Council of Science and Industrial Research, India; the Compagnia di
San Paolo (Torino); and the HOMING PLUS programme of Foundation for
Polish Science, cofinanced from European Union, Regional Development
Fund.

\bibliography{auto_generated}   

\ifthenelse{\boolean{cms@external}}{}{
\clearpage
\appendix
\section{Supplemental Material}
\label{app:suppMat}

The appendix contains additional figures such as limit contours from the
three interpretations (GGM, SMS, and UED) that are not part of the main
body of the paper but are submitted as supplemental material to the
journal.

\subsection{GGM Interpretation}
\label{app:GGM}

This section contains additional 95\% CL upper limits on the signal
cross section and exclusion contours in the interpretation of the GGM
SUSY breaking scenario.

Figure~\ref{fig:GGM_sq_gl2} shows the upper limits on the GGM production
cross section as well as exclusion contours in the squark versus gluino
mass plane for the single-photon bino-like neutralino and the diphoton
wino-like scenario.

\begin{figure*}[bp]
\begin{center}
\includegraphics[width=0.49\textwidth]{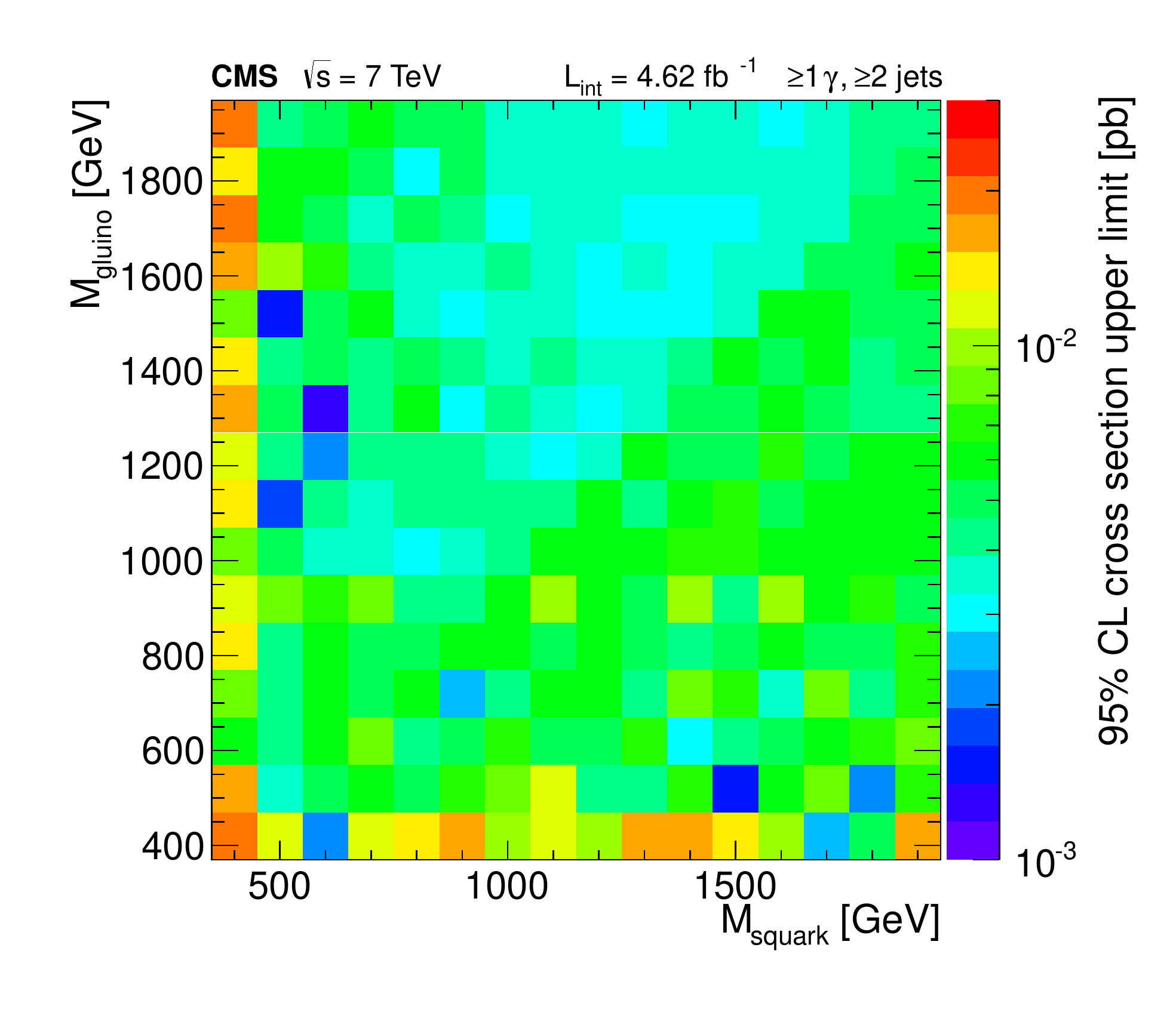}\
\includegraphics[width=0.49\textwidth]{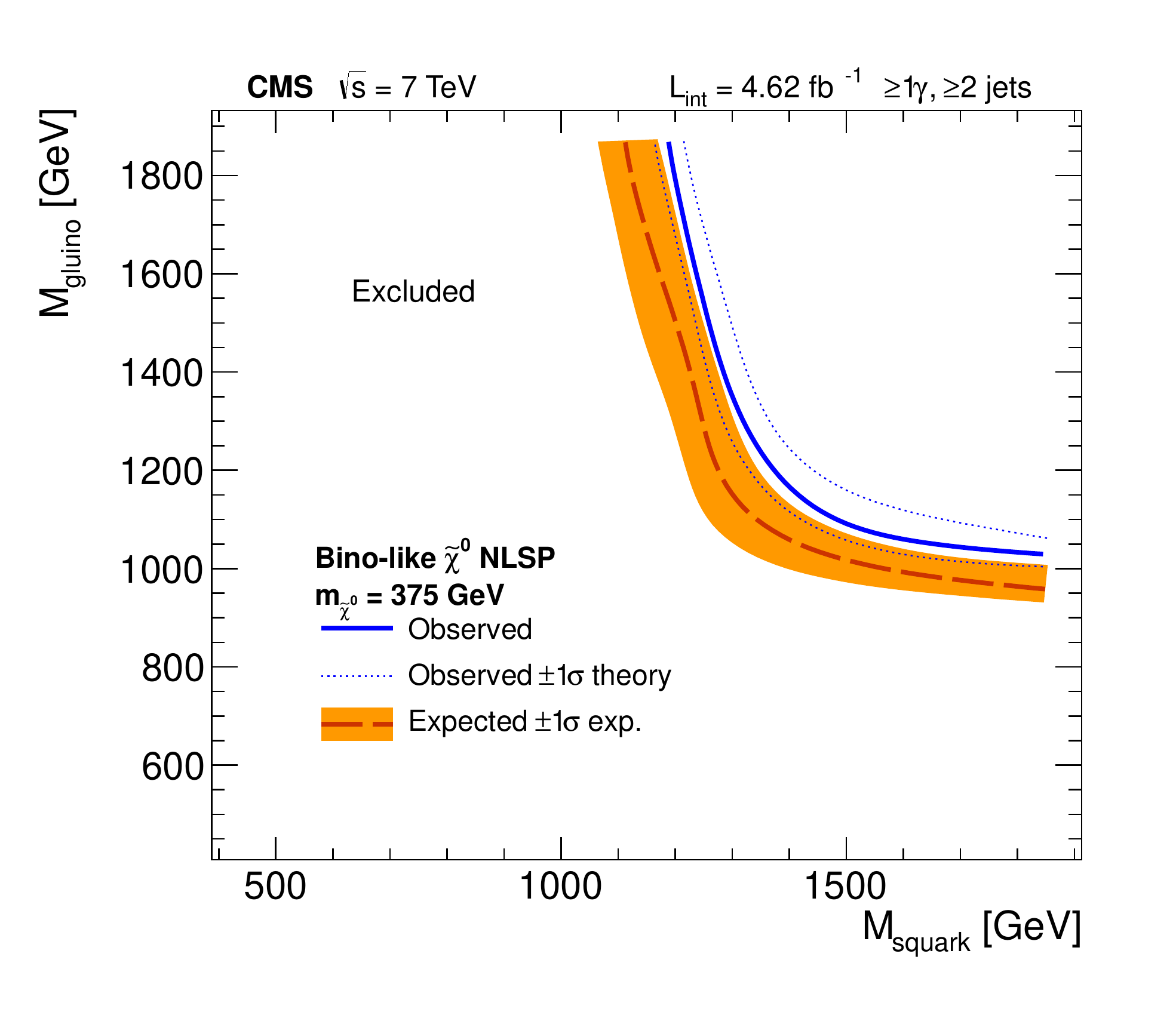}
\includegraphics[width=0.49\textwidth]{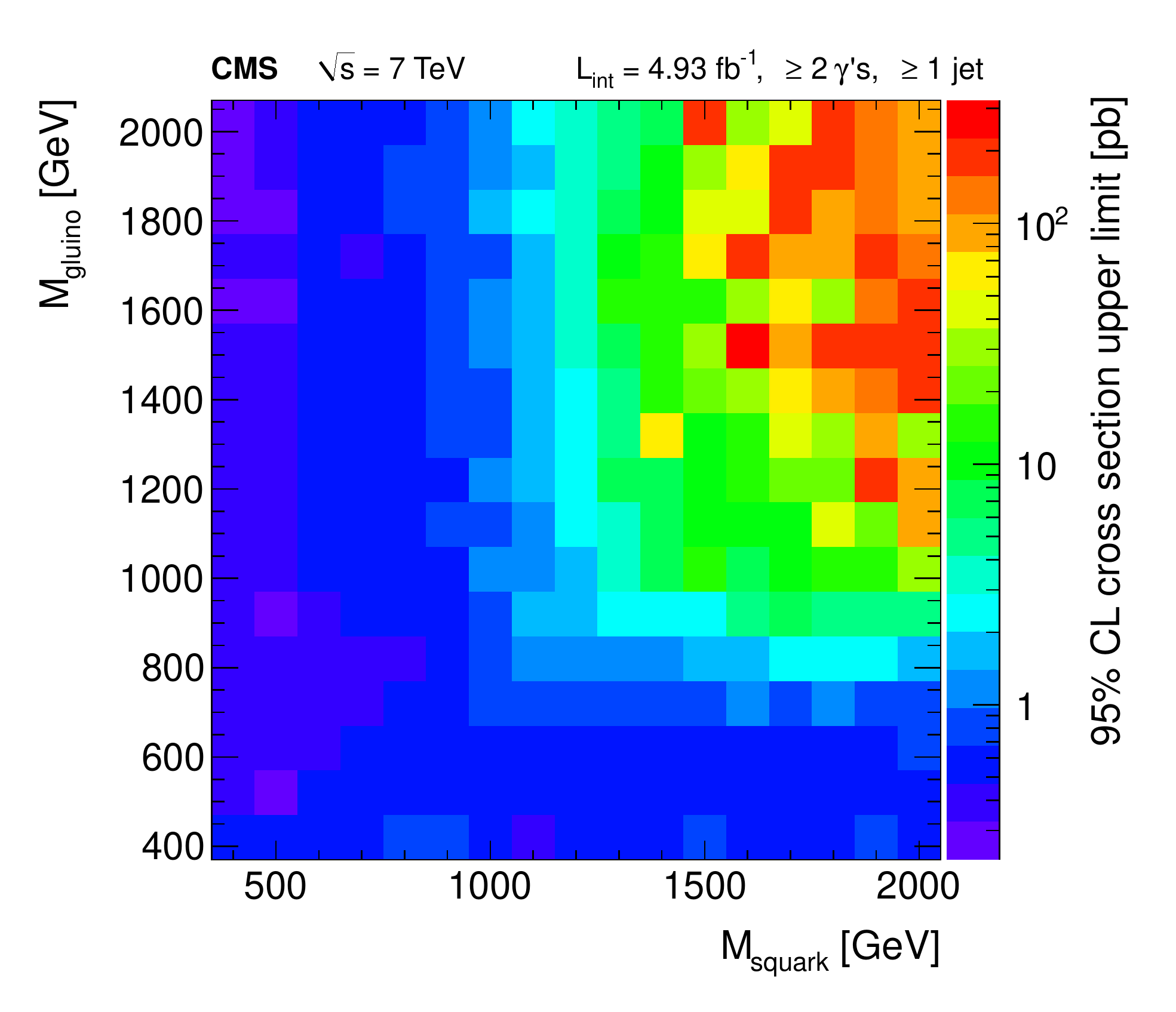}\
\includegraphics[width=0.49\textwidth]{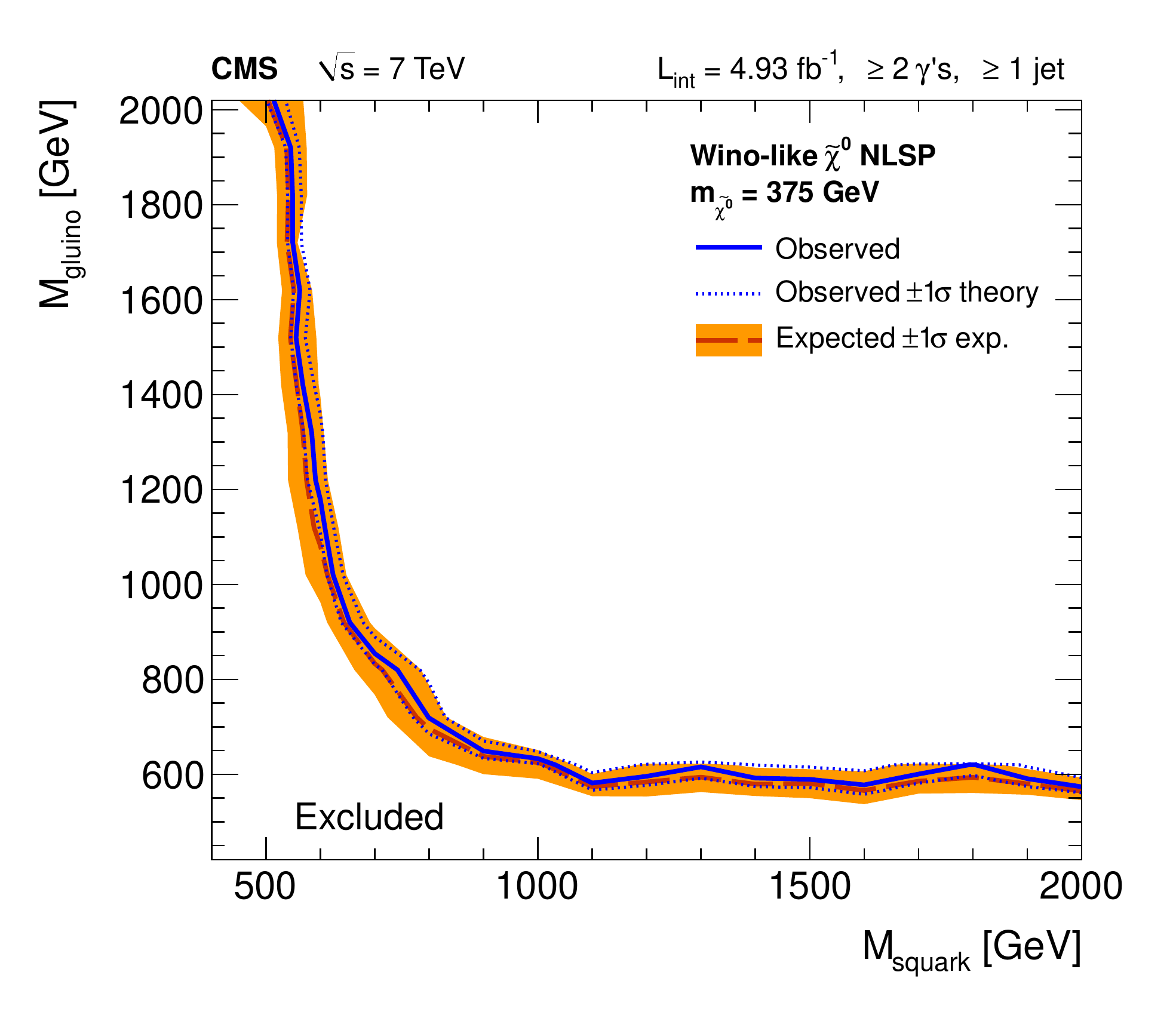}
\caption{\label{fig:GGM_sq_gl2} 
  95\% CL upper limits on the signal cross section (left) and
  corresponding exclusion contours (right) in gluino-squark mass space
  for the single-photon bino-like (top) and diphoton
  wino-like (bottom) scenario. The shaded uncertainty bands around the
  expected exclusion contours correspond to experimental uncertainties,
  while the NLO renormalization and PDF uncertainties of the signal
  cross section are indicated around the observed limit contour.
}
\end{center}
\end{figure*}

As further interpretation of the single-photon and diphoton results,
Fig.~\ref{fig:GGM_chi1_gl2} shows the 95\% CL upper limits on the
signal cross section for the single-photon (diphoton) wino-like
(bino-like) scenario in the gluino-neutralino mass plane, while
Fig.~\ref{fig:GGM_chi1_gl3} displays 95\% CL upper limits on the
signal cross section and exclusion contours for the single-photon
(diphoton) bino-like (wino-like) scenarios.

\begin{figure*}[tbp]
\begin{center}
\includegraphics[width=0.49\textwidth]{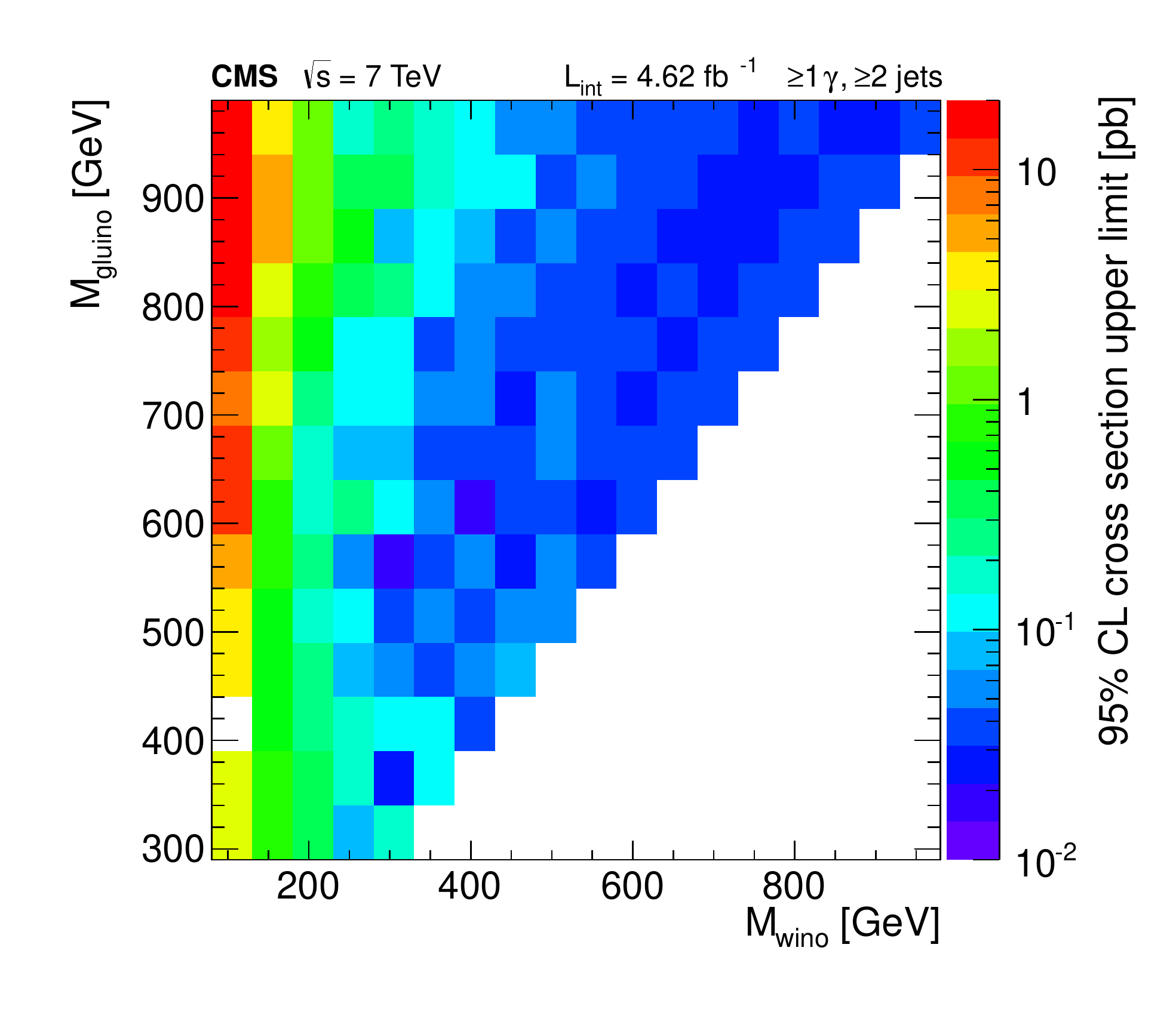}\
\includegraphics[width=0.49\textwidth]{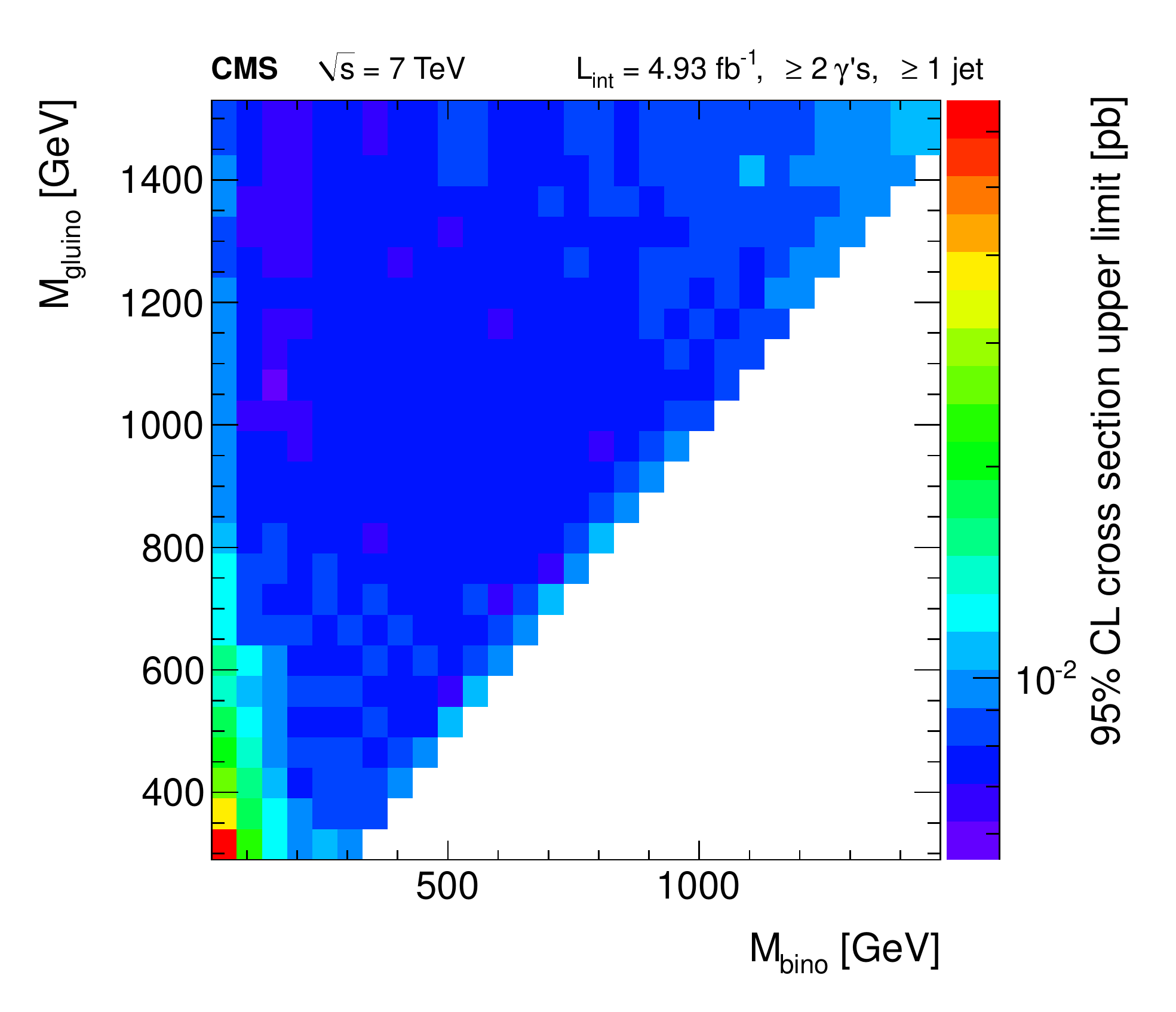}
\caption{\label{fig:GGM_chi1_gl2} 
  95\% CL upper limits on the signal cross section in the plane of
  gluino versus neutralino mass for the single-photon search in the
  wino-like scenario (left) and the diphoton analysis for a bino-like
  neutralino (right).  
}
\end{center}
\end{figure*}

\begin{figure*}[tbp]
\begin{center}
\includegraphics[width=0.49\textwidth]{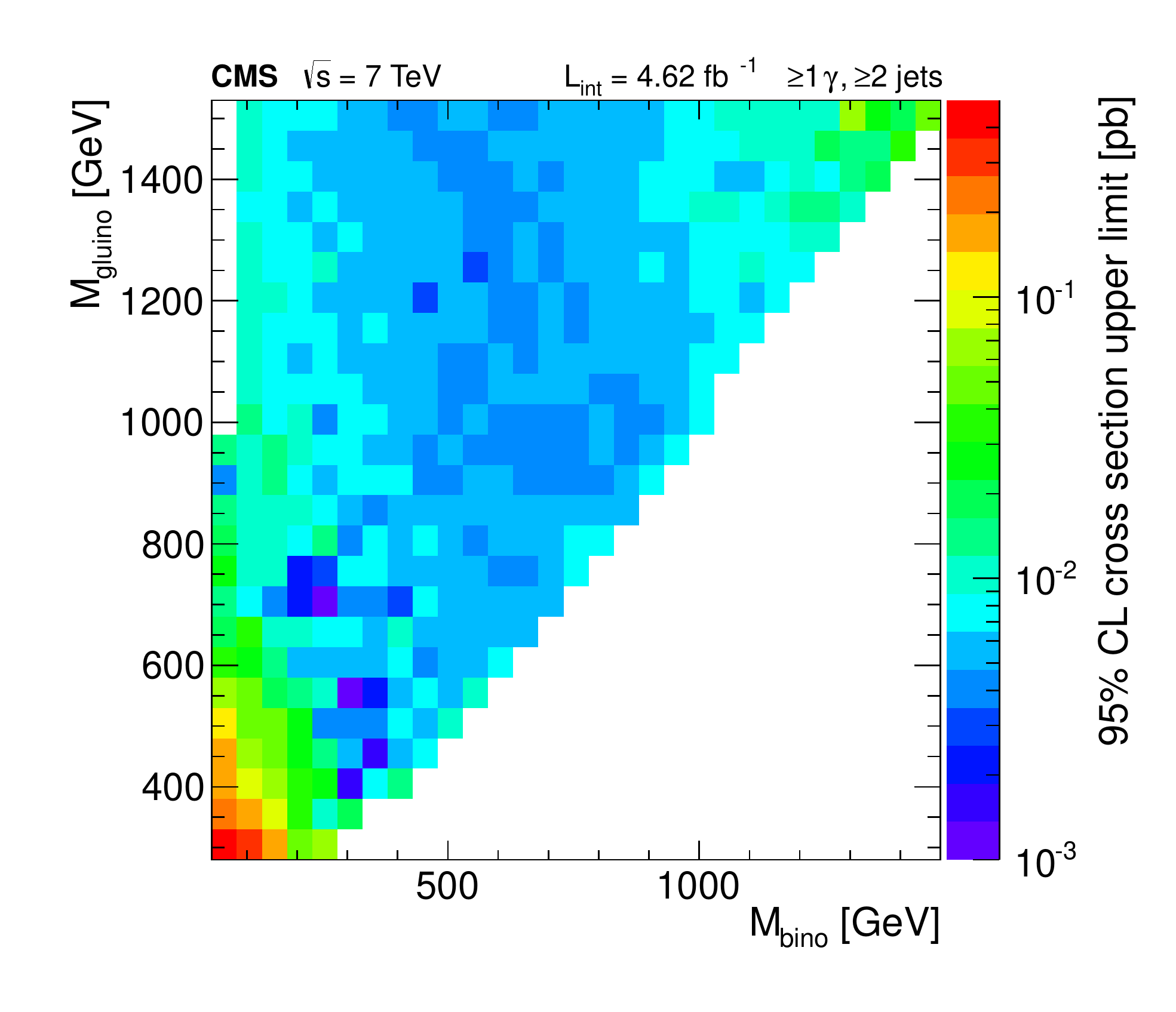}\
\includegraphics[width=0.49\textwidth]{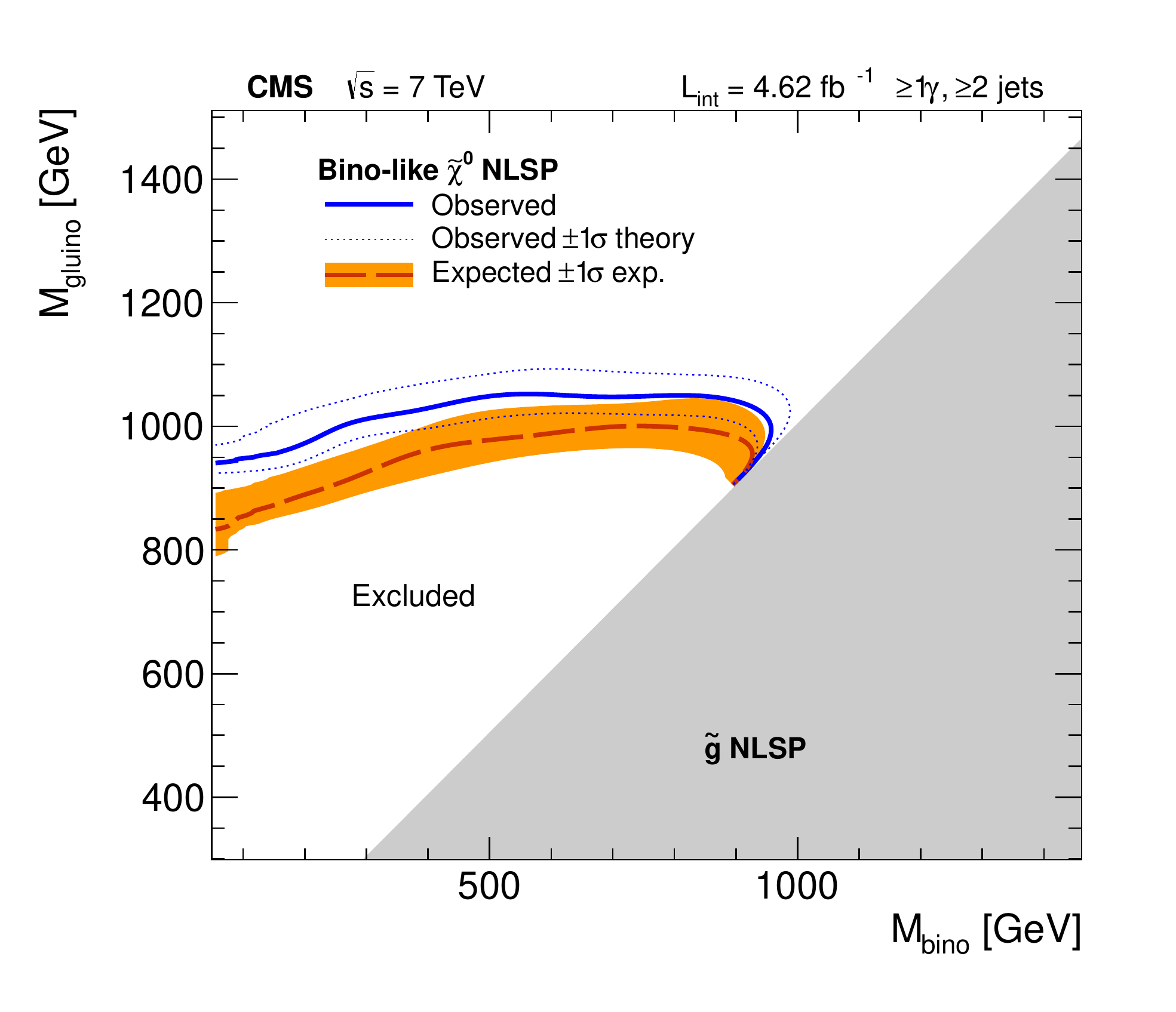}
\includegraphics[width=0.49\textwidth]{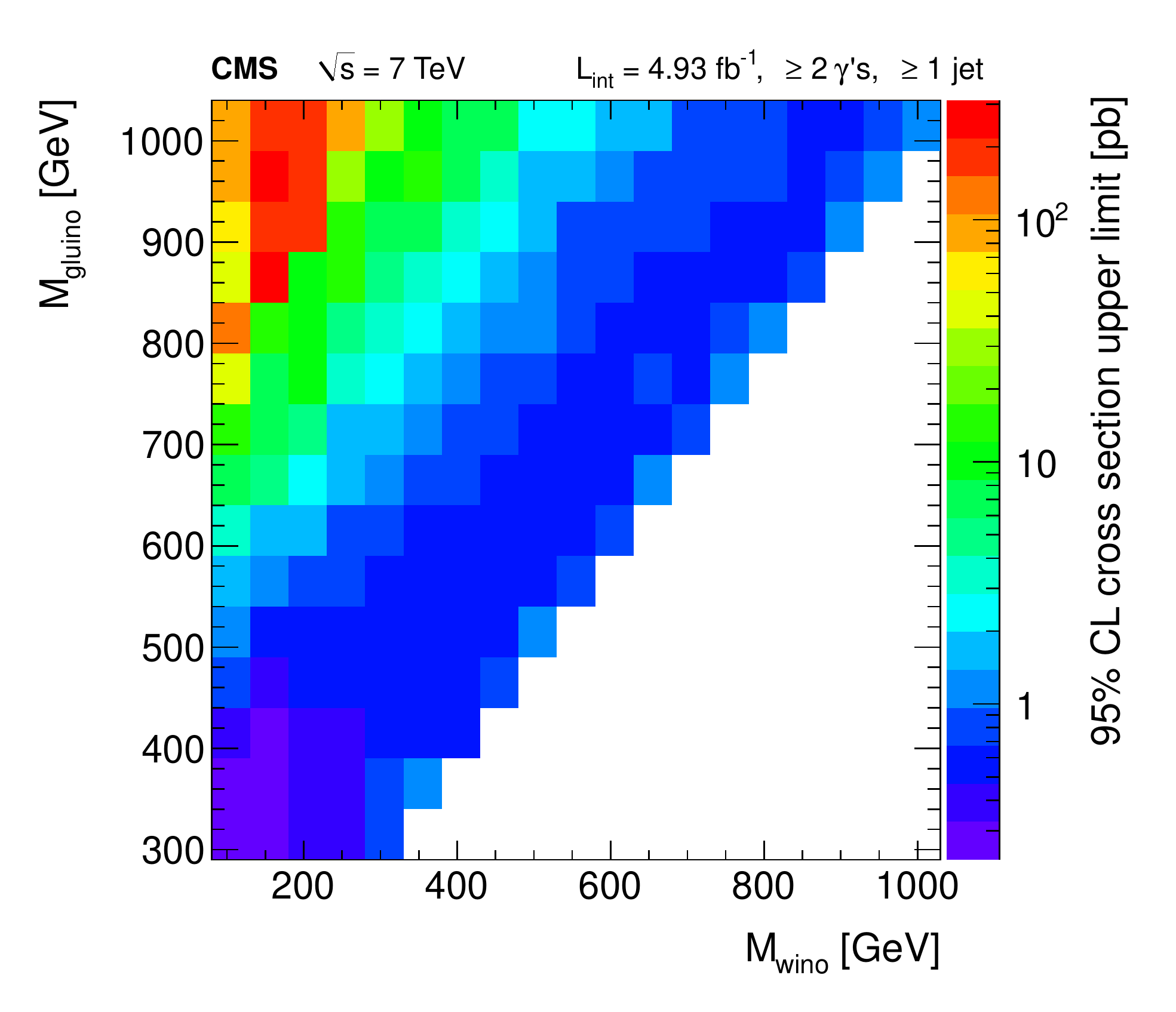}\
\includegraphics[width=0.49\textwidth]{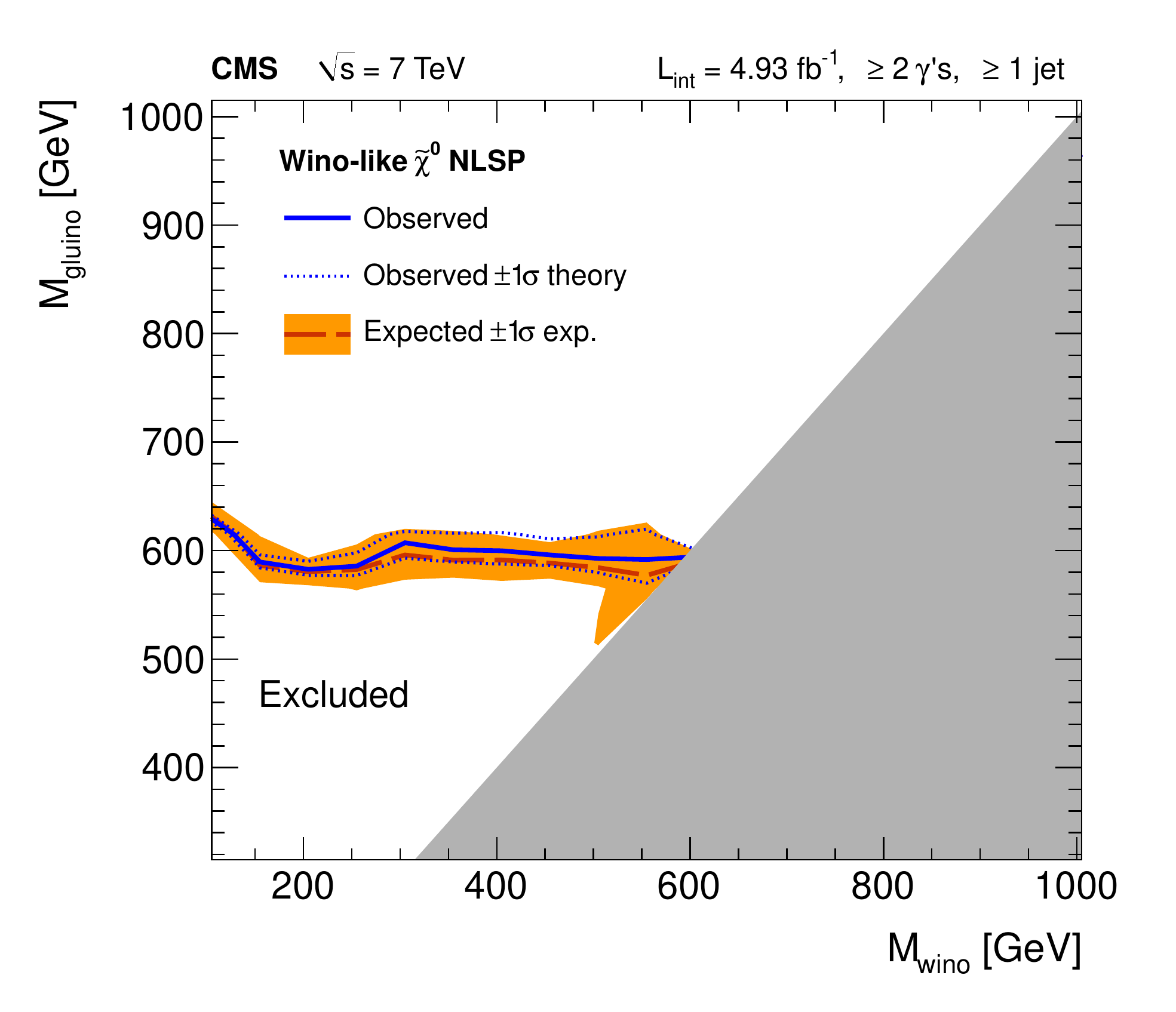}
\caption{\label{fig:GGM_chi1_gl3} 
  95\% CL upper limits on the signal cross section (left) and
  corresponding exclusion contours (right) in the plane of gluino versus
  neutralino mass for the single-photon bino-like (top) and the
  diphoton wino-like scenario (bottom). 
}
\end{center}
\end{figure*}

Figure~\ref{fig:GGM_W_B2} shows the the 95\% CL~upper limits on the
signal cross section in the plane of wino-like versus bino-like gaugino
mass for the diphoton analysis.

\begin{figure*}[tbp]
\begin{center}
\includegraphics[width=0.49\textwidth]{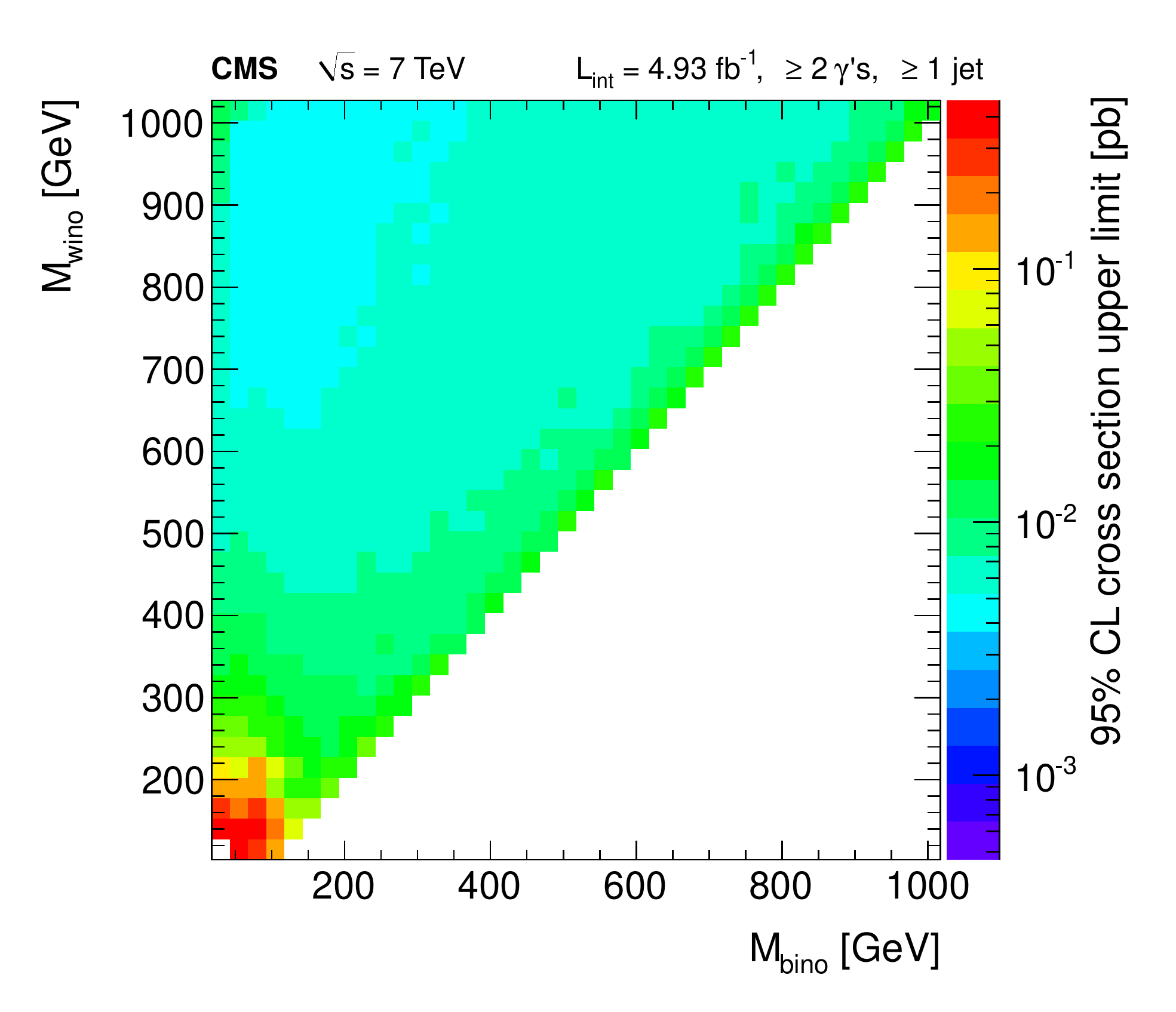}\
\caption{\label{fig:GGM_W_B2} 
  95\% CL upper limits on the signal cross section in the plane of
  bino-like versus wino-like gaugino mass for the diphoton analysis.  
}
\end{center}
\end{figure*}

\subsection{Simplified Model Interpretation}
\label{app:SMS}

This section contains additional figures from the interpretation of the
single and diphoton analyses in terms of Simplified Models.  Acceptances
for the single-photon analysis in the case of the $\PW\gamma$ Simplified
Model and for the diphoton analysis in the $\gamma\gamma$ SMS
interpretation are shown in the neutralino versus gluino mass plane in
Fig.~\ref{fig:SMSInterpretation2}.

\begin{figure*}[tbp]
\begin{center}
\includegraphics[width=0.49\textwidth]{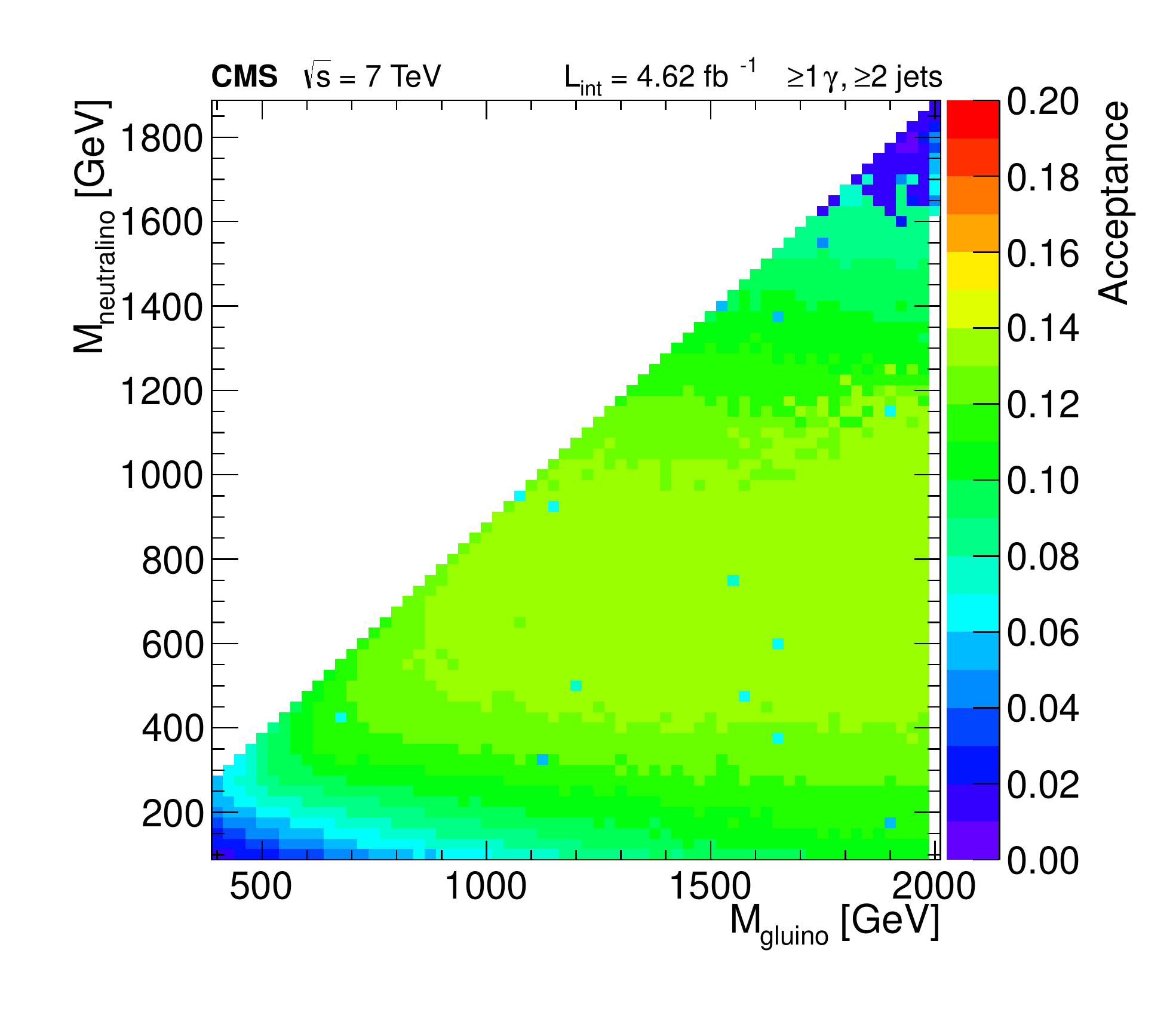}
\includegraphics[width=0.49\textwidth]{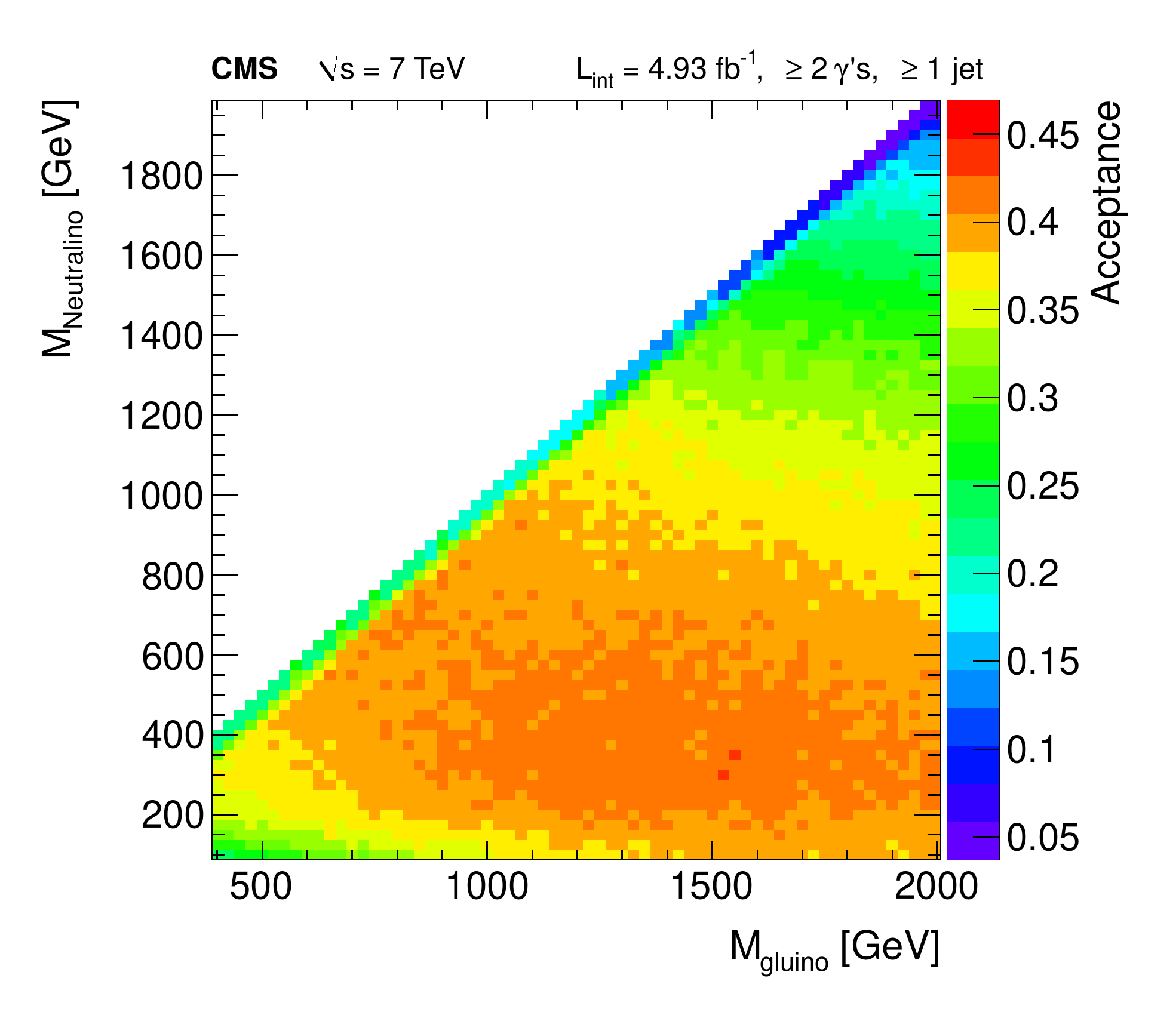}
\caption{\label{fig:SMSInterpretation2} 
  Acceptance for the single-photon analysis in the $\PW\gamma$
  Simplified Model (left) and for the diphoton analysis in the
  $\gamma\gamma$ SMS interpretation (right).  }
\end{center}
\end{figure*}

The SMS results from the single-photon analysis for the $\gamma\gamma$
Simplified Model are given in Fig.~\ref{fig:SMSInterpretation3}.  The
distribution of acceptance as well as upper limits on the cross section
plus exclusion contours are displayed.

\begin{figure*}[tbp]
\begin{center}
\includegraphics[width=0.49\textwidth]{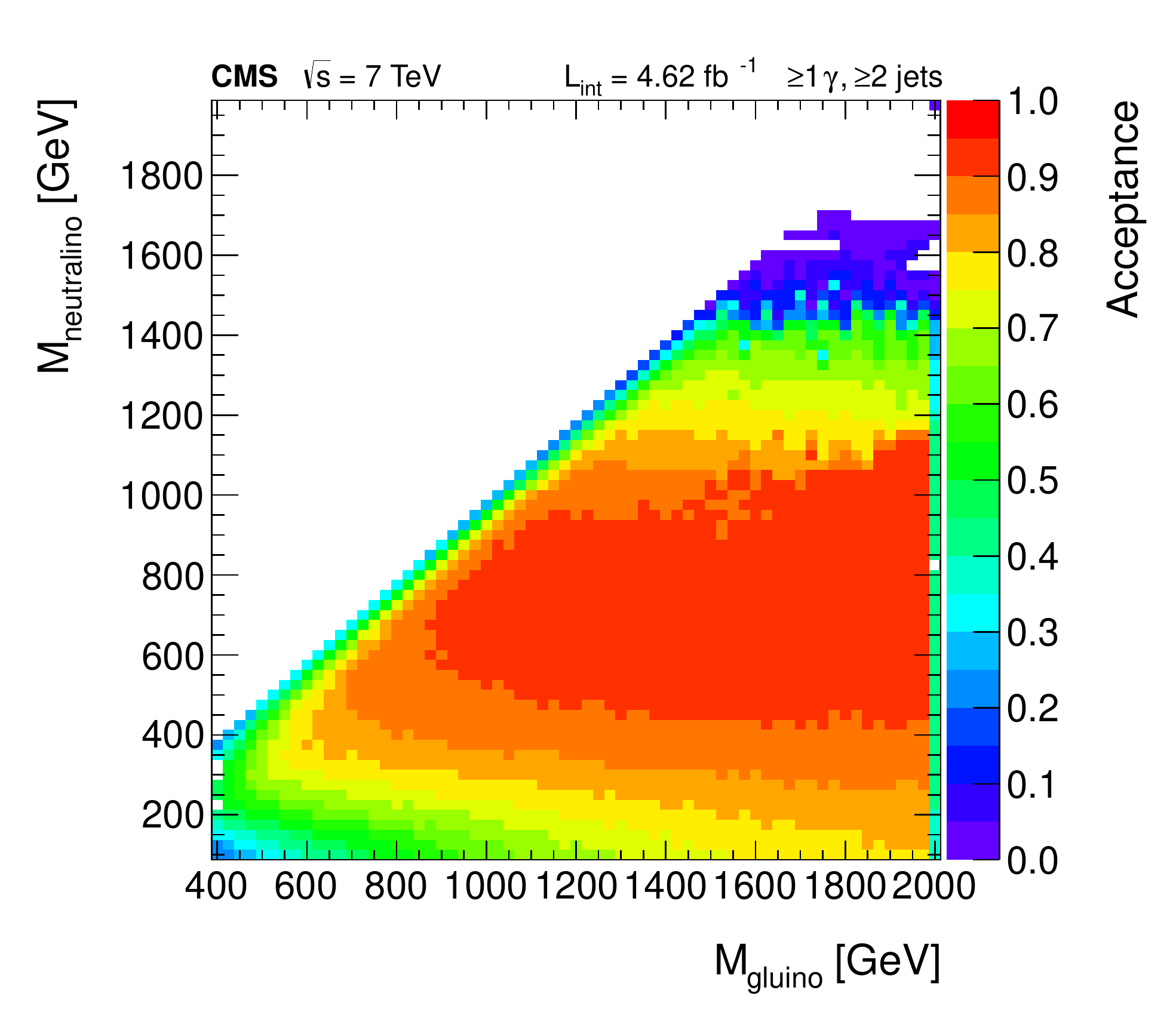}\
\includegraphics[width=0.49\textwidth]{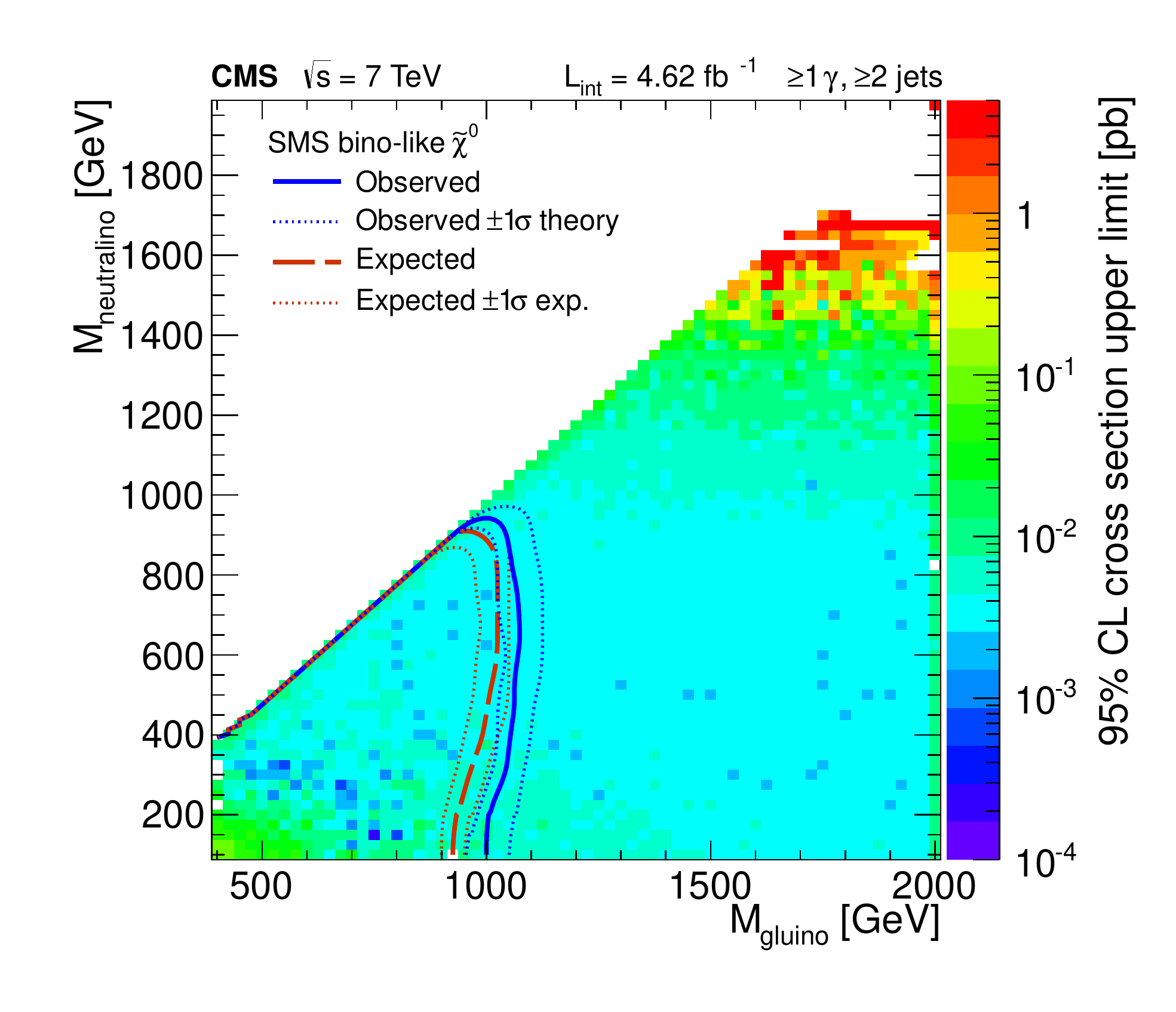}
\caption{\label{fig:SMSInterpretation3}
  Acceptance (left) and 95\%~CL observed upper limit on the cross
  section plus exclusion contours (right) in the gluino-neutralino mass
  plane for the single-photon analysis interpreted in the $\gamma\gamma$
  Simplified Model.   }
\end{center}
\end{figure*}

\subsection{UED Interpretation}
\label{app:UED}

This section contains additional figures from the interpretation of the
diphoton analysis in terms of universal extra dimension models.  
The UED
acceptance times efficiency is shown
in Fig.~\ref{fig:UED2} for $n=2$ and 6 large ED.

The UED cross section upper limit is shown
in Fig.~\ref{fig:UED3} for $n=2$ large ED. The 95\% CL~limit for
$n= 2$ is compared with the expected UED production
cross sections resulting in an exclusion limit of 1350~\GeV for an
expected limit of 1340~\GeV.

\begin{figure*}[tbp]
\begin{center}
\includegraphics[width=0.49\textwidth]{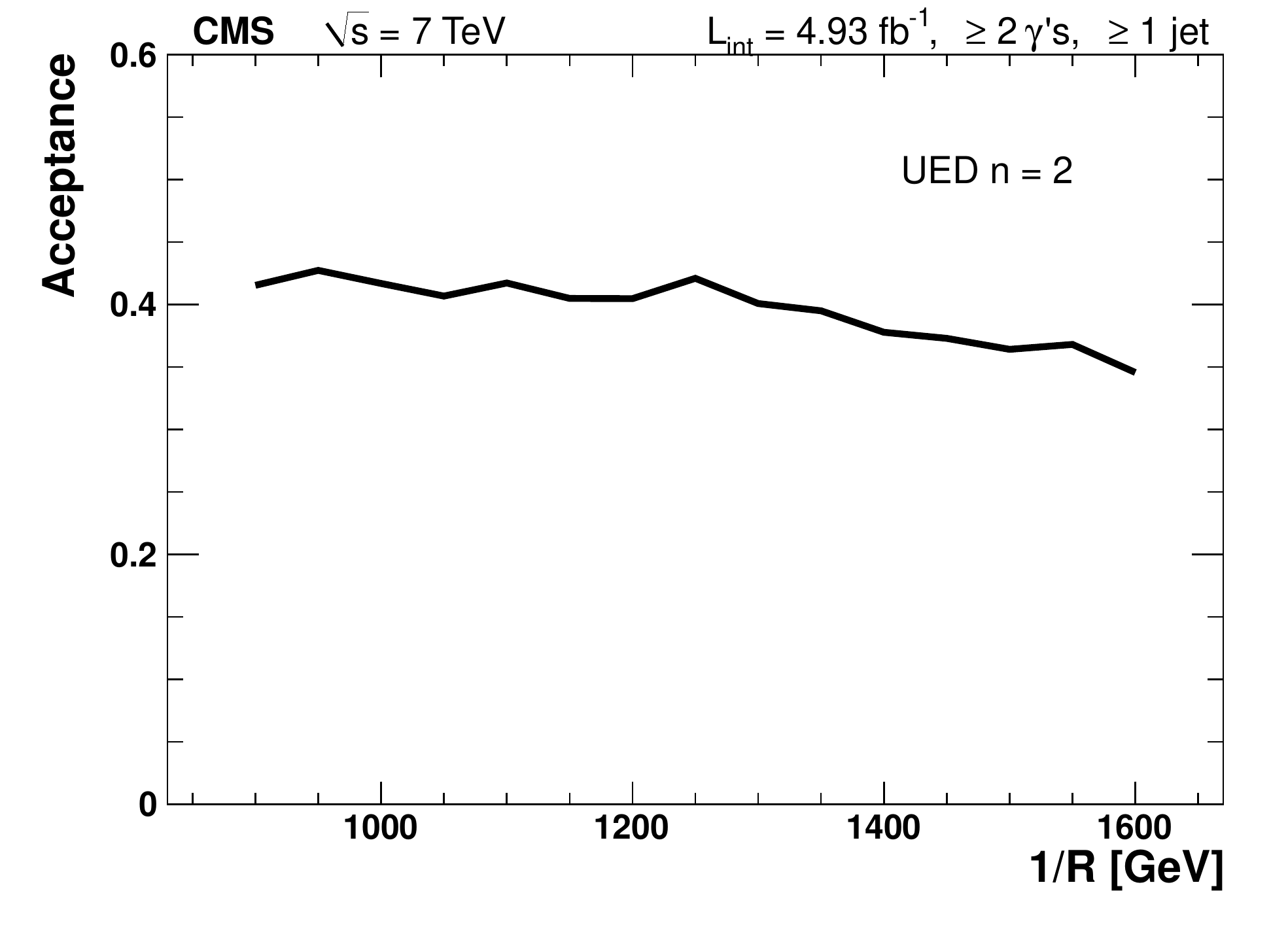}\ 
\includegraphics[width=0.49\textwidth]{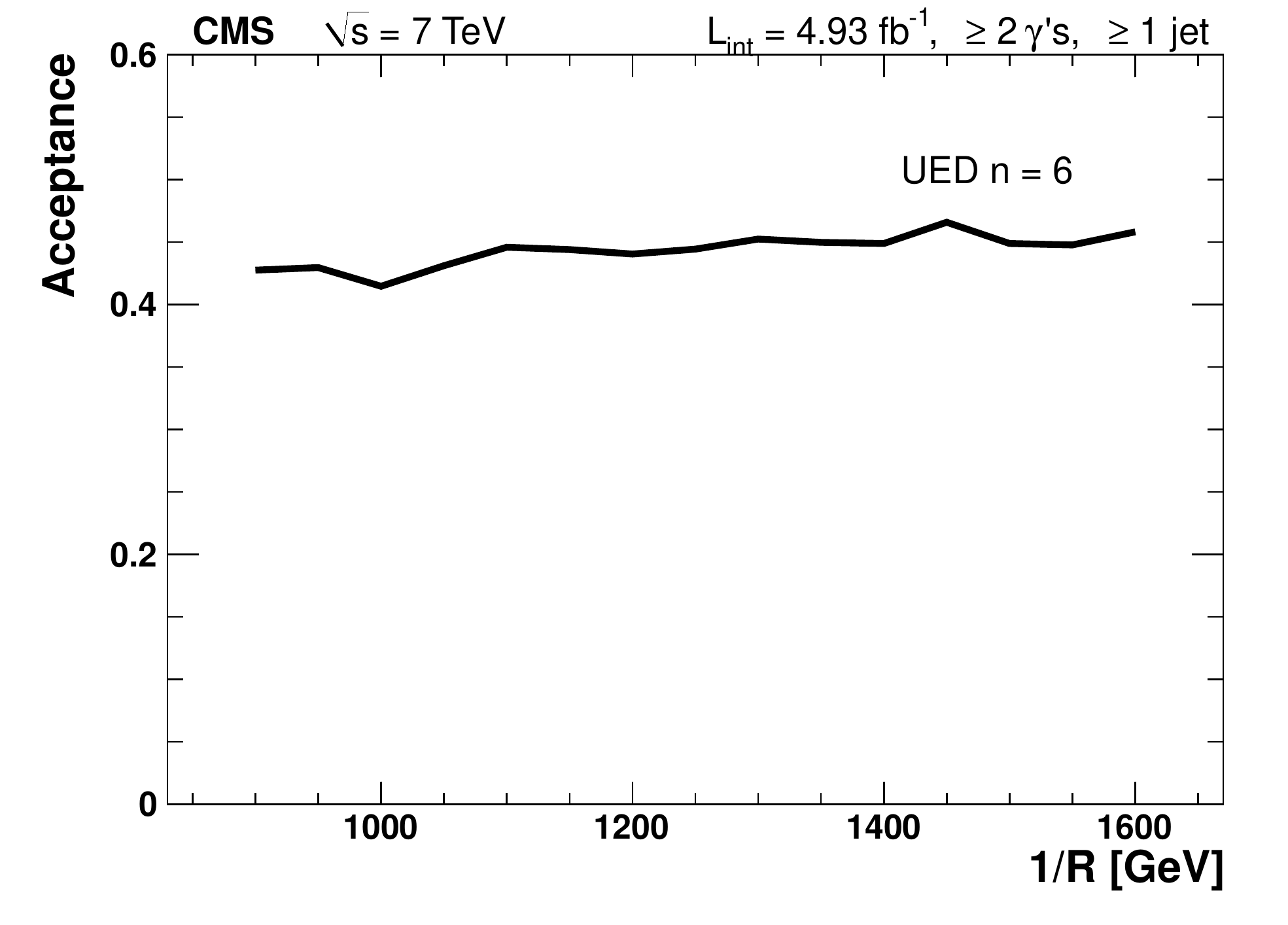}\ 
\caption{\label{fig:UED2} 
  UED acceptance times efficiency for $n=2$ large ED (left) and for $n=6$ large ED
  (right). 
}
\end{center}
\end{figure*}

\begin{figure*}[tbp]
\begin{center}
\includegraphics[width=0.49\textwidth]{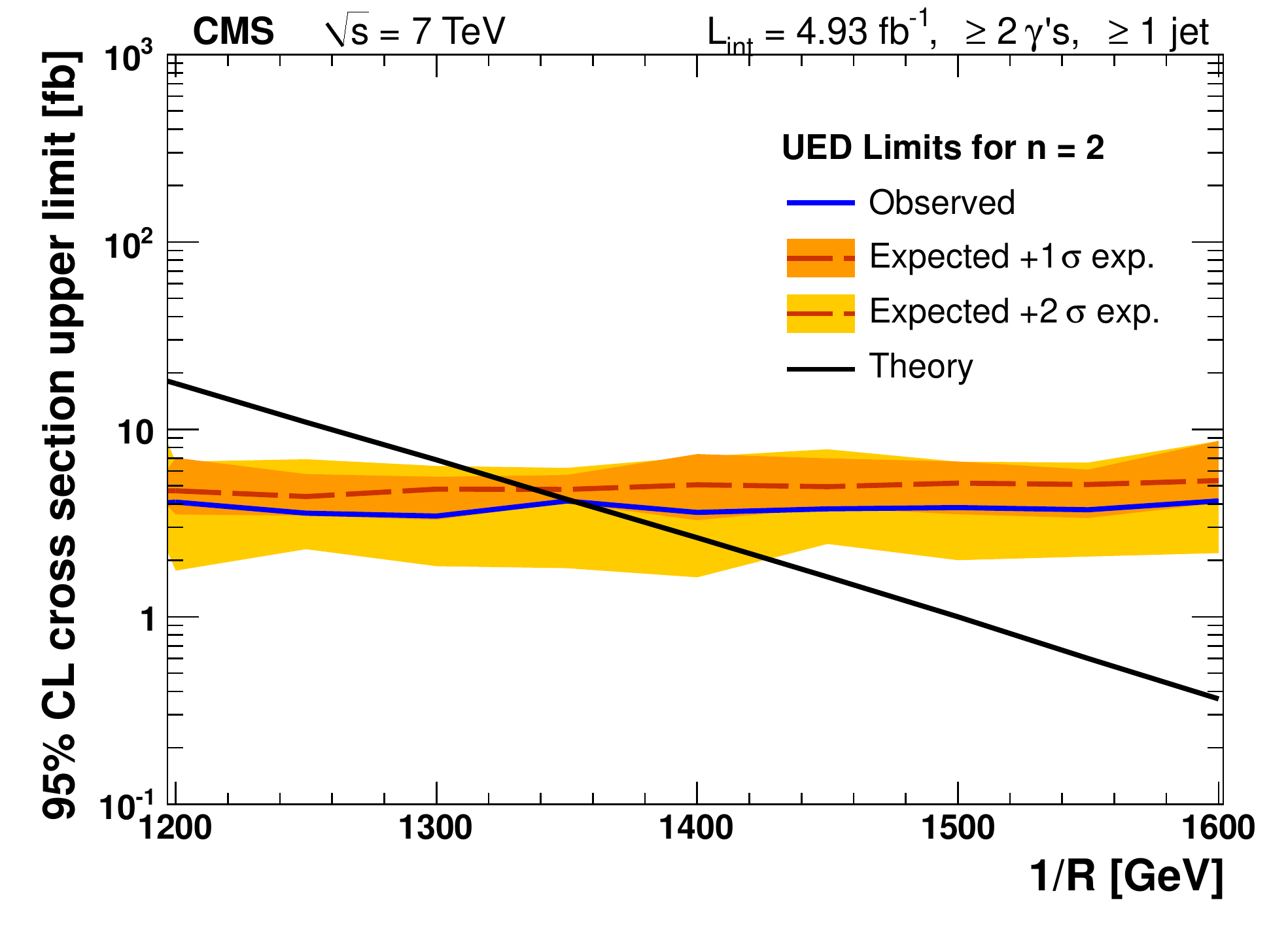}
\caption{\label{fig:UED3} 
  UED cross section upper limit for $n=2$ large ED at 95\% CL compared
  with expected UED production cross sections (black diagonal line). The
  shaded region shows the uncertainty due to PDFs and renormalization
  scale on the expected limit.  }
\end{center}
\end{figure*}

}
\cleardoublepage \appendix\section{The CMS Collaboration \label{app:collab}}\begin{sloppypar}\hyphenpenalty=5000\widowpenalty=500\clubpenalty=5000\textbf{Yerevan Physics Institute,  Yerevan,  Armenia}\\*[0pt]
S.~Chatrchyan, V.~Khachatryan, A.M.~Sirunyan, A.~Tumasyan
\vskip\cmsinstskip
\textbf{Institut f\"{u}r Hochenergiephysik der OeAW,  Wien,  Austria}\\*[0pt]
W.~Adam, E.~Aguilo, T.~Bergauer, M.~Dragicevic, J.~Er\"{o}, C.~Fabjan\cmsAuthorMark{1}, M.~Friedl, R.~Fr\"{u}hwirth\cmsAuthorMark{1}, V.M.~Ghete, J.~Hammer, N.~H\"{o}rmann, J.~Hrubec, M.~Jeitler\cmsAuthorMark{1}, W.~Kiesenhofer, V.~Kn\"{u}nz, M.~Krammer\cmsAuthorMark{1}, I.~Kr\"{a}tschmer, D.~Liko, I.~Mikulec, M.~Pernicka$^{\textrm{\dag}}$, B.~Rahbaran, C.~Rohringer, H.~Rohringer, R.~Sch\"{o}fbeck, J.~Strauss, A.~Taurok, W.~Waltenberger, G.~Walzel, E.~Widl, C.-E.~Wulz\cmsAuthorMark{1}
% ...
\vskip\cmsinstskip
\dag:~Deceased\\
1:~~Also at Vienna University of Technology, Vienna, Austria\\
% ...

\end{sloppypar}
\end{document}